\documentstyle[12pt,a4,here,epic,eepic]{article}

\makeatletter
	\def\@cite#1#2{\leavevmode\hbox{$^{\mbox{\the\scriptfont0 #1}}$}}
\makeatother

     \def\ks{k\mbox{\hspace{-0.95ex}}/}

	\def\half{ {1 \over 2} }
	\def\betamu{ {\beta\mu} }

\makeatletter
\@addtoreset{equation}{subsection}

\makeatother

\makeatletter
\long\def\@makecaption#1#2{%
   \vskip 10\p@
   \setbox\@tempboxa\hbox{#1\ \ #2}%
   \ifdim \wd\@tempboxa >\hsize
	#1\ \ #2\par		%
      \else
	\hbox to\hsize{\hfil\box\@tempboxa\hfil}%
   \fi}
\def\fnum@figure{Fig. \thefigure}
\makeatother

\begin{document}

\begin{flushright}
\begin{tabular}{l}
HUPD-9402 \hspace{0.5cm}\\
September 1994
\end{tabular}
\end{flushright}

\vspace{1cm}

\begin{center}
\Large
Phase Structure of Four-Fermion Theories \\
at Finite Temperature and Chemical Potential \\
in Arbitrary Dimensions \\[1cm]
\normalsize
T.~Inagaki
\footnote{e-mail: inagaki@theo.phys.sci.hiroshima-u.ac.jp}
, T.~Kouno and T.~Muta
\footnote{Supported in part by Monbusho Grant-in-Aid
for Scientific Research(C) under the contract \\
No. 04640301.} \\[0.5cm]
{\it
Department of Physics, Hiroshima University, \\
Higashi-Hiroshima, Hiroshima 724, Japan \\[3cm]
}
\end{center}

\begin{abstract}
     The phase structure of four-fermion theories is
thoroughly investigated with varying temperature and
chemical potential for arbitrary
space-time dimensions  $(2 \leq D < 4)$ by
using the $1/N$ expansion method. It is shown that the chiral
symmetry is restored in the theory under consideration for
sufficiently high temperature and/or chemical potential. The
critical line dividing the symmetric and broken phase is
given explicitly. It is found that for space-time dimension
$2 \leq D < 3$ both the first-order and second-order phase
transition occur depending on the value of temperature and
chemical potential while for $3 \leq D < 4$ only the second-order
phase transition exists.
\end{abstract}

\newpage


\renewcommand{\thesubsection}{\arabic{subsection}}
\renewcommand{\thesubsubsection}
	{\arabic{subsection}.\arabic{subsubsection}}
\baselineskip=20pt
\subsection{INTRODUCTION}

The idea of spontaneous symmetry breaking has played a
decisive role in recent developments of particle physics. In
fact the spontaneous breaking of the gauge symmetry
$SU(3)_{c} \otimes SU(2)_{L} \otimes U(1)_{Y}$
 is the basic ingredient of the standard
electroweak theory and grand unified theories (GUT) are
constructed on the basis of the Higgs mechanism. In these
approaches, however, the spontaneous breaking of symmetries
is introduced phenomenologically in terms of the Higgs
field. In other words the search for the dynamical origin of
the spontaneous symmetry breaking is passed over by
introducing elementary Higgs fields. Under this circumstance
it is very interesting to look for a possible dynamical
mechanism of the spontaneous symmetry breaking.

More than 30 years ago Y. Nambu and G. Jona-Lasinio first
introduced the idea of dynamical symmetry breaking in
particle physics\cite{NJL}
and the idea has attracted a vast number of researchers
in particle physics. One of the major applications of this
idea has been made on the construction of composite Higgs
models such as the technicolor model.\cite{TC} It is important to
test phenomenologically whether any of these composite Higgs
models is a candidate of the underlying theory of the
standard electroweak theory.

One of the possible environments where the composite Higgs
models may be tested is found in the early universe where
the symmetry of the primary unified theory is broken down to
yield lower-level theories which describe phenomena at
lower energy scales.  In the early universe it is not
adequate to neglect the effect of the curvature, temperature
and density. Hence it is important to examine a theory in
the finite curvature, temperature and density. In our study
we are interested in the spontaneous breaking of the
electroweak symmetry and hence we consider the scale much
smaller than the Planck scale. In such a situation the
essential feature of the early universe may be described
only with the effects of finite temperature and density, the
curvature effect being negligible. In the present
communication we would like to report our investigation in a
simple model field theory for composite Higgs fields at
finite temperature and chemical potential.

We work in a four-fermion theory in the flat space-time with
arbitrary dimensions. The theory is composed of $N$-component
fermions with four-fermion interactions. We employ the $1/N$
expansion method to estimate the effective potential for the
fermion-antifermion composite field. In two space-time
dimensions the theory reduces to the Gross-Neveu model\cite{GN}
which possesses the discrete chiral symmetry that
is broken dynamically. In three space-time dimensions the theory
defines a model which is solvable and renormalizable in the
sense of the $1/N$ expansion.\cite{D3REN} Beyond four space-time
dimensions the theory is nonrenormalizable. We confine
ourselves to the space-time dimensions greater or equal to 2
and less than 4.

We introduce the temperature and chemical potential in the
theory through the standard procedure. We investigate the
behavior of the effective potential in the leading order of
the $1/N$ expansion by varying the temperature and chemical
potential for arbitrary space-time dimensions. Through the
study of the shape of the effective potential and the
detailed analysis of the gap equation we observe the phase
transition from the phase with the broken chiral symmetry to
the phase with the chiral symmetry when the temperature and
chemical potential vary. With this analysis we shall be able
to derive explicitly the critical curve on the
temperature-chemical potential plane. We find that for
space-time dimensions less than 3 both the first-order and
second-order phase transition exist while for space-time
dimensions greater or equal to 3 only the second-order phase
transition is realized. At some specific points on the
critical curve we obtain an explicit expression for the
critical temperature and/or chemical potential. We also
calculate the dynamical fermion mass as a function of the
temperature and chemical potential.

In section 2 we briefly review the general properties of
four-fermion theories in space-time dimensions $2 \leq D < 4$
with vanishing temperature and chemical potential. In
section 3 we introduce the temperature and chemical
potential in the theory. We calculate the effective
potential in the leading order of the $1/N$ expansion and
observe the change of the phase. We derive the critical curve
which divide the chiral symmetric phase and the chiral
asymmetric phase.  The
section 4 is devoted for the concluding remarks.

\subsection{FOUR-FERMION THEORY IN ARBITRARY SPACE-TIME
DIMENSIONS}

In the present section we briefly summarize the
characteristic features of four-fermion theories in
arbitrary space-time dimensions $2 \leq D < 4$ for vanishing
temperature and chemical potential. We start with the simple
Lagrangian
\begin{equation}
     {\cal L} =
     \sum^{N}_{k=1}\bar{\psi}_{k}i\gamma_{\mu}\partial^{\mu}\psi_{k}
     -\frac{\lambda_0}{2N}\sum^{N}_{k=1}(\bar{\psi}_{k}\psi_{k})^{2}\, ,
\label{l:gn}
\end{equation}
where index $k$ represents the flavors of the fermion field
$\psi$, $N$ is the number of flavors and $\lambda_0$ is a
bare coupling constant.  In the following, for simplicity,
we neglect the flavor index. In two dimensions the theory is
nothing but the Gross-Neveu model\cite{GN} in which
Lagrangian (\ref{l:gn}) is invariant under the discrete
chiral transformation,
\begin{equation}
     \psi \longrightarrow \gamma_{5}\psi\, .
\label{t:dischi}
\end{equation}
The above discrete chiral symmetry in two
dimensions prevents the Lagrangian to have mass terms.
In arbitrary dimensions the transformation (\ref{t:dischi})
may be generalized so that $\bar\psi\psi \rightarrow -\bar\psi\psi$.

The theory has a global $SU(N)$ flavor symmetry : the
Lagrangian is invariant under the transformation,
\begin{equation}
     \psi \longrightarrow {\large e}^
{i{\scriptstyle \Sigma}_\alpha g_{a}T^{a}}\psi
\end{equation}
where $T^{a}$ are generators of the $SU(N)$ symmetry. Under the
circumstance of the global $SU(N)$ symmetry we may work in the
scheme of the $1/N$ expansion.

For practical calculations in four-fermion theories it is
convenient to introduce auxiliary field $\sigma$ and consider
the following equivalent Lagrangian
\begin{equation}
     {\cal L}_{y} = \bar{\psi}i\gamma_{\mu}\partial^{\mu}\psi
     -\frac{N}{2\lambda_0}\sigma^{2}-\bar{\psi}\sigma\psi\, .
\label{l:yukawa}
\end{equation}
If the non-vanishing vacuum expectation value is assigned to
the auxiliary field $\sigma$, then there appears a mass term
for the fermion field $\psi$ and the discrete chiral
symmetry(the $Z_2$ symmetry in odd dimensions) is eventually
broken.

We would like to find a ground state of the system described
by four-fermion theories.  For this purpose we evaluate an
effective potential for composite field $\bar\psi\psi$
in the theory described by Eq. (\ref{l:gn}).
This effective potential is essentially the
same as the one for field $\sigma$ in the theory described
by Eq. (\ref{l:yukawa}).  The effective
potential $V_0(\sigma)$ in the
leading order of the $1/N$ expansion reads (See Appendix A)
\begin{equation}
     V_{0}(\sigma ) = \frac{1}{2\lambda_0}\sigma^{2}
                  +i\, \mbox{ln\ det}(i \gamma_{\mu}\partial^{\mu}-\sigma)
                  +\mbox{O}(1/N)\, ,
\label{v:gn}
\end{equation}
where the suffix $0$ for $V_0(\sigma)$ is introduced to keep
the memory that $T=\mu=0$ with $T$ the temperature
and $\mu$ the chemical potential.

For integral dimensions the effective
potential is in general divergent.
Performing the integral in the potential (\ref{v:gn})
we obtain
\begin{equation}
     V_{0}(\sigma) = \frac{1}{2\lambda_0}\sigma^{2}
                 -\frac{1}{(2\pi)^{D/2}D}
                  \Gamma \left( 1-\frac{D}{2} \right)\sigma^{D} \, .
\label{v:nonren}
\end{equation}
We clearly see that the effective potential is
divergent in two and four dimensions. It happens to be
finite in three dimensions in the leading order of the $1/N$
expansion. If the next-to-leading order is taken into
account, it may be divergent in three dimensions $D=3$.

As is well-known, four-fermion theory is renormalizable in
two dimensions.  Therefore the potential
(\ref{v:nonren}) is made finite at $D=2$ by the usual
renormalization procedure. For $D=3$ four-fermion theory is
known to be renormalizable in the sense of the $1/N$
expansion.\cite{D3REN} The potential (\ref{v:nonren}) is
finite by itself and hence we do not need renormalization in
the leading order of the $1/N$ expansion at $D=3$.  In four
dimensions four-fermion theory is not renormalizable and the
finite effective potential can not be defined in four
dimensions. We regard the effective potential for
$D=4-\epsilon$ with $\epsilon$ sufficiently small positive
as a regularization of the one in four dimensions.

We perform renormalization in two dimensions by imposing
the renormalization condition,
\begin{equation}
     \left.
     \frac{\partial^{2}V_{0}(\sigma)}{\partial \sigma^{2}}
     \right|_{\sigma = \sigma_{0}}
     =\frac{1}{\lambda_{r}}\, .
\label{cond:ren}
\end{equation}
where $\sigma_{0}$ is the renormalization scale.  From
this renormalization condition we get
\begin{equation}
     \frac{1}{\lambda_0}=\frac{1}{\lambda_{r}}
                       +\frac{1}{(2\pi)^{D/2}}(D-1)
                        \Gamma \left( 1-\frac{D}{2} \right)
                        \sigma_{0}^{D-2}\, .
\label{eqn:ren}
\end{equation}
The renormalized effective potential reads
\begin{eqnarray}
     \frac{V_{0}(\sigma)}{\sigma_{0}^{D}}
     & = &  \frac{1}{2\lambda}\frac{\sigma^{2}}{\sigma_{0}^{2}}
                                                    \nonumber \\
     &   &  +\frac{1}{2(2\pi)^{D/2}}(D-1)
             \Gamma \left( 1-\frac{D}{2} \right)
             \frac{\sigma^{2}}{\sigma_{0}^{2}}
            -\frac{1}{(2\pi)^{D/2}D}
             \Gamma \left( 1-\frac{D}{2} \right)
             \frac{\sigma^{D}}{\sigma_{0}^{D}} \, .
\label{v:ren}
\end{eqnarray}
Here we used the dimensionless coupling constant $\lambda$
defined by
\begin{equation}
     \lambda = \lambda_{r}\sigma_{0}^{D-2}\, .
\label{eqn:ren2}
\end{equation}
The renormalized potential (\ref{v:ren}) is no longer divergent in
the whole range of $D$ considered here : $2\leq D < 4$.

\begin{figure}
\setlength{\unitlength}{0.240900pt}
\begin{picture}(1500,900)(0,0)
\tenrm
\thicklines \path(220,113)(240,113)
\thicklines \path(1436,113)(1416,113)
\put(198,113){\makebox(0,0)[r]{0}}
\thicklines \path(220,266)(240,266)
\thicklines \path(1436,266)(1416,266)
\put(198,266){\makebox(0,0)[r]{0.5}}
\thicklines \path(220,419)(240,419)
\thicklines \path(1436,419)(1416,419)
\put(198,419){\makebox(0,0)[r]{1}}
\thicklines \path(220,571)(240,571)
\thicklines \path(1436,571)(1416,571)
\put(198,571){\makebox(0,0)[r]{1.5}}
\thicklines \path(220,724)(240,724)
\thicklines \path(1436,724)(1416,724)
\put(198,724){\makebox(0,0)[r]{2}}
\thicklines \path(220,877)(240,877)
\thicklines \path(1436,877)(1416,877)
\put(198,877){\makebox(0,0)[r]{2.5}}
\thicklines \path(220,113)(220,133)
\thicklines \path(220,877)(220,857)
\put(220,68){\makebox(0,0){2}}
\thicklines \path(342,113)(342,133)
\thicklines \path(342,877)(342,857)
\put(342,68){\makebox(0,0){2.2}}
\thicklines \path(463,113)(463,133)
\thicklines \path(463,877)(463,857)
\put(463,68){\makebox(0,0){2.4}}
\thicklines \path(585,113)(585,133)
\thicklines \path(585,877)(585,857)
\put(585,68){\makebox(0,0){2.6}}
\thicklines \path(706,113)(706,133)
\thicklines \path(706,877)(706,857)
\put(706,68){\makebox(0,0){2.8}}
\thicklines \path(828,113)(828,133)
\thicklines \path(828,877)(828,857)
\put(828,68){\makebox(0,0){3}}
\thicklines \path(950,113)(950,133)
\thicklines \path(950,877)(950,857)
\put(950,68){\makebox(0,0){3.2}}
\thicklines \path(1071,113)(1071,133)
\thicklines \path(1071,877)(1071,857)
\put(1071,68){\makebox(0,0){3.4}}
\thicklines \path(1193,113)(1193,133)
\thicklines \path(1193,877)(1193,857)
\put(1193,68){\makebox(0,0){3.6}}
\thicklines \path(1314,113)(1314,133)
\thicklines \path(1314,877)(1314,857)
\put(1314,68){\makebox(0,0){3.8}}
\thicklines \path(1436,113)(1436,133)
\thicklines \path(1436,877)(1436,857)
\put(1436,68){\makebox(0,0){4}}
\thicklines \path(220,113)(1436,113)(1436,877)(220,877)(220,113)
\put(45,495){\makebox(0,0)[l]{\shortstack{$\lambda_c$}}}
\put(894,23){\makebox(0,0){$D$}}
\thinlines \path(221,115)(221,115)(222,117)(224,119)(225,121)(226,123)
(227,124)(229,126)(230,128)(231,130)(232,132)(233,134)(235,136)(236,138)
(237,140)(238,142)(239,143)(241,145)(242,147)(243,149)(244,151)(246,153)
(247,155)(248,156)(249,158)(250,160)(252,162)(253,164)(254,166)(255,168)
(257,169)(258,171)(259,173)(260,175)(261,177)(263,179)(264,180)(265,182)
(266,184)(267,186)(269,188)(270,190)(271,191)(272,193)(274,195)(275,197)
(276,199)(277,200)(278,202)(280,204)(281,206)
\thinlines \path(281,206)(282,208)(283,209)(285,211)(286,213)(287,215)
(288,217)(289,218)(291,220)(292,222)(293,224)(294,225)(295,227)(297,229)
(298,231)(299,233)(300,234)(302,236)(303,238)(304,240)(305,241)(306,243)
(308,245)(309,247)(310,248)(311,250)(313,252)(314,254)(315,255)(316,257)
(317,259)(319,261)(320,262)(321,264)(322,266)(323,267)(325,269)(326,271)
(327,273)(328,274)(330,276)(331,278)(332,279)(333,281)(334,283)(336,285)
(337,286)(338,288)(339,290)(341,291)(342,293)
\thinlines \path(342,293)(343,295)(344,297)(345,298)(347,300)(348,302)
(349,303)(350,305)(351,307)(353,308)(354,310)(355,312)(356,313)(358,315)
(359,317)(360,318)(361,320)(362,322)(364,323)(365,325)(366,327)(367,328)
(369,330)(370,332)(371,333)(372,335)(373,337)(375,338)(376,340)(377,342)
(378,343)(379,345)(381,346)(382,348)(383,350)(384,351)(386,353)(387,355)
(388,356)(389,358)(390,359)(392,361)(393,363)(394,364)(395,366)(396,368)
(398,369)(399,371)(400,372)(401,374)(403,376)
\thinlines \path(403,376)(404,377)(405,379)(406,380)(407,382)(409,384)
(410,385)(411,387)(412,388)(414,390)(415,392)(416,393)(417,395)(418,396)
(420,398)(421,399)(422,401)(423,403)(424,404)(426,406)(427,407)(428,409)
(429,410)(431,412)(432,414)(433,415)(434,417)(435,418)(437,420)(438,421)
(439,423)(440,424)(442,426)(443,428)(444,429)(445,431)(446,432)(448,434)
(449,435)(450,437)(451,438)(452,440)(454,441)(455,443)(456,444)(457,446)
(459,448)(460,449)(461,451)(462,452)(463,454)
\thinlines \path(463,454)(465,455)(466,457)(467,458)(468,460)(470,461)
(471,463)(472,464)(473,466)(474,467)(476,469)(477,470)(478,472)(479,473)
(480,475)(482,476)(483,478)(484,479)(485,481)(487,482)(488,483)(489,485)
(490,486)(491,488)(493,489)(494,491)(495,492)(496,494)(498,495)(499,497)
(500,498)(501,500)(502,501)(504,502)(505,504)(506,505)(507,507)(508,508)
(510,510)(511,511)(512,513)(513,514)(515,515)(516,517)(517,518)(518,520)
(519,521)(521,523)(522,524)(523,525)(524,527)
\thinlines \path(524,527)(526,528)(527,530)(528,531)(529,532)(530,534)
(532,535)(533,537)(534,538)(535,539)(536,541)(538,542)(539,544)(540,545)
(541,546)(543,548)(544,549)(545,551)(546,552)(547,553)(549,555)(550,556)
(551,557)(552,559)(554,560)(555,562)(556,563)(557,564)(558,566)(560,567)
(561,568)(562,570)(563,571)(564,572)(566,574)(567,575)(568,576)(569,578)
(571,579)(572,580)(573,582)(574,583)(575,584)(577,586)(578,587)(579,588)
(580,590)(582,591)(583,592)(584,593)(585,595)
\thinlines \path(585,595)(586,596)(588,597)(589,599)(590,600)(591,601)
(592,603)(594,604)(595,605)(596,606)(597,608)(599,609)(600,610)(601,611)
(602,613)(603,614)(605,615)(606,617)(607,618)(608,619)(610,620)(611,622)
(612,623)(613,624)(614,625)(616,627)(617,628)(618,629)(619,630)(620,631)
(622,633)(623,634)(624,635)(625,636)(627,638)(628,639)(629,640)(630,641)
(631,642)(633,644)(634,645)(635,646)(636,647)(638,648)(639,650)(640,651)
(641,652)(642,653)(644,654)(645,655)(646,657)
\thinlines \path(646,657)(647,658)(648,659)(650,660)(651,661)(652,662)
(653,664)(655,665)(656,666)(657,667)(658,668)(659,669)(661,670)(662,671)
(663,673)(664,674)(666,675)(667,676)(668,677)(669,678)(670,679)(672,680)
(673,682)(674,683)(675,684)(676,685)(678,686)(679,687)(680,688)(681,689)
(683,690)(684,691)(685,692)(686,693)(687,694)(689,696)(690,697)(691,698)
(692,699)(693,700)(695,701)(696,702)(697,703)(698,704)(700,705)(701,706)
(702,707)(703,708)(704,709)(706,710)(707,711)
\thinlines \path(707,711)(708,712)(709,713)(711,714)(712,715)(713,716)
(714,717)(715,718)(717,719)(718,720)(719,721)(720,722)(721,723)(723,724)
(724,725)(725,726)(726,727)(728,728)(729,729)(730,730)(731,730)(732,731)
(734,732)(735,733)(736,734)(737,735)(739,736)(740,737)(741,738)(742,739)
(743,740)(745,741)(746,741)(747,742)(748,743)(749,744)(751,745)(752,746)
(753,747)(754,748)(756,748)(757,749)(758,750)(759,751)(760,752)(762,753)
(763,753)(764,754)(765,755)(767,756)(768,757)
\thinlines \path(768,757)(769,758)(770,758)(771,759)(773,760)(774,761)
(775,762)(776,762)(777,763)(779,764)(780,765)(781,766)(782,766)(784,767)
(785,768)(786,769)(787,769)(788,770)(790,771)(791,772)(792,772)(793,773)
(795,774)(796,774)(797,775)(798,776)(799,777)(801,777)(802,778)(803,779)
(804,779)(805,780)(807,781)(808,781)(809,782)(810,783)(812,783)(813,784)
(814,785)(815,785)(816,786)(818,787)(819,787)(820,788)(821,789)(823,789)
(824,790)(825,790)(826,791)(827,792)(829,792)
\thinlines \path(829,792)(830,793)(831,793)(832,794)(833,794)(835,795)
(836,796)(837,796)(838,797)(840,797)(841,798)(842,798)(843,799)(844,799)
(846,800)(847,800)(848,801)(849,802)(851,802)(852,803)(853,803)(854,803)
(855,804)(857,804)(858,805)(859,805)(860,806)(861,806)(863,807)(864,807)
(865,808)(866,808)(868,809)(869,809)(870,809)(871,810)(872,810)(874,811)
(875,811)(876,811)(877,812)(879,812)(880,813)(881,813)(882,813)(883,814)
(885,814)(886,814)(887,815)(888,815)(889,815)
\thinlines \path(889,815)(891,816)(892,816)(893,816)(894,817)(896,817)
(897,817)(898,817)(899,818)(900,818)(902,818)(903,819)(904,819)(905,819)
(907,819)(908,820)(909,820)(910,820)(911,820)(913,820)(914,821)(915,821)
(916,821)(917,821)(919,821)(920,822)(921,822)(922,822)(924,822)(925,822)
(926,822)(927,823)(928,823)(930,823)(931,823)(932,823)(933,823)(935,823)
(936,823)(937,824)(938,824)(939,824)(941,824)(942,824)(943,824)(944,824)
(945,824)(947,824)(948,824)(949,824)(950,824)
\thinlines \path(950,824)(952,824)(953,824)(954,824)(955,824)(956,824)
(958,824)(959,824)(960,824)(961,824)(963,824)(964,824)(965,824)(966,824)
(967,824)(969,824)(970,824)(971,824)(972,823)(973,823)(975,823)(976,823)
(977,823)(978,823)(980,823)(981,823)(982,822)(983,822)(984,822)(986,822)
(987,822)(988,822)(989,821)(990,821)(992,821)(993,821)(994,821)(995,820)
(997,820)(998,820)(999,820)(1000,819)(1001,819)(1003,819)(1004,818)(1005,818)
(1006,818)(1008,818)(1009,817)(1010,817)(1011,817)
\thinlines \path(1011,817)(1012,816)(1014,816)(1015,816)(1016,815)(1017,815)
(1018,814)(1020,814)(1021,814)(1022,813)(1023,813)(1025,812)(1026,812)
(1027,812)(1028,811)(1029,811)(1031,810)(1032,810)(1033,809)(1034,809)
(1036,808)(1037,808)(1038,807)(1039,807)(1040,806)(1042,806)(1043,805)
(1044,805)(1045,804)(1046,804)(1048,803)(1049,803)(1050,802)(1051,801)
(1053,801)(1054,800)(1055,800)(1056,799)(1057,798)(1059,798)(1060,797)
(1061,796)(1062,796)(1064,795)(1065,794)(1066,794)(1067,793)(1068,792)
(1070,792)(1071,791)(1072,790)
\thinlines \path(1072,790)(1073,789)(1074,789)(1076,788)(1077,787)
(1078,786)(1079,786)(1081,785)(1082,784)(1083,783)(1084,782)(1085,782)
(1087,781)(1088,780)(1089,779)(1090,778)(1092,777)(1093,777)(1094,776)
(1095,775)(1096,774)(1098,773)(1099,772)(1100,771)(1101,770)(1102,769)
(1104,768)(1105,767)(1106,766)(1107,765)(1109,764)(1110,763)(1111,762)
(1112,761)(1113,760)(1115,759)(1116,758)(1117,757)(1118,756)(1120,755)
(1121,754)(1122,753)(1123,752)(1124,751)(1126,750)(1127,748)(1128,747)
(1129,746)(1130,745)(1132,744)(1133,743)
\thinlines \path(1133,743)(1134,741)(1135,740)(1137,739)(1138,738)
(1139,737)(1140,735)(1141,734)(1143,733)(1144,732)(1145,730)(1146,729)
(1148,728)(1149,727)(1150,725)(1151,724)(1152,723)(1154,721)(1155,720)
(1156,719)(1157,717)(1158,716)(1160,714)(1161,713)(1162,712)(1163,710)
(1165,709)(1166,707)(1167,706)(1168,704)(1169,703)(1171,701)(1172,700)
(1173,699)(1174,697)(1176,695)(1177,694)(1178,692)(1179,691)(1180,689)
(1182,688)(1183,686)(1184,685)(1185,683)(1186,681)(1188,680)(1189,678)
(1190,677)(1191,675)(1193,673)(1194,672)
\thinlines \path(1194,672)(1195,670)(1196,668)(1197,666)(1199,665)
(1200,663)(1201,661)(1202,660)(1204,658)(1205,656)(1206,654)(1207,653)
(1208,651)(1210,649)(1211,647)(1212,645)(1213,643)(1214,642)(1216,640)
(1217,638)(1218,636)(1219,634)(1221,632)(1222,630)(1223,628)(1224,626)
(1225,625)(1227,623)(1228,621)(1229,619)(1230,617)(1232,615)(1233,613)
(1234,611)(1235,609)(1236,607)(1238,604)(1239,602)(1240,600)(1241,598)
(1242,596)(1244,594)(1245,592)(1246,590)(1247,588)(1249,586)(1250,583)
(1251,581)(1252,579)(1253,577)(1255,575)
\thinlines \path(1255,575)(1256,572)(1257,570)(1258,568)(1260,566)
(1261,563)(1262,561)(1263,559)(1264,557)(1266,554)(1267,552)(1268,550)
(1269,547)(1270,545)(1272,543)(1273,540)(1274,538)(1275,535)(1277,533)
(1278,531)(1279,528)(1280,526)(1281,523)(1283,521)(1284,518)(1285,516)
(1286,513)(1287,511)(1289,508)(1290,506)(1291,503)(1292,501)(1294,498)
(1295,495)(1296,493)(1297,490)(1298,488)(1300,485)(1301,482)(1302,480)
(1303,477)(1305,474)(1306,472)(1307,469)(1308,466)(1309,463)(1311,461)
(1312,458)(1313,455)(1314,452)(1315,450)
\thinlines \path(1315,450)(1317,447)(1318,444)(1319,441)(1320,438)
(1322,436)(1323,433)(1324,430)(1325,427)(1326,424)(1328,421)(1329,418)
(1330,415)(1331,412)(1333,409)(1334,406)(1335,403)(1336,400)(1337,397)
(1339,394)(1340,391)(1341,388)(1342,385)(1343,382)(1345,379)(1346,376)
(1347,373)(1348,370)(1350,367)(1351,364)(1352,361)(1353,357)(1354,354)
(1356,351)(1357,348)(1358,345)(1359,341)(1361,338)(1362,335)(1363,332)
(1364,328)(1365,325)(1367,322)(1368,318)(1369,315)(1370,312)(1371,308)
(1373,305)(1374,302)(1375,298)(1376,295)
\thinlines \path(1376,295)(1378,292)(1379,288)(1380,285)(1381,281)
(1382,278)(1384,274)(1385,271)(1386,267)(1387,264)(1389,260)(1390,257)
(1391,253)(1392,250)(1393,246)(1395,242)(1396,239)(1397,235)(1398,232)
(1399,228)(1401,224)(1402,221)(1403,217)(1404,213)(1406,210)(1407,206)
(1408,202)(1409,198)(1410,195)(1412,191)(1413,187)(1414,183)(1415,180)
(1417,176)(1418,172)(1419,168)(1420,164)(1421,160)(1423,157)(1424,153)
(1425,149)(1426,145)(1427,141)(1429,137)(1430,133)(1431,129)(1432,125)
(1434,121)(1435,117)
\end{picture}

\caption{Critical coupling $\lambda_c$ of the four-fermion theory as a
function of dimension $D$ .}
\label{fig:crcpl}
\end{figure}
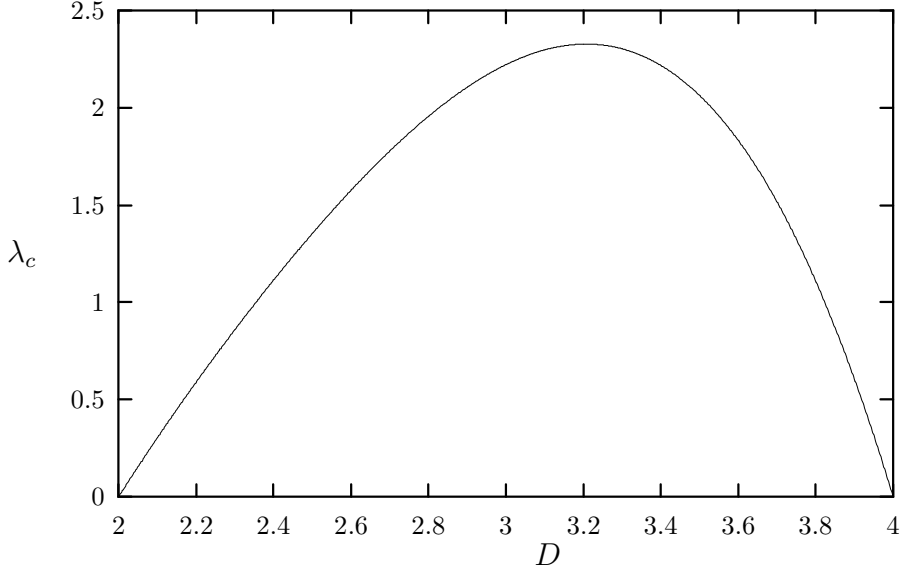

The ground state of the theory is determined by observing
the minimum of the effective potential. The necessary
condition for the minimum is given by
\begin{equation}
     \left.
     \frac{\partial V_{0}(\sigma)}{\partial \sigma}
     \right|_{\sigma = m} =
	m\sigma_0^{D-2}
	\left[
		{1 \over \lambda} - {1 \over \lambda_c}
		- \frac{\Gamma\left(1-{\displaystyle{ D \over 2}}\right)}
			{(2\pi)^{D/2}}
		\left({m \over \sigma_0}\right)^{D-2}
	\right] =
	0\, .
\label{eqn:gap}
\end{equation}
where $m$ is the dynamical fermion mass
and  $\lambda_{c}$ is defined by
\begin{equation}
     \lambda_{c} = (2\pi)^{D/2}
                        \left[
                        (1-D)\Gamma \left( 1-\frac{D}{2} \right)
                        \right]^{-1}\, .
\end{equation}
If the coupling constant $\lambda$ is no less than
 a critical value $\lambda_{c}$, the gap equation allows a non-trivial
solution which is given by
\begin{equation}
     m = \sigma_0
	\left[
                \frac{(2\pi)^{D/2}}
			{\Gamma \left( 1-{\displaystyle{ D \over 2}} \right)}
                \left(
                \frac{1}{\lambda}-\frac{1}{\lambda_{c}}
                \right)
          \right]^{1/(2-D)} \, .
\label{mass:d}
\end{equation}
In Fig. \ref{fig:crcpl}. we plot the critical coupling
constant $\lambda_c$ as a function of dimension $D$ .
The non-trivial solution $m$ of the gap equation
(\ref{eqn:gap}) corresponds to the nonvanishing vacuum
expectation value of the composite field $\sigma$ and is
equal to the dynamically generated fermion mass.
For some special values of $D$ the solution $m$ simplifies :
\begin{eqnarray}
     m = \sigma_0 {\,\large e}^{1-{\pi/\lambda}}
		\hspace{3em}; D=2\, , \\
     m = \sigma_0 \left(2-\frac{ \sqrt{\pi} }{ \lambda }\right)
		\hspace{2em}; D=3\, .
\end{eqnarray}
and $\lambda_c$ also simplifies :
\begin{eqnarray}
	\lambda_c = 0
		\hspace{3em}; D=2\, , \\
	\lambda_c = \frac{\pi}{\sqrt{2}}
		\hspace{2em}; D=3\, .
\end{eqnarray}
\noindent
As is well-known, the shape of the effective potential
$V_0(\sigma)$ is of single-well for $\lambda<\lambda_c$
while it is of double-well for $\lambda>\lambda_c$.

The renormalization group $\beta$ function is found to be\cite{GNRENG}
\begin{eqnarray}
\beta(\lambda) & = &
	\sigma_0 \left.
		\frac{ d\lambda }{ d\sigma_0 }
	\right|_{\lambda_0}\, , \\
& = &
	{D-2 \over \lambda_c}\lambda
	( \lambda_c - \lambda )\, ,
\label{eq:rgbeta}
\end{eqnarray}
in the leading order of the $1/N$ expansion. It simplifies
for some special values of $D$ :
\begin{eqnarray}
\beta(\lambda) & = &
	- {\lambda^2 \over \pi}
		\hspace{7em}; D=2\, , \\
& = &
	-\lambda\left(
		\frac{ \sqrt{2}\lambda }{\pi}-1
	\right)
		\hspace{2em}; D=3\, .
\end{eqnarray}
We find from Eq. (\ref{eq:rgbeta}) that $\lambda=\lambda_c$ is
the ultraviolet stable fixed point. In particular for $D=2$
the theory is asymptotically free and for $D=3$ the theory
has a nontrivial ultraviolet fixed point
at $\lambda=\displaystyle\frac{\pi}{\sqrt{2}}$ .

We turn our attention to the theory governed by the
following Lagrangian,
\begin{equation}
     {\cal L} =
     \bar{\psi}i\gamma_{\mu}\partial^{\mu}\psi
     -\frac{\lambda_0}{2N}[(\bar{\psi}\psi)^{2}
                        +(\bar{\psi}i\gamma_{5}\psi)^{2}]\, .
\label{l:njl}
\end{equation}
In four dimensions this Lagrangian defines the
Nambu-Jona-Lasinio model.\cite{NJL} The Lagrangian is
invariant under the chiral
$U(1)$ transformation in even dimensions,
\begin{equation}
     \psi \longrightarrow {\large e}^{i\theta\gamma_{5}}\psi\, .
\end{equation}
The chiral $U(1)$ symmetry prevents the Lagrangian to have
mass terms. Using the auxiliary
field method, we obtain the effective potential
\begin{equation}
     V_{0}(\sigma' ) = \frac{1}{2\lambda_0}{\sigma'}^{2}
                  +i \mbox{ln det}(i \gamma_{\mu}\partial^{\mu}-\sigma')
                  +O(1/N)\, ,
\label{v:njl}
\end{equation}
\begin{equation}
     \sigma'=\sqrt{\sigma^{2}+\pi^{2}}\, .
\end{equation}
Corresponding to the two four-fermion
interaction terms in the Lagrangian (\ref{l:njl}),
we need two kinds of auxiliary
fields $\sigma$ and $\pi$. In two dimensions it is
well-known that $\pi$ loop has an infra-red divergence and we can
not neglect the next to leading order terms of
the $1/N$ expansion.\cite{D2NJL} Therefore our present analysis is not
sufficient to deal with the
Lagrangian (\ref{l:njl}) in two dimensions.

The effective potentials (\ref{v:gn}) and (\ref{v:njl}) have
the same form in the leading order of the $1/N$ expansion.
Thus using the effective potential (\ref{v:gn}), we can
discuss the phase structure of both theories except for the
case in two dimensions.

\subsection{PHASE STRUCTURE AT FINITE TEMPERATURE AND CHEMICAL POTENTIAL}

\subsubsection{Effective potential and gap equation}

As we have seen in the previous section the chiral symmetry is
broken above the critical coupling. The
symmetry broken spontaneously may be restored
in an environment of high
temperature and density. To see whether this situation is
realized we would like to study the phase
structure at finite temperature and chemical potential.

The $n$-point Green function at finite temperature and chemical
potential is defined by
\begin{equation}
     G^{\beta \mu}_{n}
     = \frac{\sum_{\alpha}{\large e}^{-\beta(E_{\alpha}-\mu N_{\alpha})}
             \langle \alpha | \mbox{T}(\phi(x_{1})\,\cdots\,\phi(x_{n}))
             | \alpha \rangle}
            {\sum_{\alpha}{\large e}^{-\beta(E_{\alpha}-\mu N_{\alpha})}}\, ,
\label{eqn:gfunc}
\end{equation}
where $\phi(x)$ represents a component of
fermion field $\psi(x)$ or $\bar\psi(x)$, $E_\alpha$ and
\ $N_\alpha$ are the energy and particle number in the state
specified by quantum number $\alpha$ respectively,
$\beta=\displaystyle\frac{1}{kT}$ with $k$ the Boltzmann
Constant and $T$ the temperature
and $\mu$ is the chemical potential. The generating
functional for Green functions (\ref{eqn:gfunc}) reads
\begin{equation}
    Z^{\beta\mu}[J]
	= {\rm Tr}\left[
		{\large e}^{-\beta(H-\mu N)}{\rm T}
		\exp i\!\int\! dx J(x)\phi(x)
	\right]\, ,
\label{def:gfunc}
\end{equation}
where $H$ is the Hamiltonian, $N$ is the number operator
and $J(x)$ is the source function for field $\phi(x)$.

Following the standard procedure of the Matsubara Green
function\cite{EFFT} (See Appendix B) we calculate the effective potential in
our theory
in the leading order of the 1/N expansion as described
already in the case of \ $T=\mu=0$\  in Eq. (\ref{v:gn}). We find
\begin{eqnarray}
     V(\sigma) & = &
     \frac{1}{2\lambda_0}\sigma^{2}
     + i\frac{1}{\beta}\sum_{n}\int\frac{d^{D-1}k}{(2\pi)^{D-1}}
     \mbox{ln\ det}\frac{\ks - \sigma}{\ks}\, ,
\label{v:lndet}
\end{eqnarray}
where the four-momentum $k^\mu$ is given by
\begin{equation}
     k^{\mu} = (i\omega_{n}-\mu,\vec{k})\, ,
\end{equation}
and the discrete variable $\omega_n$ given by
\begin{equation}
     \omega_{n}=\frac{(2n+1)\pi}{\beta}\, ,
\end{equation}
according to the anti-periodic boundary condition.

Performing a summation and integrating over angle variables
(See Appendix B) we get
\begin{equation}
     V(\sigma) = V_{0}(\sigma) + V^{\beta \mu}(\sigma)\, ,
\label{v:full}
\end{equation}
where $V_{0}(\sigma)$ is the effective potential
for $T=\mu=0$ shown in Eq. (\ref{v:nonren}) and
$V^{\beta \mu}(\sigma)$ is given by
\begin{eqnarray}
     V^{\beta \mu}(\sigma) & = &
     -\frac{1}{\beta}\frac{\sqrt{2}}{(2\pi)^{(D-1)/2}}
      \frac{1}{\Gamma \left(\displaystyle\frac{D-1}{2}\right)}
		\sigma_{0}^{D}
                                                  \nonumber \\
      & & \times\int dk k^{D-2}\left\{ \mbox{ln}
      \frac{1+{\large e}^{-\beta(\sqrt{k^{2}+\sigma^{2}}+\mu)}}
           {1+{\large e}^{-\beta(k+\mu)}}
	 + \ln
      \frac{1+{\large e}^{-\beta(\sqrt{k^{2}+\sigma^{2}}-\mu)}}
           {1+{\large e}^{-\beta(k-\mu)}} \right\} \, .
\label{v:exp}
\end{eqnarray}
It is found that $V^{\beta \mu}(\sigma)$ is finite
in the space-time dimension $2 \leq D < 4$. Thus
we need not renormalize the parts including the effects of
finite temperature and density.  We apply the same
renormalization condition
as shown in Eq. (\ref{cond:ren})
and obtain the renormalized effective potential
as in the $T=\mu=0$ case. The renormalized
effective potential $V(\sigma)$ is given by
replacing $V_0$ in Eq. (\ref{v:full}) by the one
in Eq. (\ref{v:ren}).

If we perform the integration first
in Eq. (\ref{v:lndet}) and leave the
infinite sum we find
\begin{eqnarray}
    V(\sigma)
	&=& \frac{\sigma^2}{2\lambda_0}
	+ \frac{2^{ D/2 -1 }}{\beta}
	\frac{\Gamma\left( {\displaystyle{1-D \over 2}} \right)}
		{(4\pi)^{(D-1)/2}}
						\nonumber \\
    & &	\mbox{}\times\sum^\infty_{n=-\infty}
	\left[
		\{ (\omega_n + i\mu)^2 + \sigma^2 \}^{(D-1)/2}
		-\{ (\omega_n + i\mu)^2 \}^{(D-1)/2}
	\right]\, .
\label{v:sum}
\end{eqnarray}
Although the above expression is essentially equivalent to
the one in Eq. (\ref{v:exp}), we shall often use
the expression (\ref{v:sum}) in
the later argument of the gap equation.

     The gap equation is obtained through the stationary condition
\begin{equation}
     \left.
     \frac{\partial V(\sigma)}{\partial \sigma}
     \right|_{\sigma = m_{\beta \mu}} = 0\, .
\label{eq:gap}
\end{equation}
where $m_{\beta\mu}$ is a possible dynamical mass of the fermion.
If we apply the
expression (\ref{v:exp}) in Eq. (\ref{eq:gap}), we find
\begin{eqnarray}
      \frac{1}{\lambda}&-&\frac{1}{\lambda_c}
      -\frac{1}{(2 \pi)^{D/2}}\Gamma \left(
      1-\frac{D}{2}\right)\left( \frac{m_{\beta \mu}}{\sigma_{0}}
      \right)^{D-2}
					    \nonumber \\
      &+&\frac{\sqrt{2}}{\Gamma((D-1)/2)(2 \pi)^{(D-1)/2}}
      \sigma_{0}^{2-D}
					\nonumber \\
      &\times&\int^{\infty}_{0}dk k^{D-2}
      \frac{1}{\sqrt{k^{2}+m_{\beta \mu}^{2}}}
      \left\{
      \frac{1}{1+{\large e}^{\beta(\sqrt{k^{2}+m_{\beta \mu}^{2}}+\mu)}}
      +\frac{1}{1+{\large e}^{\beta(\sqrt{k^{2}+m_{\beta \mu}^{2}}-\mu)}}
      \right\}                              \nonumber \\
      &=& 0\, .
\label{gapeqn:tm}
\end{eqnarray}
In Eq. (\ref{gapeqn:tm}) we neglected the trivial
solution $m_{\beta\mu}=0$. If Eq. (\ref{v:sum}) is employed, the
following expression is obtained,
\begin{eqnarray}
	\frac{1}{\lambda}-\frac{1}{\lambda_c}
	&-& \frac{2^{D/2}}{\beta\sigma_0}
	\frac{\Gamma\left({\displaystyle{3-D \over 2}}\right)}
		{(4 \pi)^{(D-1)/2}}
	\sum_{n=-\infty}^\infty
	\left\{
		\frac{ (\omega_n+i\mu)^2 + m_{\beta\mu}^2 }
			{\sigma_0^2}
	\right\}
	^{(D-3)/2} = 0 \, .
\label{gap:nontri}
\end{eqnarray}

\subsubsection{Phase boundary}

 The expression (\ref{v:exp}) is useful to calculate
the effective potential numerically as a function
of the composite field $\sigma$ . Using Eq. (\ref{v:exp}) we
calculate the effective potential for $D=2$ to find
the behavior of the effective potential\cite{GNTM}
as shown in Fig. \ref{fig:2D}.
\begin{figure}
\hspace*{-2em}
	\begin{minipage}[t]{.47\linewidth}
\setlength{\unitlength}{0.240900pt}
\begin{picture}(1049,900)(0,0)
\tenrm
\thicklines \path(220,113)(240,113)
\thicklines \path(985,113)(965,113)
\put(198,113){\makebox(0,0)[r]{0}}
\thicklines \path(220,266)(240,266)
\thicklines \path(985,266)(965,266)
\put(198,266){\makebox(0,0)[r]{0.2}}
\thicklines \path(220,419)(240,419)
\thicklines \path(985,419)(965,419)
\put(198,419){\makebox(0,0)[r]{0.4}}
\thicklines \path(220,571)(240,571)
\thicklines \path(985,571)(965,571)
\put(198,571){\makebox(0,0)[r]{0.6}}
\thicklines \path(220,724)(240,724)
\thicklines \path(985,724)(965,724)
\put(198,724){\makebox(0,0)[r]{0.8}}
\thicklines \path(220,877)(240,877)
\thicklines \path(985,877)(965,877)
\put(198,877){\makebox(0,0)[r]{1}}
\thicklines \path(220,113)(220,133)
\thicklines \path(220,877)(220,857)
\put(220,68){\makebox(0,0){0}}
\thicklines \path(373,113)(373,133)
\thicklines \path(373,877)(373,857)
\put(373,68){\makebox(0,0){0.2}}
\thicklines \path(526,113)(526,133)
\thicklines \path(526,877)(526,857)
\put(526,68){\makebox(0,0){0.4}}
\thicklines \path(679,113)(679,133)
\thicklines \path(679,877)(679,857)
\put(679,68){\makebox(0,0){0.6}}
\thicklines \path(832,113)(832,133)
\thicklines \path(832,877)(832,857)
\put(832,68){\makebox(0,0){0.8}}
\thicklines \path(985,113)(985,133)
\thicklines \path(985,877)(985,857)
\put(985,68){\makebox(0,0){1}}
\thicklines \path(220,113)(985,113)(985,877)(220,877)(220,113)
\put(133,945){\makebox(0,0)[l]{\shortstack{$\mu/m$}}}
\put(602,23){\makebox(0,0){$kT/ m$}}
\put(909,319){\makebox(0,0)[l]{A}}
\put(335,801){\makebox(0,0)[l]{B}}
\put(526,266){\vector(1,0){383}}
\put(373,419){\vector(0,1){382}}
\end{picture}

		\hspace*{4em}(a) The $T$-$\mu$ plane showing\\
		\hspace*{5em}the fixed $\mu$ direction A\\
		\hspace*{5em}and the fixed $T$ direction B.
		\vspace{5ex}
\setlength{\unitlength}{0.240900pt}
\begin{picture}(1049,900)(0,0)
\tenrm
\thinlines \dashline[-10]{25}(220,495)(985,495)
\thicklines \path(220,189)(240,189)
\thicklines \path(985,189)(965,189)
\put(198,189){\makebox(0,0)[r]{-0.004}}
\thicklines \path(220,342)(240,342)
\thicklines \path(985,342)(965,342)
\put(198,342){\makebox(0,0)[r]{-0.002}}
\thicklines \path(220,495)(240,495)
\thicklines \path(985,495)(965,495)
\put(198,495){\makebox(0,0)[r]{0}}
\thicklines \path(220,648)(240,648)
\thicklines \path(985,648)(965,648)
\put(198,648){\makebox(0,0)[r]{0.002}}
\thicklines \path(220,801)(240,801)
\thicklines \path(985,801)(965,801)
\put(198,801){\makebox(0,0)[r]{0.004}}
\thicklines \path(220,113)(220,133)
\thicklines \path(220,877)(220,857)
\put(220,68){\makebox(0,0){0}}
\thicklines \path(373,113)(373,133)
\thicklines \path(373,877)(373,857)
\put(373,68){\makebox(0,0){0.2}}
\thicklines \path(526,113)(526,133)
\thicklines \path(526,877)(526,857)
\put(526,68){\makebox(0,0){0.4}}
\thicklines \path(679,113)(679,133)
\thicklines \path(679,877)(679,857)
\put(679,68){\makebox(0,0){0.6}}
\thicklines \path(832,113)(832,133)
\thicklines \path(832,877)(832,857)
\put(832,68){\makebox(0,0){0.8}}
\thicklines \path(985,113)(985,133)
\thicklines \path(985,877)(985,857)
\put(985,68){\makebox(0,0){1}}
\thicklines \path(220,113)(985,113)(985,877)(220,877)(220,113)
\put(133,945){\makebox(0,0)[l]{\shortstack{$V/m^2$}}}
\put(602,23){\makebox(0,0){$\sigma/m$}}
\thinlines \path(220,495)(220,495)(221,495)(221,495)(222,495)(223,495)
(224,495)(226,495)(227,495)(229,495)(232,495)(235,495)(238,496)(243,496)
(255,497)(267,499)(290,504)(314,511)(337,519)(361,528)(384,538)(407,549)
(431,559)(454,570)(478,579)(501,587)(513,590)(524,592)(536,595)(542,596)
(548,596)(554,597)(557,597)(560,597)(561,597)(562,597)(564,597)(565,597)
(566,597)(567,597)(568,597)(568,597)(569,597)(570,598)(571,598)(571,598)
(572,598)(573,597)(573,597)(574,597)(576,597)
\thinlines \path(576,597)(577,597)(580,597)(583,597)(589,597)(595,596)
(601,595)(606,594)(618,592)(642,585)(665,575)(688,564)(712,550)(735,536)
(759,522)(782,509)(794,503)(805,498)(817,494)(823,492)(829,491)(832,490)
(835,490)(838,489)(841,489)(842,489)(844,489)(845,489)(846,489)(846,489)
(847,488)(848,488)(849,488)(849,488)(850,488)(851,488)(852,488)(852,489)
(854,489)(855,489)(857,489)(858,489)(861,489)(864,490)(870,491)(876,493)
(882,495)(887,498)(899,505)(911,515)(923,527)
\thinlines \path(923,527)(946,561)(969,607)(985,647)
\thinlines \path(220,495)(220,495)(221,495)(221,495)(222,495)(223,495)
(224,495)(226,495)(227,495)(229,495)(232,495)(235,495)(238,496)(243,496)
(249,497)(255,497)(267,499)(290,505)(314,512)(337,521)(361,531)(384,542)
(407,554)(431,566)(454,578)(478,589)(501,598)(524,606)(536,609)(548,612)
(554,613)(560,614)(565,615)(571,616)(574,616)(577,616)(580,616)(582,616)
(583,616)(584,616)(586,616)(587,617)(588,617)(589,617)(590,617)(590,617)
(591,617)(592,617)(592,617)(593,617)(595,616)
\thinlines \path(595,616)(596,616)(598,616)(601,616)(603,616)(606,616)
(612,615)(618,615)(630,613)(642,610)(665,603)(688,594)(712,583)(735,571)
(759,559)(782,548)(794,544)(805,540)(811,538)(817,537)(823,535)(826,535)
(829,534)(832,534)(833,534)(835,534)(836,534)(838,534)(838,534)(839,534)
(840,534)(841,534)(841,534)(842,534)(843,534)(844,534)(844,534)(845,534)
(846,534)(848,534)(849,534)(852,534)(855,535)(858,535)(864,536)(870,538)
(876,540)(887,546)(899,555)(911,565)(923,578)
\thinlines \path(923,578)(946,613)(969,660)(985,701)
\thinlines \path(220,495)(220,495)(221,495)(221,495)(222,495)(223,495)
(224,495)(226,495)(227,495)(229,495)(232,495)(235,495)(238,495)(243,496)
(255,497)(267,498)(290,502)(314,507)(337,512)(361,519)(384,526)(407,533)
(431,539)(454,545)(466,547)(478,549)(483,549)(489,550)(495,550)(498,551)
(501,551)(504,551)(505,551)(507,551)(508,551)(510,551)(511,551)(511,551)
(512,551)(513,551)(513,551)(514,551)(515,551)(516,551)(517,551)(519,551)
(520,551)(522,551)(524,551)(527,551)(530,550)
\thinlines \path(530,550)(536,550)(542,549)(548,548)(560,546)(571,543)
(595,534)(618,523)(642,509)(665,492)(688,473)(712,452)(735,431)(759,410)
(782,390)(805,373)(817,366)(829,360)(835,357)(841,355)(846,353)(849,353)
(852,352)(855,351)(858,351)(860,351)(861,351)(863,351)(863,351)(864,351)
(865,351)(865,351)(866,351)(867,351)(868,351)(868,351)(869,351)(870,351)
(871,351)(871,351)(873,351)(874,351)(876,351)(879,352)(882,352)(887,354)
(893,356)(899,359)(911,366)(923,376)(934,389)
\thinlines \path(934,389)(946,406)(969,448)(985,486)
\end{picture}

		\hspace*{4em}(c) Behavior of the effective\\
		\hspace*{5em}potential as  $\mu$ varies\\
		\hspace*{5em}in the fixed $T$ direction B.
	\end{minipage}
\hfill
	\begin{minipage}[t]{.47\linewidth}
\setlength{\unitlength}{0.240900pt}
\begin{picture}(1049,900)(0,0)
\tenrm
\thinlines \dashline[-10]{25}(220,495)(985,495)
\thicklines \path(220,189)(240,189)
\thicklines \path(985,189)(965,189)
\put(198,189){\makebox(0,0)[r]{-0.004}}
\thicklines \path(220,342)(240,342)
\thicklines \path(985,342)(965,342)
\put(198,342){\makebox(0,0)[r]{-0.002}}
\thicklines \path(220,495)(240,495)
\thicklines \path(985,495)(965,495)
\put(198,495){\makebox(0,0)[r]{0}}
\thicklines \path(220,648)(240,648)
\thicklines \path(985,648)(965,648)
\put(198,648){\makebox(0,0)[r]{0.002}}
\thicklines \path(220,801)(240,801)
\thicklines \path(985,801)(965,801)
\put(198,801){\makebox(0,0)[r]{0.004}}
\thicklines \path(220,113)(220,133)
\thicklines \path(220,877)(220,857)
\put(220,68){\makebox(0,0){0}}
\thicklines \path(373,113)(373,133)
\thicklines \path(373,877)(373,857)
\put(373,68){\makebox(0,0){0.2}}
\thicklines \path(526,113)(526,133)
\thicklines \path(526,877)(526,857)
\put(526,68){\makebox(0,0){0.4}}
\thicklines \path(679,113)(679,133)
\thicklines \path(679,877)(679,857)
\put(679,68){\makebox(0,0){0.6}}
\thicklines \path(832,113)(832,133)
\thicklines \path(832,877)(832,857)
\put(832,68){\makebox(0,0){0.8}}
\thicklines \path(985,113)(985,133)
\thicklines \path(985,877)(985,857)
\put(985,68){\makebox(0,0){1}}
\thicklines \path(220,113)(985,113)(985,877)(220,877)(220,113)
\put(133,945){\makebox(0,0)[l]{\shortstack{$V/m^2$}}}
\put(602,23){\makebox(0,0){$\sigma/m$}}
\thinlines \path(220,495)(220,495)(221,495)(222,495)(223,495)(224,495)
(226,495)(228,495)(230,495)(232,495)(236,495)(240,494)(243,494)(251,493)
(267,491)(282,488)(314,479)(345,467)(376,452)(407,436)(439,419)(470,402)
(501,386)(532,372)(548,367)(563,362)(571,360)(579,359)(583,358)(587,358)
(591,357)(593,357)(595,357)(597,357)(598,357)(599,357)(600,357)(601,357)
(602,357)(603,357)(603,357)(604,357)(605,357)(606,357)(607,357)(608,357)
(609,357)(610,357)(614,357)(616,357)(618,358)
\thinlines \path(618,358)(626,359)(630,359)(634,360)(642,362)(657,368)
(673,377)(688,388)(704,402)(720,419)(751,464)(782,523)(813,600)(844,695)
(876,812)(890,877)
\thinlines \path(220,495)(220,495)(221,495)(222,495)(223,495)(224,495)
(226,495)(228,495)(232,495)(236,495)(243,495)(251,495)(267,495)(275,495)
(282,495)(286,495)(288,495)(290,495)(292,495)(293,495)(294,495)(295,495)
(296,495)(297,495)(298,495)(299,495)(300,495)(301,495)(302,495)(304,495)
(305,495)(306,495)(310,495)(312,495)(314,495)(318,495)(321,495)(325,495)
(329,495)(337,495)(345,495)(353,496)(361,496)(368,496)(376,497)(392,498)
(407,500)(423,502)(439,505)(454,508)(470,513)
\thinlines \path(470,513)(501,525)(532,542)(563,565)(595,596)(626,635)
(657,684)(688,745)(720,818)(740,877)
\thinlines \path(220,495)(220,495)(221,495)(222,495)(223,495)(224,495)
(226,495)(228,495)(230,495)(232,495)(236,495)(240,495)(243,496)(251,496)
(259,497)(267,498)(282,500)(298,503)(314,507)(345,516)(376,529)(407,546)
(439,567)(470,594)(501,627)(532,667)(563,715)(595,772)(626,839)(641,877)
\end{picture}

		\hspace*{4em}(b) Behavior of the effective\\
		\hspace*{5em}potential as $T$ varies\\
		\hspace*{5em}in the fixed $\mu$ direction A.
		\vspace{5ex}
\setlength{\unitlength}{0.240900pt}
\begin{picture}(1049,900)(0,0)
\tenrm
\thicklines \path(220,113)(240,113)
\thicklines \path(985,113)(965,113)
\put(198,113){\makebox(0,0)[r]{0}}
\thicklines \path(220,240)(240,240)
\thicklines \path(985,240)(965,240)
\put(198,240){\makebox(0,0)[r]{0.2}}
\thicklines \path(220,368)(240,368)
\thicklines \path(985,368)(965,368)
\put(198,368){\makebox(0,0)[r]{0.4}}
\thicklines \path(220,495)(240,495)
\thicklines \path(985,495)(965,495)
\put(198,495){\makebox(0,0)[r]{0.6}}
\thicklines \path(220,622)(240,622)
\thicklines \path(985,622)(965,622)
\put(198,622){\makebox(0,0)[r]{0.8}}
\thicklines \path(220,750)(240,750)
\thicklines \path(985,750)(965,750)
\put(198,750){\makebox(0,0)[r]{1}}
\thicklines \path(220,877)(240,877)
\thicklines \path(985,877)(965,877)
\put(198,877){\makebox(0,0)[r]{1.2}}
\thicklines \path(220,113)(220,133)
\thicklines \path(220,877)(220,857)
\put(220,68){\makebox(0,0){0}}
\thicklines \path(373,113)(373,133)
\thicklines \path(373,877)(373,857)
\put(373,68){\makebox(0,0){0.2}}
\thicklines \path(526,113)(526,133)
\thicklines \path(526,877)(526,857)
\put(526,68){\makebox(0,0){0.4}}
\thicklines \path(679,113)(679,133)
\thicklines \path(679,877)(679,857)
\put(679,68){\makebox(0,0){0.6}}
\thicklines \path(832,113)(832,133)
\thicklines \path(832,877)(832,857)
\put(832,68){\makebox(0,0){0.8}}
\thicklines \path(985,113)(985,133)
\thicklines \path(985,877)(985,857)
\put(985,68){\makebox(0,0){1}}
\thicklines \path(220,113)(985,113)(985,877)(220,877)(220,113)
\put(133,945){\makebox(0,0)[l]{\shortstack{$m_{\beta\mu}/m$}}}
\put(602,23){\makebox(0,0){$\mu/m$}}
\thinlines \path(220,745)(220,745)(228,745)(230,745)(232,745)(237,745)
(242,745)(246,745)(255,745)(265,745)(274,744)(283,744)(293,744)(302,744)
(311,744)(320,744)(330,743)(339,743)(348,743)(357,743)(367,742)(376,742)
(385,742)(395,741)(404,741)(413,740)(422,740)(432,739)(441,738)(450,738)
(459,737)(469,736)(478,735)(487,734)(497,733)(506,732)(515,731)(524,730)
(534,729)(543,727)(552,726)(561,724)(571,722)(580,720)(589,718)(599,715)
(608,713)(617,710)(626,707)(636,703)(645,699)
\thinlines \path(645,699)(654,695)(663,689)(673,684)(682,677)(716,636)
\thinlines \dashline[-10]{25}(716,636)(716,113)
\end{picture}

		\hspace*{4em}(d) The dynamical fermion mass\\
		\hspace*{5em}as a function\\
		\hspace*{5em}of $\mu$ with $T$ fixed.
	\end{minipage}
\vspace{1.5ex}
\caption{Behaviors of the effective  potential and the
dynamical fermion mass for $D=2$}
\label{fig:2D}
\end{figure}

\begin{figure}[H]
\setlength{\unitlength}{0.240900pt}
\begin{picture}(1500,900)(0,0)
\tenrm
\thinlines \dashline[-10]{25}(220,495)(1436,495)
\thicklines \path(220,113)(240,113)
\thicklines \path(1436,113)(1416,113)
\put(198,113){\makebox(0,0)[r]{-0.002}}
\thicklines \path(220,304)(240,304)
\thicklines \path(1436,304)(1416,304)
\put(198,304){\makebox(0,0)[r]{-0.001}}
\thicklines \path(220,495)(240,495)
\thicklines \path(1436,495)(1416,495)
\put(198,495){\makebox(0,0)[r]{0}}
\thicklines \path(220,686)(240,686)
\thicklines \path(1436,686)(1416,686)
\put(198,686){\makebox(0,0)[r]{0.001}}
\thicklines \path(220,877)(240,877)
\thicklines \path(1436,877)(1416,877)
\put(198,877){\makebox(0,0)[r]{0.002}}
\thicklines \path(220,113)(220,133)
\thicklines \path(220,877)(220,857)
\put(220,68){\makebox(0,0){0}}
\thicklines \path(463,113)(463,133)
\thicklines \path(463,877)(463,857)
\put(463,68){\makebox(0,0){0.2}}
\thicklines \path(706,113)(706,133)
\thicklines \path(706,877)(706,857)
\put(706,68){\makebox(0,0){0.4}}
\thicklines \path(950,113)(950,133)
\thicklines \path(950,877)(950,857)
\put(950,68){\makebox(0,0){0.6}}
\thicklines \path(1193,113)(1193,133)
\thicklines \path(1193,877)(1193,857)
\put(1193,68){\makebox(0,0){0.8}}
\thicklines \path(1436,113)(1436,133)
\thicklines \path(1436,877)(1436,857)
\put(1436,68){\makebox(0,0){1}}
\thicklines \path(220,113)(1436,113)(1436,877)(220,877)(220,113)
\put(45,945){\makebox(0,0)[l]{\shortstack{$V/m^3$}}}
\put(828,23){\makebox(0,0){$\sigma/m$}}
\thinlines \path(220,495)(220,495)(221,495)(222,495)(222,495)(223,495)
(225,495)(226,495)(228,495)(229,495)(232,495)(236,495)(239,495)(245,495)
(257,495)(270,495)(294,495)(319,495)(344,495)(369,495)(394,495)(406,495)
(412,495)(419,495)(425,495)(428,495)(431,495)(434,495)(437,495)(440,495)
(442,495)(443,495)(443,495)(444,495)(445,495)(446,495)(446,495)(447,495)
(448,495)(449,495)(450,495)(450,495)(451,495)(453,495)(454,495)(456,495)
(459,495)(462,495)(468,495)(474,495)(481,495)
\thinlines \path(481,495)(493,495)(505,495)(518,495)(530,495)(543,495)
(567,495)(592,496)(617,497)(642,498)(667,499)(692,501)(716,503)(741,506)
(766,509)(791,513)(816,517)(840,522)(865,529)(890,536)(915,545)(940,555)
(964,567)(989,580)(1014,596)(1039,614)(1064,634)(1089,658)(1113,684)
(1138,714)(1163,749)(1188,787)(1213,831)(1236,877)
\thinlines \path(220,495)(220,495)(221,495)(222,495)(222,495)(223,495)
(225,495)(226,495)(228,495)(229,495)(232,495)(236,495)(239,495)(245,495)
(251,494)(257,494)(270,494)(282,493)(294,492)(319,490)(344,487)(369,483)
(394,479)(419,474)(443,469)(468,463)(493,456)(518,449)(543,442)(567,434)
(592,425)(617,417)(642,407)(667,398)(692,388)(716,378)(741,369)(766,359)
(791,349)(816,340)(840,330)(865,322)(890,314)(915,307)(940,301)(952,298)
(964,296)(977,294)(983,294)(989,293)(996,292)
\thinlines \path(996,292)(1002,292)(1005,292)(1008,292)(1011,291)(1013,291)
(1014,291)(1016,291)(1016,291)(1017,291)(1018,291)(1019,291)(1020,291)
(1020,291)(1021,291)(1022,291)(1023,291)(1023,291)(1025,291)(1027,291)
(1028,291)(1030,291)(1033,292)(1036,292)(1039,292)(1045,292)(1051,293)
(1064,295)(1076,297)(1089,301)(1113,309)(1138,322)(1163,338)(1188,358)
(1213,384)(1237,415)(1262,452)(1287,496)(1312,547)(1337,607)(1362,675)
(1386,752)(1411,840)(1420,877)
\thinlines \path(220,495)(220,495)(221,495)(222,495)(222,495)(223,495)
(225,495)(226,495)(228,495)(229,495)(232,495)(236,495)(239,495)(245,495)
(251,496)(257,496)(270,497)(282,497)(294,498)(319,501)(344,504)(369,509)
(394,514)(419,519)(443,526)(468,533)(493,542)(518,551)(543,560)(567,571)
(592,583)(617,596)(642,609)(667,624)(692,639)(716,656)(741,674)(766,693)
(791,714)(816,736)(840,759)(865,784)(890,810)(915,839)(940,869)(946,877)
\end{picture}

\caption{Behavior of the effective potential at $D=3$
along the B line in Fig. 2 (a).}
\label{fig:pt3d}
\end{figure}
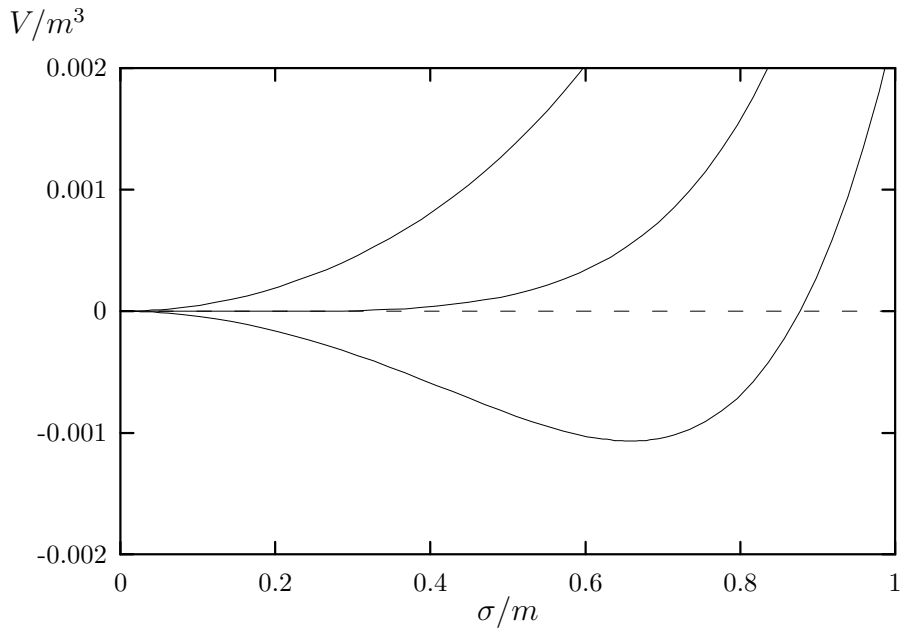

As is seen in Fig. \ref{fig:2D}, we observe that
\begin{enumerate}
	\item There is a second-order phase transition
		as $T$ is increased with $\mu$ kept small fixed.
	\item There is a first-order phase transition
		as $\mu$ is increased with $T$ kept small fixed.
\end{enumerate}
Thus we expect that the first- and second-order phase transition coexist
on the $T-\mu$ plane. For $D$ just above 2 this property is expected to
persist. On the other hand at $D = 3$ we observe only the second-order
phase transition for varying $T$ and $\mu$.\cite{D3NJL}
This situation is illustrated in Fig. \ref{fig:pt3d}.
\vspace{3ex}
The above preliminary analysis suggests that the first-order phase
transition observed at $D = 2$ disappears at some critical dimension
$D_c \leq 3$.

\begin{figure}
\setlength{\unitlength}{0.240900pt}
\begin{picture}(1500,900)(0,0)
\tenrm
\thicklines \path(220,113)(240,113)
\thicklines \path(1436,113)(1416,113)
\put(198,113){\makebox(0,0)[r]{0}}
\thicklines \path(220,240)(240,240)
\thicklines \path(1436,240)(1416,240)
\put(198,240){\makebox(0,0)[r]{0.2}}
\thicklines \path(220,368)(240,368)
\thicklines \path(1436,368)(1416,368)
\put(198,368){\makebox(0,0)[r]{0.4}}
\thicklines \path(220,495)(240,495)
\thicklines \path(1436,495)(1416,495)
\put(198,495){\makebox(0,0)[r]{0.6}}
\thicklines \path(220,622)(240,622)
\thicklines \path(1436,622)(1416,622)
\put(198,622){\makebox(0,0)[r]{0.8}}
\thicklines \path(220,750)(240,750)
\thicklines \path(1436,750)(1416,750)
\put(198,750){\makebox(0,0)[r]{1}}
\thicklines \path(220,877)(240,877)
\thicklines \path(1436,877)(1416,877)
\put(198,877){\makebox(0,0)[r]{1.2}}
\thicklines \path(220,113)(220,133)
\thicklines \path(220,877)(220,857)
\put(220,68){\makebox(0,0){0}}
\thicklines \path(394,113)(394,133)
\thicklines \path(394,877)(394,857)
\put(394,68){\makebox(0,0){0.1}}
\thicklines \path(567,113)(567,133)
\thicklines \path(567,877)(567,857)
\put(567,68){\makebox(0,0){0.2}}
\thicklines \path(741,113)(741,133)
\thicklines \path(741,877)(741,857)
\put(741,68){\makebox(0,0){0.3}}
\thicklines \path(915,113)(915,133)
\thicklines \path(915,877)(915,857)
\put(915,68){\makebox(0,0){0.4}}
\thicklines \path(1089,113)(1089,133)
\thicklines \path(1089,877)(1089,857)
\put(1089,68){\makebox(0,0){0.5}}
\thicklines \path(1262,113)(1262,133)
\thicklines \path(1262,877)(1262,857)
\put(1262,68){\makebox(0,0){0.6}}
\thicklines \path(1436,113)(1436,133)
\thicklines \path(1436,877)(1436,857)
\put(1436,68){\makebox(0,0){0.7}}
\thicklines \path(220,113)(1436,113)(1436,877)(220,877)(220,113)
\put(45,945){\makebox(0,0)[l]{\shortstack{$m_{\beta\mu}/m$}}}
\put(828,23){\makebox(0,0){$kT/m$}}
\thinlines \path(220,750)(220,750)(237,750)(243,750)(246,750)(249,750)
(252,750)(254,750)(256,750)(257,750)(257,750)(258,750)(259,750)(260,750)
(271,750)(283,750)(294,750)(305,750)(311,750)(317,750)(322,750)(328,750)
(334,750)(337,750)(339,750)(341,750)(342,750)(343,750)(344,750)(345,750)
(348,750)(351,750)(354,750)(356,750)(362,750)(365,750)(366,750)(368,750)
(369,750)(369,750)(370,750)(371,750)(371,750)(372,750)(374,750)(379,749)
(385,749)(396,749)(408,749)(419,749)(430,749)
\thinlines \path(430,749)(442,748)(464,747)(487,746)(510,744)(532,742)
(555,739)(578,736)(600,732)(623,728)(646,723)(668,718)(691,712)(714,705)
(737,698)(759,690)(782,681)(805,672)(827,662)(850,651)(873,639)(895,626)
(918,612)(941,597)(963,581)(986,563)(1009,543)(1031,522)(1054,499)(1077,473)
(1100,443)(1122,410)(1145,371)(1168,323)(1179,293)(1190,257)(1196,234)
(1202,206)(1207,166)(1208,159)(1209,148)(1209,134)(1210,113)
\end{picture}

\caption{Behavior of the dynamical fermion mass
$m_{\beta\mu}$
in the case of the second-order phase transition as $T$
varies with $\mu$ fixed .}
\label{fig:mt25d0}
\end{figure}
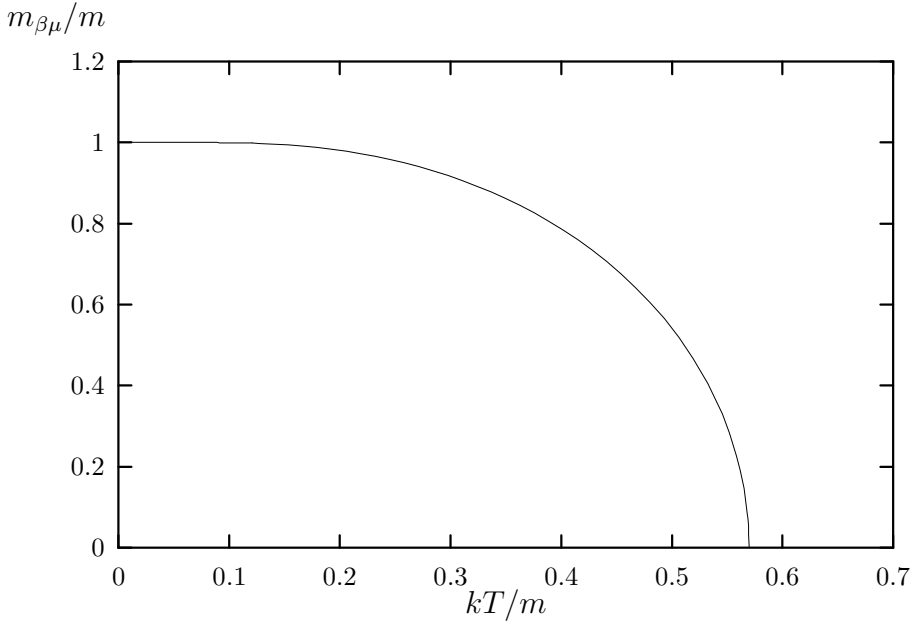

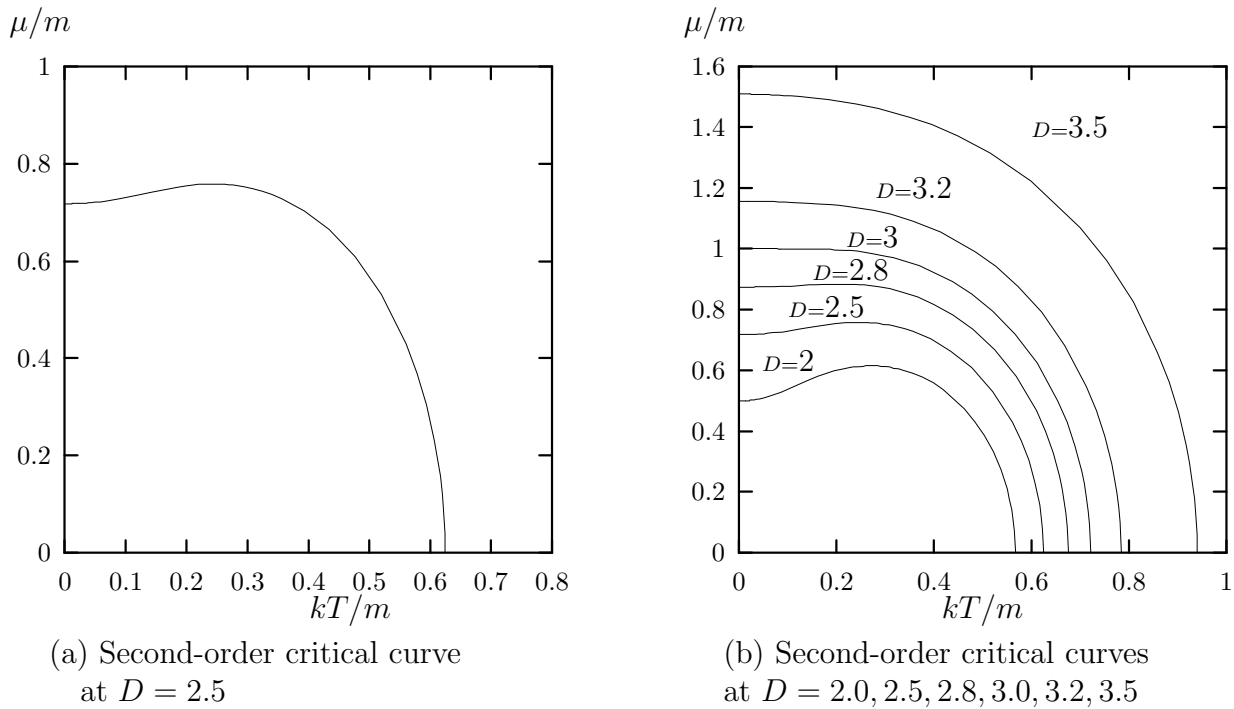
\begin{figure}
\hspace*{-2em}
	\begin{minipage}[t]{.47\linewidth}
\setlength{\unitlength}{0.240900pt}
\begin{picture}(1049,900)(0,0)
\tenrm
\thicklines \path(220,113)(240,113)
\thicklines \path(985,113)(965,113)
\put(198,113){\makebox(0,0)[r]{0}}
\thicklines \path(220,266)(240,266)
\thicklines \path(985,266)(965,266)
\put(198,266){\makebox(0,0)[r]{0.2}}
\thicklines \path(220,419)(240,419)
\thicklines \path(985,419)(965,419)
\put(198,419){\makebox(0,0)[r]{0.4}}
\thicklines \path(220,571)(240,571)
\thicklines \path(985,571)(965,571)
\put(198,571){\makebox(0,0)[r]{0.6}}
\thicklines \path(220,724)(240,724)
\thicklines \path(985,724)(965,724)
\put(198,724){\makebox(0,0)[r]{0.8}}
\thicklines \path(220,877)(240,877)
\thicklines \path(985,877)(965,877)
\put(198,877){\makebox(0,0)[r]{1}}
\thicklines \path(220,113)(220,133)
\thicklines \path(220,877)(220,857)
\put(220,68){\makebox(0,0){0}}
\thicklines \path(316,113)(316,133)
\thicklines \path(316,877)(316,857)
\put(316,68){\makebox(0,0){0.1}}
\thicklines \path(411,113)(411,133)
\thicklines \path(411,877)(411,857)
\put(411,68){\makebox(0,0){0.2}}
\thicklines \path(507,113)(507,133)
\thicklines \path(507,877)(507,857)
\put(507,68){\makebox(0,0){0.3}}
\thicklines \path(603,113)(603,133)
\thicklines \path(603,877)(603,857)
\put(603,68){\makebox(0,0){0.4}}
\thicklines \path(698,113)(698,133)
\thicklines \path(698,877)(698,857)
\put(698,68){\makebox(0,0){0.5}}
\thicklines \path(794,113)(794,133)
\thicklines \path(794,877)(794,857)
\put(794,68){\makebox(0,0){0.6}}
\thicklines \path(889,113)(889,133)
\thicklines \path(889,877)(889,857)
\put(889,68){\makebox(0,0){0.7}}
\thicklines \path(985,113)(985,133)
\thicklines \path(985,877)(985,857)
\put(985,68){\makebox(0,0){0.8}}
\thicklines \path(220,113)(985,113)(985,877)(220,877)(220,113)
\put(133,945){\makebox(0,0)[l]{\shortstack{$\mu/m$}}}
\put(668,23){\makebox(0,0){$kT/m$}}
\thinlines \path(817,113)(817,113)(817,143)(815,174)(813,204)(810,233)
(805,262)(800,290)(788,345)(773,395)(756,441)(717,519)(676,578)(635,621)
(596,650)(561,669)(545,676)(530,681)(515,685)(502,688)(496,689)(490,690)
(484,691)(478,691)(472,692)(467,692)(462,692)(457,692)(452,692)(447,692)
(443,692)(439,692)(434,692)(430,692)(423,691)(408,689)(396,687)(359,680)
(335,675)(305,669)(288,666)(276,664)(268,664)(262,663)(257,663)(254,662)
(251,662)(248,662)(246,662)(244,662)(242,662)
\thinlines \path(242,662)(241,662)(240,662)(239,662)(238,662)(237,662)
(236,662)(235,662)(235,662)(234,662)(233,662)(233,662)(232,662)(232,662)
(232,662)(231,661)(231,661)(231,661)(230,661)(230,661)(230,661)(229,661)
(229,661)(229,661)(229,661)(228,661)(228,661)(228,661)(228,661)(228,661)
(227,661)(227,661)(227,661)(227,661)(227,661)(220,661)
\end{picture}

		\hspace*{4em}(a) Second-order critical curve\\
		\hspace*{5em}at $D=2.5$
	\end{minipage}
\hfill
	\begin{minipage}[t]{.47\linewidth}
\setlength{\unitlength}{0.240900pt}
\begin{picture}(1049,900)(0,0)
\tenrm
\thicklines \path(220,113)(240,113)
\thicklines \path(985,113)(965,113)
\put(198,113){\makebox(0,0)[r]{0}}
\thicklines \path(220,209)(240,209)
\thicklines \path(985,209)(965,209)
\put(198,209){\makebox(0,0)[r]{0.2}}
\thicklines \path(220,304)(240,304)
\thicklines \path(985,304)(965,304)
\put(198,304){\makebox(0,0)[r]{0.4}}
\thicklines \path(220,400)(240,400)
\thicklines \path(985,400)(965,400)
\put(198,400){\makebox(0,0)[r]{0.6}}
\thicklines \path(220,495)(240,495)
\thicklines \path(985,495)(965,495)
\put(198,495){\makebox(0,0)[r]{0.8}}
\thicklines \path(220,591)(240,591)
\thicklines \path(985,591)(965,591)
\put(198,591){\makebox(0,0)[r]{1}}
\thicklines \path(220,686)(240,686)
\thicklines \path(985,686)(965,686)
\put(198,686){\makebox(0,0)[r]{1.2}}
\thicklines \path(220,782)(240,782)
\thicklines \path(985,782)(965,782)
\put(198,782){\makebox(0,0)[r]{1.4}}
\thicklines \path(220,877)(240,877)
\thicklines \path(985,877)(965,877)
\put(198,877){\makebox(0,0)[r]{1.6}}
\thicklines \path(220,113)(220,133)
\thicklines \path(220,877)(220,857)
\put(220,68){\makebox(0,0){0}}
\thicklines \path(373,113)(373,133)
\thicklines \path(373,877)(373,857)
\put(373,68){\makebox(0,0){0.2}}
\thicklines \path(526,113)(526,133)
\thicklines \path(526,877)(526,857)
\put(526,68){\makebox(0,0){0.4}}
\thicklines \path(679,113)(679,133)
\thicklines \path(679,877)(679,857)
\put(679,68){\makebox(0,0){0.6}}
\thicklines \path(832,113)(832,133)
\thicklines \path(832,877)(832,857)
\put(832,68){\makebox(0,0){0.8}}
\thicklines \path(985,113)(985,133)
\thicklines \path(985,877)(985,857)
\put(985,68){\makebox(0,0){1}}
\thicklines \path(220,113)(985,113)(985,877)(220,877)(220,113)
\put(133,945){\makebox(0,0)[l]{\shortstack{$\mu/m$}}}
\put(602,23){\makebox(0,0){$kT/m$}}
\put(256,414){\makebox(0,0)[l]{${\scriptstyle D=}2$}}
\put(297,500){\makebox(0,0)[l]{${\scriptstyle D=}2.5$}}
\put(335,557){\makebox(0,0)[l]{${\scriptstyle D=}2.8$}}
\put(388,610){\makebox(0,0)[l]{${\scriptstyle D=}3$}}
\put(434,686){\makebox(0,0)[l]{${\scriptstyle D=}3.2$}}
\put(679,782){\makebox(0,0)[l]{${\scriptstyle D=}3.5$}}
\thinlines \path(654,113)(654,113)(653,130)(652,147)(650,164)(648,181)
(644,197)(641,213)(631,244)(619,272)(605,297)(590,319)(574,339)(541,369)
(525,380)(508,389)(493,395)(478,400)(470,402)(463,403)(456,405)(450,406)
(443,406)(437,407)(431,407)(425,407)(419,407)(414,406)(404,405)(399,405)
(394,404)(385,402)(370,399)(344,391)(326,383)(286,365)(268,359)(258,356)
(252,355)(247,354)(244,353)(241,353)(239,353)(237,353)(236,352)(234,352)
(233,352)(233,352)(232,352)(231,352)(230,352)
\thinlines \path(230,352)(230,352)(229,352)(229,352)(229,352)(228,352)
(228,352)(228,352)(227,352)(227,352)(227,352)(226,352)(226,352)(226,352)
(226,352)(226,352)(226,352)(225,352)(225,352)(225,352)(225,352)(225,352)
(225,352)(225,352)(224,352)(224,352)(224,352)(224,352)(224,352)(224,352)
(224,352)(224,352)(220,352)
\thinlines \path(698,113)(698,113)(697,132)(696,151)(694,170)(692,188)
(688,206)(684,224)(675,258)(663,289)(649,318)(618,366)(585,404)(552,430)
(521,449)(493,461)(480,465)(468,468)(456,471)(446,472)(441,473)(436,474)
(431,474)(426,474)(422,475)(418,475)(414,475)(410,475)(406,475)(402,475)
(398,475)(395,475)(391,475)(388,475)(382,474)(371,473)(361,472)(331,467)
(312,464)(288,460)(274,459)(265,458)(259,457)(254,457)(250,457)(247,456)
(244,456)(242,456)(241,456)(239,456)(238,456)
\thinlines \path(238,456)(237,456)(236,456)(235,456)(234,456)(233,456)
(233,456)(232,456)(232,456)(231,456)(231,456)(230,456)(230,456)(230,456)
(229,456)(229,456)(229,456)(228,456)(228,456)(228,456)(228,456)(227,456)
(227,456)(227,456)(227,456)(227,456)(227,456)(226,456)(226,456)(226,456)
(226,456)(226,456)(226,456)(226,456)(225,456)(220,456)
\thinlines \path(737,113)(737,113)(736,134)(735,154)(733,174)(730,194)
(727,214)(723,233)(713,270)(701,304)(687,336)(655,390)(621,432)(587,464)
(556,487)(526,503)(500,515)(477,522)(457,528)(439,531)(423,533)(416,534)
(409,534)(403,535)(397,535)(394,535)(391,535)(388,535)(386,535)(383,535)
(381,535)(378,535)(376,535)(374,535)(372,535)(369,535)(367,535)(363,535)
(360,535)(353,535)(340,534)(330,534)(302,532)(286,532)(275,531)(267,531)
(261,531)(256,531)(253,531)(250,531)(247,531)
\thinlines \path(247,531)(245,531)(243,530)(242,530)(240,530)(239,530)
(238,530)(237,530)(236,530)(236,530)(235,530)(234,530)(234,530)(233,530)
(233,530)(232,530)(232,530)(231,530)(231,530)(231,530)(230,530)(230,530)
(230,530)(229,530)(229,530)(229,530)(229,530)(228,530)(228,530)(228,530)
(228,530)(228,530)(227,530)(227,530)(227,530)(227,530)(227,530)(227,530)
(220,530)
\thinlines \path(772,113)(772,113)(771,135)(770,157)(768,178)(765,200)
(762,221)(758,241)(747,281)(735,318)(720,352)(688,411)(653,457)(618,493)
(585,520)(555,539)(503,564)(482,571)(463,577)(446,580)(431,583)(406,587)
(395,588)(386,588)(370,589)(363,590)(356,590)(350,590)(345,590)(335,590)
(331,590)(327,590)(323,590)(320,590)(317,590)(314,590)(311,590)(308,590)
(306,590)(303,590)(301,590)(299,590)(297,590)(295,590)(293,590)(291,590)
(290,590)(288,590)(287,590)(285,590)(284,590)
\thinlines \path(284,590)(282,590)(281,590)(280,591)(279,591)(278,591)
(277,591)(276,591)(275,591)(274,591)(273,591)(272,591)(271,591)(270,591)
(269,591)(268,591)(268,591)(267,591)(266,591)(265,591)(265,591)(264,591)
(263,591)(263,591)(262,591)(262,591)(261,591)(261,591)(260,591)(259,591)
(259,591)(258,591)(258,591)(257,591)(257,591)(257,591)(256,591)(256,591)
(255,591)(255,591)(254,591)(254,591)(254,591)(253,591)(253,591)(253,591)
(252,591)(252,591)(252,591)(251,591)(251,591)
\thinlines \path(251,591)(251,591)(250,591)(250,591)(250,591)(249,591)
(249,591)(249,591)(249,591)(248,591)(248,591)(248,591)(248,591)(247,591)
(247,591)(247,591)(247,591)(246,591)(246,591)(246,591)(246,591)(245,591)
(245,591)(245,591)(245,591)(244,591)(244,591)(244,591)(244,591)(244,591)
(244,591)(244,591)(244,591)(244,591)(244,591)(244,591)(244,591)(244,591)
(244,591)(244,591)(244,591)(244,591)(244,591)(244,591)(243,591)(243,591)
(243,591)(243,591)(243,591)(243,591)(243,591)
\thinlines \path(243,591)(243,591)(243,591)(243,591)(243,591)(243,591)
(243,591)(243,591)(242,591)(242,591)(242,591)(241,591)(240,591)(239,591)
(238,591)(238,591)(238,591)(237,591)(237,591)(237,591)(237,591)(237,591)
(237,591)(237,591)(237,591)(237,591)(237,591)(237,591)(237,591)(237,591)
(237,591)(236,591)(236,591)(236,591)(236,591)(236,591)(236,591)(236,591)
(236,591)(236,591)(236,591)(236,591)(236,591)(236,591)(236,591)(236,591)
(236,591)(236,591)(236,591)(236,591)(236,591)
\thinlines \path(236,591)(235,591)(235,591)(235,591)(235,591)(235,591)
(235,591)(235,591)(235,591)(235,591)(235,591)(235,591)(235,591)(234,591)
(234,591)(233,591)(233,591)(233,591)(233,591)(233,591)(232,591)(232,591)
(232,591)(232,591)(232,591)(232,591)(232,591)(232,591)(232,591)(232,591)
(232,591)(232,591)(232,591)(232,591)(232,591)(232,591)(232,591)(232,591)
(232,591)(232,591)(232,591)(232,591)(232,591)(232,591)(232,591)(232,591)
(232,591)(232,591)(232,591)(231,591)(231,591)
\thinlines \path(231,591)(231,591)(230,591)(230,591)(230,591)(230,591)
(230,591)(230,591)(230,591)(230,591)(230,591)(230,591)(230,591)(230,591)
(230,591)(230,591)(230,591)(230,591)(230,591)(230,591)(230,591)(230,591)
(230,591)(229,591)(229,591)(229,591)(229,591)(229,591)(229,591)(229,591)
(229,591)(229,591)(229,591)(229,591)(229,591)(229,591)(229,591)(229,591)
(229,591)(229,591)(229,591)(229,591)(229,591)(229,591)(229,591)(229,591)
(229,591)(229,591)(229,591)(229,591)(229,591)
\thinlines \path(229,591)(229,591)(229,591)(229,591)(229,591)(229,591)
(229,591)(229,591)(229,591)(229,591)(228,591)(228,591)(228,591)(228,591)
(228,591)(228,591)(228,591)(228,591)(228,591)(228,591)(228,591)(228,591)
(228,591)(228,591)(228,591)(228,591)(228,591)(228,591)(228,591)(228,591)
\thinlines \path(820,113)(820,113)(819,137)(818,161)(816,184)(813,207)
(809,230)(805,253)(794,296)(781,336)(765,374)(731,439)(695,491)(658,531)
(623,563)(591,586)(536,617)(493,634)(458,645)(431,651)(391,657)(376,659)
(363,660)(343,661)(328,662)(306,663)(292,664)(282,664)(274,665)(268,665)
(263,665)(259,665)(256,665)(253,665)(251,665)(249,665)(247,665)(245,665)
(244,665)(243,665)(242,665)(241,665)(240,665)(239,665)(238,665)(237,665)
(237,665)(236,665)(235,665)(235,665)(234,665)
\thinlines \path(234,665)(234,665)(234,665)(233,665)(233,665)(232,665)
(232,665)(232,665)(231,665)(231,665)(231,665)(231,665)(230,665)(230,665)
(230,665)(230,665)(229,665)(229,665)(229,665)(229,665)(220,665)
\thinlines \path(939,113)(939,113)(939,142)(937,170)(935,198)(931,226)
(927,254)(922,281)(910,333)(895,382)(877,427)(839,507)(797,572)(755,624)
(678,697)(614,741)(563,768)(522,786)(488,797)(439,811)(404,818)(360,825)
(332,828)(314,830)(300,831)(290,831)(283,832)(276,832)(271,832)(267,833)
(263,833)(260,833)(258,833)(255,833)(253,833)(251,833)(250,833)(248,833)
(247,833)(246,833)(245,833)(244,833)(243,833)(242,833)(241,833)(240,833)
(240,833)(239,833)(238,834)(238,834)(237,834)
\thinlines \path(237,834)(237,834)(236,834)(236,834)(235,834)(235,834)
(235,834)(234,834)(234,834)(233,834)(233,834)(233,834)(233,834)(232,834)
(232,834)(232,834)(232,834)(220,834)
\end{picture}

		\hspace*{4em}(b) Second-order critical curves\\
		\hspace*{4em}at $D=2.0,2.5,2.8,3.0,3.2,3.5$
	\end{minipage}
\vspace{1.5ex}
\caption{Second-order critical curves}
\label{fig:cl2nd25d}
\end{figure}

In order to see the situation more precisely we would like to
perform a rigorous analysis on the critical temperature $T_c$ and chemical
potential $\mu_c$. We would like to find the critical line
on the $T-\mu$ plane
in an analytic form if possible. Let us first assume that the phase
transition is of the second order. Since for the second-order phase
transition the nontrivial dynamical fermion mass develops continuously
at the critical temperature and chemical potential as illustrated in
Fig. \ref{fig:mt25d0},
the critical temperature and chemical potential are determined
by letting $m_{\beta\mu}\rightarrow 0$ in the gap equation. Thus
we set $m_{\beta\mu}=0$ in Eq. (\ref{gap:nontri})
and obtain
\begin{eqnarray}
     \frac{1}{\lambda}-\frac{1}{\lambda_c}
     -\frac{2(2\pi)^{D/2-2}}{\sqrt{\pi}}\Gamma \left(
     \frac{3-D}{2}\right)(\sigma_{0}\beta)^{2-D}
     \mbox{Re \,}\zeta
     \left(
     3-D,\,\frac{1}{2}+i\frac{\beta\mu}{2\pi}
     \right) = 0\, .
\label{eq:2ndcrln}
\end{eqnarray}
where we used the following relation
\begin{eqnarray}
\sum_{n=-\infty}^\infty \left\{
		(\omega_n+i\mu)^2 \right\}^{-\nu}
	&=& \left({\beta \over 2\pi}\right)^{2\nu}
	\left\{
		\zeta\left(2\nu,\,{1 \over 2}
			-i{\beta\mu \over 2\pi}\right)
		+\zeta\left(2\nu,\,{1 \over 2}
			+i{\beta\mu \over 2\pi}\right)
	\right\} \, ,
\label{eq:sum2zeta}
\end{eqnarray}
with the definition of the generalized
zeta function $\zeta(z,a)$,\cite{ZETA}
\begin{equation}
	\zeta(z,a) =
		\sum_{n=0}^\infty {1 \over (n+a)^z}
		\hspace*{1em}; {\rm Re\,}z>1 \, ,
\label{eq:zeta}
\end{equation}
Eq. (\ref{eq:2ndcrln}) provides us with the relation
between the critical temperature $T_c$ and
chemical potential $\mu_c$, i. e. the critical curve
on the $T-\mu$ plane
if the transition is of the second-order. It should be noted that the
relation (\ref{eq:sum2zeta}) holds only for $D<2$ according to
the condition ${\rm Re\,}(2\nu)> 1$ as given
in Eq. (\ref{eq:zeta}). In Eq. (\ref{eq:2ndcrln}),
however, we analytically continue the variable $D$ to
the region ${\rm Re\,}D \geq 2$. As a typical example the
second-order critical curve at $D=2.5$ is presented
in Fig. \ref{fig:cl2nd25d} using
Eq. (\ref{eq:2ndcrln}).

Eq. (\ref{eq:2ndcrln}) simplifies when $D=2$ and $D=3$.
We note that the generalized
zeta function $\zeta(z,a)$ has a pole at $z=1$ and its behavior
near $z=1$ is given by
\begin{equation}
	\zeta(z,a)\rightarrow
		{1 \over z-1}-\psi(a)
	\hspace*{2em} ; z\rightarrow 1 \, ,
\label{eq:zetalimit}
\end{equation}
where $\psi(a)$ is the digamma function.
Using Eq. (\ref{eq:zetalimit}) for $z=3-D$ and
applying some formulae for the digamma function
we find for $D=2$
\begin{equation}
	2{\rm Re\,}\psi\left(1+i{\beta\mu \over \pi}\right)
		- {\rm Re\,}\psi\left(1+i{\beta\mu \over 2\pi}\right)
	= \ln {\beta m \over \pi} \, ,
\label{crln:2d2nd}
\end{equation}
where we employed Eq. (\ref{mass:d}) with $m$ the dynamical
fermion mass for $T=\mu=0$. Eq. (\ref{crln:2d2nd})
exactly reproduces the result obtained in Ref. 11.
To derive the corresponding formula for $D=3$ we employ the following
expansion formula for the generalized zeta function:
\begin{equation}
	\zeta(x,a)={1 \over 2} -a
		+ x \left\{
			\ln\Gamma(a)-\half \ln 2\pi
		\right\} + O(x^2) \, .
\label{zeta:expand}
\end{equation}
Using Eq. (\ref{zeta:expand})
with $x=3-D$ and $a={\displaystyle \half+i{\beta\mu \over 2\pi}}$ we
rewrite Eq. (\ref{eq:2ndcrln}) as follows,
\begin{equation}
	{1 \over \lambda} - {1 \over \lambda_c}
	- \frac{\sqrt{2}}{\pi\beta\sigma_0}\ln
	\frac{2 \left| \Gamma\left(i
			{\displaystyle{\beta\mu \over \pi}}\right) \right|^2}
		{\left| \Gamma\left(i
			{\displaystyle{\beta\mu \over 2\pi}}\right) \right|^2}
	=0 \, ,
\end{equation}
which reduces to the simple formula
\begin{equation}
	{\large e}^{\beta m / 2}
	= 2\cosh{\beta\mu \over 2} \, .
\label{eq:3dcrln}
\end{equation}
Eq. (\ref{eq:3dcrln}) exactly agrees with the one obtained
in Ref. 9.

Our formula (\ref{eq:2ndcrln}) gives a correct critical curve
on the $T-\mu$ plane
as far as the transition is of the second order. If the transition is of
the first order, the curve given by the formula (\ref{eq:2ndcrln})
corresponds to the
local maximum (first extremum) of the effective potential and does not
give a phase boundary. In such a case we have to find the critical curve
corresponding to the true minimum (second extremum) of the effective
potential. The critical curve for the first-order phase transition is
obtained by eliminating the variable $\sigma$ in the following two equations:
\begin{equation}
	V(\sigma)=0 \, , \hspace{2em}
	{\partial V(\sigma) \over \partial \sigma}=0 \, ,
\label{cond:1st}
\end{equation}
where we assume that $\sigma \neq 0$.

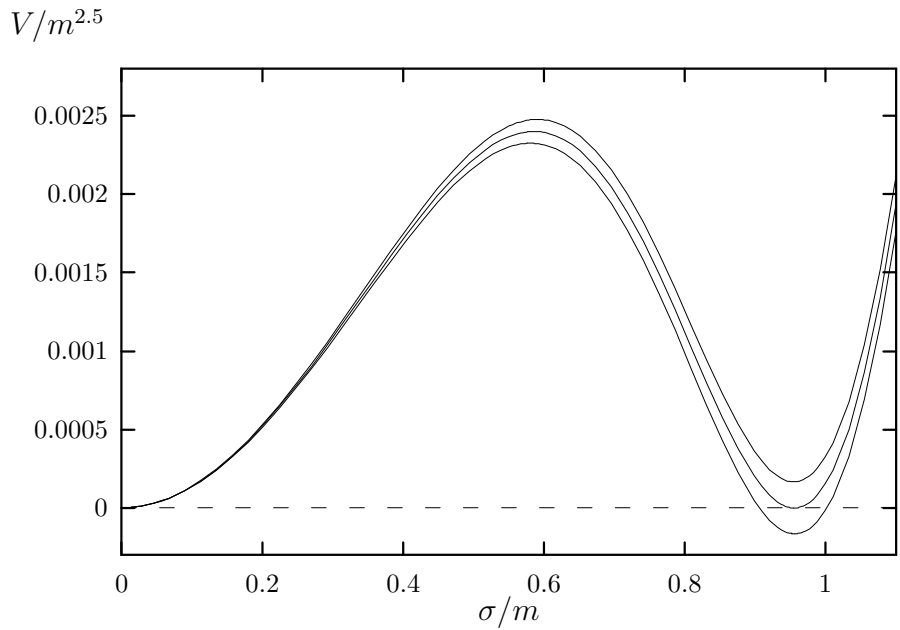
\begin{figure}
\setlength{\unitlength}{0.240900pt}
\begin{picture}(1500,900)(0,0)
\tenrm
\thinlines \dashline[-10]{25}(220,187)(1436,187)
\thicklines \path(220,187)(240,187)
\thicklines \path(1436,187)(1416,187)
\put(198,187){\makebox(0,0)[r]{0}}
\thicklines \path(220,310)(240,310)
\thicklines \path(1436,310)(1416,310)
\put(198,310){\makebox(0,0)[r]{0.0005}}
\thicklines \path(220,433)(240,433)
\thicklines \path(1436,433)(1416,433)
\put(198,433){\makebox(0,0)[r]{0.001}}
\thicklines \path(220,557)(240,557)
\thicklines \path(1436,557)(1416,557)
\put(198,557){\makebox(0,0)[r]{0.0015}}
\thicklines \path(220,680)(240,680)
\thicklines \path(1436,680)(1416,680)
\put(198,680){\makebox(0,0)[r]{0.002}}
\thicklines \path(220,803)(240,803)
\thicklines \path(1436,803)(1416,803)
\put(198,803){\makebox(0,0)[r]{0.0025}}
\thicklines \path(220,113)(220,133)
\thicklines \path(220,877)(220,857)
\put(220,68){\makebox(0,0){0}}
\thicklines \path(441,113)(441,133)
\thicklines \path(441,877)(441,857)
\put(441,68){\makebox(0,0){0.2}}
\thicklines \path(662,113)(662,133)
\thicklines \path(662,877)(662,857)
\put(662,68){\makebox(0,0){0.4}}
\thicklines \path(883,113)(883,133)
\thicklines \path(883,877)(883,857)
\put(883,68){\makebox(0,0){0.6}}
\thicklines \path(1104,113)(1104,133)
\thicklines \path(1104,877)(1104,857)
\put(1104,68){\makebox(0,0){0.8}}
\thicklines \path(1325,113)(1325,133)
\thicklines \path(1325,877)(1325,857)
\put(1325,68){\makebox(0,0){1}}
\thicklines \path(220,113)(1436,113)(1436,877)(220,877)(220,113)
\put(45,945){\makebox(0,0)[l]{\shortstack{$V/m^{2.5}$}}}
\put(828,23){\makebox(0,0){$\sigma/m$}}
\thinlines \path(220,187)(220,187)(221,187)(222,187)(222,187)(223,187)
(225,187)(226,187)(228,187)(229,187)(232,187)(236,188)(239,188)(245,189)
(251,190)(257,191)(270,194)(282,198)(294,202)(319,214)(344,229)(369,247)
(394,268)(419,292)(443,318)(468,347)(493,377)(518,409)(543,443)(567,477)
(592,512)(617,547)(642,582)(667,616)(692,648)(716,678)(741,705)(766,729)
(791,749)(803,758)(816,765)(828,770)(834,772)(840,774)(847,776)(850,776)
(853,777)(856,777)(857,778)(859,778)(861,778)
\thinlines \path(861,778)(862,778)(863,778)(864,778)(864,778)(865,778)
(866,778)(867,778)(868,778)(868,778)(869,778)(870,778)(871,778)(871,778)
(873,778)(875,778)(878,778)(881,777)(884,777)(890,775)(896,773)(902,771)
(915,765)(927,758)(940,748)(964,723)(989,691)(1014,651)(1039,606)(1064,554)
(1089,499)(1113,442)(1138,384)(1163,329)(1188,280)(1213,238)(1225,221)
(1237,207)(1250,196)(1256,192)(1259,191)(1262,189)(1265,188)(1267,188)
(1268,188)(1270,187)(1272,187)(1272,187)(1273,187)(1274,187)(1275,187)
\thinlines \path(1275,187)(1275,187)(1276,187)(1277,187)(1278,187)(1279,187)
(1279,187)(1281,187)(1282,187)(1284,188)(1287,189)(1290,190)(1293,191)
(1300,195)(1306,200)(1312,207)(1324,224)(1337,246)(1362,310)(1386,400)
(1411,517)(1436,665)
\thinlines \path(220,187)(220,187)(221,187)(222,187)(222,187)(223,187)
(225,187)(226,187)(228,187)(229,187)(232,187)(236,188)(239,188)(245,189)
(251,190)(257,191)(270,194)(282,197)(294,202)(319,214)(344,228)(369,246)
(394,267)(419,290)(443,316)(468,344)(493,374)(518,406)(543,438)(567,472)
(592,506)(617,541)(642,574)(667,607)(692,638)(716,667)(741,694)(766,716)
(791,735)(803,743)(816,749)(828,754)(834,756)(840,757)(847,759)(850,759)
(853,759)(854,759)(856,760)(857,760)(857,760)
\thinlines \path(857,760)(858,760)(859,760)(860,760)(861,760)(861,760)
(862,760)(863,760)(864,760)(864,760)(865,760)(867,760)(868,759)(871,759)
(875,759)(878,758)(884,757)(890,755)(902,751)(915,744)(927,735)(940,725)
(964,699)(989,665)(1014,623)(1039,576)(1064,523)(1089,467)(1113,408)
(1138,349)(1163,293)(1188,242)(1213,199)(1225,182)(1237,167)(1250,156)
(1256,152)(1259,151)(1262,149)(1265,148)(1267,148)(1268,147)(1270,147)
(1272,147)(1272,146)(1273,146)(1274,146)(1275,146)(1275,146)(1276,146)
(1277,146)
\thinlines \path(1277,146)(1278,146)(1279,146)(1279,146)(1280,146)
(1281,146)(1282,147)(1284,147)(1287,148)(1290,149)(1293,150)(1300,154)
(1306,159)(1312,165)(1324,182)(1337,204)(1362,268)(1386,357)(1411,475)
(1436,622)
\thinlines \path(220,187)(220,187)(221,187)(222,187)(222,187)(223,187)
(225,187)(226,187)(228,187)(229,187)(232,187)(236,188)(239,188)(245,189)
(251,190)(257,191)(270,194)(282,198)(294,202)(319,214)(344,230)(369,248)
(394,269)(419,293)(443,320)(468,349)(493,380)(518,413)(543,447)(567,482)
(592,518)(617,554)(642,590)(667,624)(692,658)(716,689)(741,717)(766,743)
(791,764)(803,773)(816,780)(828,787)(834,789)(840,792)(847,793)(853,795)
(856,796)(859,796)(862,797)(864,797)(865,797)
\thinlines \path(865,797)(867,797)(868,797)(869,797)(870,797)(871,797)
(871,797)(872,797)(873,797)(874,797)(875,797)(875,797)(876,797)(878,797)
(879,797)(881,797)(884,796)(887,796)(890,796)(896,794)(902,792)(915,787)
(927,780)(940,771)(964,748)(989,718)(1014,680)(1039,636)(1064,586)(1089,532)
(1113,476)(1138,420)(1163,367)(1188,318)(1213,277)(1225,261)(1237,247)
(1250,237)(1256,233)(1259,232)(1262,230)(1265,229)(1267,229)(1268,229)
(1270,228)(1271,228)(1272,228)(1272,228)(1273,228)(1274,228)(1275,228)
\thinlines \path(1275,228)(1275,228)(1276,228)(1277,228)(1278,228)(1279,228)
(1279,228)(1281,228)(1282,229)(1284,229)(1287,230)(1290,231)(1293,233)
(1300,237)(1306,242)(1312,249)(1324,266)(1337,289)(1362,353)(1386,443)
(1411,561)(1436,709)
\end{picture}

\caption{Searching for the first order critical point by
directly observing the behavior of the effective potential.}
\label{fig:find1st}
\end{figure}

     As we have seen before there exists a first-order phase transition
for $2 \leq D<3$. Here we have to perform
an analysis based on Eq. (\ref{cond:1st}).
Unfortunately we are unable to solve this problem analytically. We then
employ expression (\ref{v:full})
with (\ref{v:exp}) and (\ref{gapeqn:tm}) to find
the critical $T$ and $\mu$ numerically through the use of
MATHEMATICA on our workstation. For example by numerically
calculating the effective potential
at $D=2.5$ for $\beta m=10$ we obtain a figure of
the effective potential as shown in Fig. \ref{fig:find1st}.

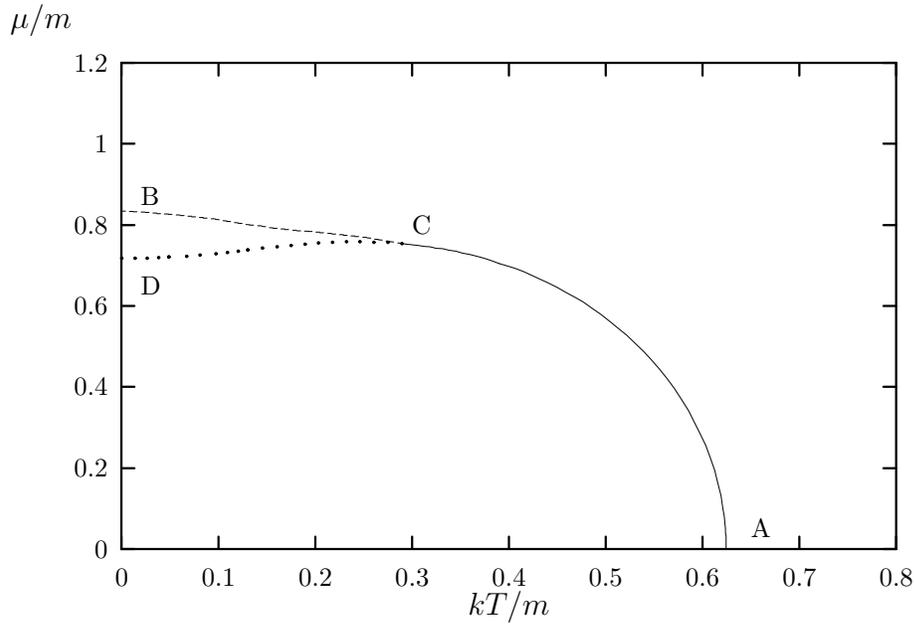
\begin{figure}
\setlength{\unitlength}{0.240900pt}
\begin{picture}(1500,900)(0,0)
\tenrm
\thicklines \path(220,113)(240,113)
\thicklines \path(1436,113)(1416,113)
\put(198,113){\makebox(0,0)[r]{0}}
\thicklines \path(220,240)(240,240)
\thicklines \path(1436,240)(1416,240)
\put(198,240){\makebox(0,0)[r]{0.2}}
\thicklines \path(220,368)(240,368)
\thicklines \path(1436,368)(1416,368)
\put(198,368){\makebox(0,0)[r]{0.4}}
\thicklines \path(220,495)(240,495)
\thicklines \path(1436,495)(1416,495)
\put(198,495){\makebox(0,0)[r]{0.6}}
\thicklines \path(220,622)(240,622)
\thicklines \path(1436,622)(1416,622)
\put(198,622){\makebox(0,0)[r]{0.8}}
\thicklines \path(220,750)(240,750)
\thicklines \path(1436,750)(1416,750)
\put(198,750){\makebox(0,0)[r]{1}}
\thicklines \path(220,877)(240,877)
\thicklines \path(1436,877)(1416,877)
\put(198,877){\makebox(0,0)[r]{1.2}}
\thicklines \path(220,113)(220,133)
\thicklines \path(220,877)(220,857)
\put(220,68){\makebox(0,0){0}}
\thicklines \path(372,113)(372,133)
\thicklines \path(372,877)(372,857)
\put(372,68){\makebox(0,0){0.1}}
\thicklines \path(524,113)(524,133)
\thicklines \path(524,877)(524,857)
\put(524,68){\makebox(0,0){0.2}}
\thicklines \path(676,113)(676,133)
\thicklines \path(676,877)(676,857)
\put(676,68){\makebox(0,0){0.3}}
\thicklines \path(828,113)(828,133)
\thicklines \path(828,877)(828,857)
\put(828,68){\makebox(0,0){0.4}}
\thicklines \path(980,113)(980,133)
\thicklines \path(980,877)(980,857)
\put(980,68){\makebox(0,0){0.5}}
\thicklines \path(1132,113)(1132,133)
\thicklines \path(1132,877)(1132,857)
\put(1132,68){\makebox(0,0){0.6}}
\thicklines \path(1284,113)(1284,133)
\thicklines \path(1284,877)(1284,857)
\put(1284,68){\makebox(0,0){0.7}}
\thicklines \path(1436,113)(1436,133)
\thicklines \path(1436,877)(1436,857)
\put(1436,68){\makebox(0,0){0.8}}
\thicklines \path(220,113)(1436,113)(1436,877)(220,877)(220,113)
\put(45,945){\makebox(0,0)[l]{\shortstack{$\mu/m$}}}
\put(828,23){\makebox(0,0){$kT/m$}}
\put(1208,145){\makebox(0,0)[l]{A}}
\put(250,667){\makebox(0,0)[l]{B}}
\put(250,527){\makebox(0,0)[l]{D}}
\put(676,622){\makebox(0,0)[l]{C}}
\thinlines \path(1169,113)(1169,113)(1169,114)(1169,114)(1169,115)(1169,116)
(1169,116)(1169,117)(1169,118)(1169,120)(1169,121)(1169,124)(1169,126)
(1169,129)(1169,134)(1168,139)(1168,145)(1167,155)(1166,166)(1164,176)
(1161,197)(1156,217)(1151,237)(1144,257)(1137,276)(1128,295)(1119,313)
(1110,331)(1099,348)(1088,365)(1077,380)(1065,395)(1052,410)(1040,423)
(1027,436)(1013,449)(1000,460)(986,471)(972,482)(959,491)(945,501)(931,509)
(917,517)(904,524)(890,531)(877,537)(864,543)(851,549)(838,554)(825,558)
(813,562)
\thinlines \path(813,562)(801,566)(789,570)(778,573)(766,576)(755,578)
(745,581)(734,583)(724,585)(714,586)(704,588)(695,589)(686,590)(677,591)
(668,592)(660,593)
\thinlines \dashline[-10]{25}(220,644)(238,643)(257,642)(275,640)(293,639)
(312,637)(330,635)(348,633)(367,631)(385,628)(403,625)(421,622)(440,620)
(458,617)(476,615)(495,613)(513,612)(531,610)(550,608)(568,606)(586,604)
(605,602)(623,599)(641,596)(660,593)
\put(441,586){\circle*{4}}
\put(418,583){\circle*{4}}
\put(403,581){\circle*{4}}
\put(386,579){\circle*{4}}
\put(322,573){\circle*{4}}
\put(294,572){\circle*{4}}
\put(278,571){\circle*{4}}
\put(260,570){\circle*{4}}
\put(241,570){\circle*{4}}
\put(220,570){\circle*{4}}
\put(660,593){\circle*{4}}
\put(638,595){\circle*{4}}
\put(616,595){\circle*{4}}
\put(594,596){\circle*{4}}
\put(573,596){\circle*{4}}
\put(551,595){\circle*{4}}
\put(529,594){\circle*{4}}
\put(508,592){\circle*{4}}
\put(486,590){\circle*{4}}
\put(464,588){\circle*{4}}
\put(418,583){\circle*{4}}
\put(368,577){\circle*{4}}
\put(343,575){\circle*{4}}
\put(294,572){\circle*{4}}
\end{picture}

\caption{Critical curve in $D=2.5$. The dashed line
represents the first-order phase transition while the full
line represents the second-order phase transition line.
The dotted line signals an appearance of the first extremum
(local maximum) in the effective potential and has nothing to
do with the phase transition.}
\label{fig:phase25d}
\end{figure}

\begin{figure}
\setlength{\unitlength}{0.240900pt}
\begin{picture}(1500,900)(0,0)
\tenrm
\thicklines \path(220,113)(240,113)
\thicklines \path(1436,113)(1416,113)
\put(198,113){\makebox(0,0)[r]{0}}
\thicklines \path(220,209)(240,209)
\thicklines \path(1436,209)(1416,209)
\put(198,209){\makebox(0,0)[r]{0.2}}
\thicklines \path(220,304)(240,304)
\thicklines \path(1436,304)(1416,304)
\put(198,304){\makebox(0,0)[r]{0.4}}
\thicklines \path(220,400)(240,400)
\thicklines \path(1436,400)(1416,400)
\put(198,400){\makebox(0,0)[r]{0.6}}
\thicklines \path(220,495)(240,495)
\thicklines \path(1436,495)(1416,495)
\put(198,495){\makebox(0,0)[r]{0.8}}
\thicklines \path(220,591)(240,591)
\thicklines \path(1436,591)(1416,591)
\put(198,591){\makebox(0,0)[r]{1}}
\thicklines \path(220,686)(240,686)
\thicklines \path(1436,686)(1416,686)
\put(198,686){\makebox(0,0)[r]{1.2}}
\thicklines \path(220,782)(240,782)
\thicklines \path(1436,782)(1416,782)
\put(198,782){\makebox(0,0)[r]{1.4}}
\thicklines \path(220,877)(240,877)
\thicklines \path(1436,877)(1416,877)
\put(198,877){\makebox(0,0)[r]{1.6}}
\thicklines \path(220,113)(220,133)
\thicklines \path(220,877)(220,857)
\put(220,68){\makebox(0,0){0}}
\thicklines \path(463,113)(463,133)
\thicklines \path(463,877)(463,857)
\put(463,68){\makebox(0,0){0.2}}
\thicklines \path(706,113)(706,133)
\thicklines \path(706,877)(706,857)
\put(706,68){\makebox(0,0){0.4}}
\thicklines \path(950,113)(950,133)
\thicklines \path(950,877)(950,857)
\put(950,68){\makebox(0,0){0.6}}
\thicklines \path(1193,113)(1193,133)
\thicklines \path(1193,877)(1193,857)
\put(1193,68){\makebox(0,0){0.8}}
\thicklines \path(1436,113)(1436,133)
\thicklines \path(1436,877)(1436,857)
\put(1436,68){\makebox(0,0){1}}
\thicklines \path(220,113)(1436,113)(1436,877)(220,877)(220,113)
\put(45,945){\makebox(0,0)[l]{\shortstack{$\mu/m$}}}
\put(828,23){\makebox(0,0){$kT/m$}}
\put(342,462){\makebox(0,0)[l]{${\scriptstyle D=}2$}}
\put(389,514){\makebox(0,0)[l]{${\scriptstyle D=}2.5$}}
\put(427,562){\makebox(0,0)[l]{${\scriptstyle D=}2.8$}}
\put(524,614){\makebox(0,0)[l]{${\scriptstyle D=}3$}}
\put(609,686){\makebox(0,0)[l]{${\scriptstyle D=}3.2$}}
\put(950,782){\makebox(0,0)[l]{${\scriptstyle D=}3.5$}}
\thinlines \path(909,113)(909,113)(909,113)(909,114)(909,114)(909,114)
(909,115)(909,115)(909,116)(909,116)(909,117)(909,118)(909,120)(909,121)
(909,124)(909,126)(909,129)(908,134)(908,139)(907,145)(906,155)(904,165)
(901,176)(899,186)(895,196)(892,206)(888,215)(883,225)(879,234)(874,243)
(868,252)(862,261)(856,269)(850,277)(844,285)(837,292)(830,300)(823,307)
(815,314)(808,320)(800,326)(793,332)(785,338)(777,343)(769,348)(761,353)
(753,358)(745,362)(737,366)(728,370)(720,373)
\thinlines \path(720,373)(712,377)(704,380)(696,383)(688,386)(681,388)
(673,390)(665,392)(657,394)(650,396)(643,398)(635,399)(628,400)(621,402)
(614,403)(607,403)
\thinlines \dashline{6}[3](220,451)(263,450)(306,445)(349,439)(392,433)
(435,427)(478,421)(521,414)(564,409)(607,403)
\thinlines \path(979,113)(979,113)(979,113)(979,114)(979,114)(979,115)
(979,115)(979,116)(979,117)(979,118)(979,119)(979,121)(979,123)(979,125)
(979,129)(979,133)(978,137)(978,145)(977,153)(976,160)(973,176)(969,191)
(964,206)(959,221)(953,236)(947,250)(940,263)(932,277)(923,289)(915,302)
(905,313)(896,325)(886,336)(876,346)(865,356)(855,365)(844,374)(833,382)
(822,390)(811,397)(800,404)(789,410)(778,416)(767,421)(756,427)(745,431)
(735,436)(725,440)(714,443)(704,447)(694,450)
\thinlines \path(694,450)(685,453)(675,455)(666,458)(657,460)(648,462)
(640,464)(631,465)(623,467)(615,468)(607,469)(600,470)(593,471)(585,472)
(578,472)(572,473)
\thinlines \dashline{6}[3](220,511)(259,509)(298,506)(337,501)(376,496)
(415,491)(454,487)(494,484)(533,479)(572,473)
\thinlines \path(1041,113)(1041,113)(1041,114)(1041,114)(1041,115)(1041,116)
(1041,117)(1041,119)(1041,120)(1041,122)(1041,125)(1041,128)(1041,131)
(1040,137)(1040,143)(1039,149)(1038,161)(1036,172)(1034,184)(1028,207)
(1021,230)(1012,252)(1002,273)(991,293)(979,312)(966,330)(952,348)(938,364)
(923,379)(907,394)(892,407)(876,419)(860,430)(845,441)(829,450)(813,459)
(798,467)(783,474)(769,481)(754,487)(740,492)(727,497)(713,501)(701,505)
(688,509)(676,512)(665,515)(654,517)(643,520)(632,522)(622,524)
\thinlines \path(622,524)(613,525)(603,527)(594,528)(586,529)(578,530)
(570,531)(562,531)(554,532)(547,533)(540,533)(534,534)(527,534)(521,534)
(515,535)
\thinlines \dashline{6}[3](220,556)(253,555)(286,553)(318,550)(351,546)
(384,543)(417,539)(449,537)(482,535)(515,535)
\thinlines \path(1097,113)(1097,113)(1097,135)(1095,157)(1091,178)(1087,200)
(1081,221)(1075,241)(1058,281)(1038,318)(1015,352)(963,411)(908,457)
(853,493)(800,520)(752,539)(670,564)(636,571)(606,577)(579,580)(555,583)
(515,587)(499,588)(484,588)(458,589)(447,590)(436,590)(427,590)(418,590)
(403,590)(396,590)(390,590)(384,590)(379,590)(374,590)(369,590)(364,590)
(360,590)(356,590)(352,590)(349,590)(345,590)(342,590)(339,590)(336,590)
(333,590)(331,590)(328,590)(326,590)(324,590)(321,590)
\thinlines \path(321,590)(319,590)(317,590)(315,591)(313,591)(312,591)
(310,591)(308,591)(307,591)(305,591)(304,591)(302,591)(301,591)(299,591)
(298,591)(297,591)(296,591)(294,591)(293,591)(292,591)(291,591)(290,591)
(289,591)(288,591)(287,591)(286,591)(285,591)(284,591)(284,591)(283,591)
(282,591)(281,591)(280,591)(280,591)(279,591)(278,591)(277,591)(277,591)
(276,591)(275,591)(275,591)(274,591)(274,591)(273,591)(272,591)(272,591)
(271,591)(271,591)(270,591)(270,591)(269,591)
\thinlines \path(269,591)(269,591)(268,591)(268,591)(267,591)(267,591)
(266,591)(266,591)(265,591)(265,591)(265,591)(264,591)(264,591)(263,591)
(263,591)(263,591)(262,591)(262,591)(261,591)(261,591)(261,591)(260,591)
(260,591)(260,591)(259,591)(259,591)(259,591)(259,591)(259,591)(259,591)
(258,591)(258,591)(258,591)(258,591)(258,591)(258,591)(258,591)(258,591)
(258,591)(258,591)(258,591)(258,591)(258,591)(257,591)(257,591)(257,591)
(257,591)(257,591)(257,591)(257,591)(257,591)
\thinlines \path(257,591)(256,591)(256,591)(256,591)(256,591)(256,591)
(256,591)(256,591)(256,591)(256,591)(255,591)(253,591)(251,591)(250,591)
(249,591)(248,591)(248,591)(248,591)(248,591)(247,591)(247,591)(247,591)
(247,591)(247,591)(247,591)(247,591)(247,591)(247,591)(247,591)(247,591)
(246,591)(246,591)(246,591)(246,591)(246,591)(246,591)(245,591)(245,591)
(245,591)(245,591)(245,591)(245,591)(245,591)(245,591)(245,591)(245,591)
(245,591)(245,591)(245,591)(245,591)(245,591)
\thinlines \path(245,591)(244,591)(244,591)(244,591)(244,591)(244,591)
(244,591)(244,591)(244,591)(244,591)(244,591)(244,591)(243,591)(243,591)
(242,591)(241,591)(241,591)(240,591)(240,591)(240,591)(240,591)(240,591)
(240,591)(240,591)(240,591)(240,591)(240,591)(240,591)(240,591)(240,591)
(240,591)(240,591)(240,591)(240,591)(239,591)(239,591)(239,591)(239,591)
(239,591)(239,591)(239,591)(239,591)(239,591)(239,591)(239,591)(239,591)
(239,591)(239,591)(239,591)(238,591)(238,591)
\thinlines \path(238,591)(237,591)(237,591)(236,591)(236,591)(236,591)
(236,591)(235,591)(235,591)(235,591)(235,591)(235,591)(235,591)(235,591)
(235,591)(235,591)(235,591)(235,591)(235,591)(235,591)(235,591)(235,591)
(235,591)(235,591)(235,591)(235,591)(235,591)(235,591)(235,591)(235,591)
(235,591)(235,591)(235,591)(235,591)(235,591)(235,591)(235,591)(234,591)
(234,591)(234,591)(234,591)(234,591)(234,591)(234,591)(234,591)(234,591)
(234,591)(234,591)(234,591)(234,591)(234,591)
\thinlines \path(234,591)(234,591)(234,591)(234,591)(234,591)(234,591)
(234,591)(234,591)(234,591)(234,591)(234,591)(233,591)(233,591)(233,591)
(233,591)(233,591)(233,591)(233,591)(233,591)(233,591)(233,591)(233,591)
(233,591)(233,591)(233,591)(233,591)(233,591)(233,591)(232,591)(232,591)
\thinlines \path(1173,113)(1173,113)(1172,137)(1170,161)(1167,184)(1162,207)
(1156,230)(1149,253)(1132,296)(1111,336)(1087,374)(1033,439)(974,491)
(916,531)(861,563)(810,586)(723,617)(654,634)(599,645)(555,651)(492,657)
(468,659)(447,660)(416,661)(391,662)(357,663)(335,664)(318,664)(306,665)
(296,665)(289,665)(283,665)(277,665)(273,665)(269,665)(266,665)(263,665)
(261,665)(258,665)(256,665)(254,665)(253,665)(251,665)(250,665)(249,665)
(248,665)(246,665)(246,665)(245,665)(244,665)(243,665)
\thinlines \path(243,665)(242,665)(242,665)(241,665)(240,665)(240,665)
(239,665)(239,665)(238,665)(238,665)(237,665)(237,665)(236,665)(236,665)
(236,665)(235,665)(235,665)(235,665)(234,665)(234,665)(220,665)
\thinlines \path(1363,113)(1363,113)(1362,142)(1360,170)(1356,198)(1351,226)
(1344,254)(1336,281)(1317,333)(1293,382)(1265,427)(1203,507)(1137,572)
(1070,624)(948,697)(847,741)(765,768)(700,786)(647,797)(568,811)(513,818)
(442,825)(398,828)(369,830)(348,831)(332,831)(320,832)(310,832)(302,832)
(295,833)(289,833)(284,833)(280,833)(276,833)(273,833)(270,833)(267,833)
(265,833)(263,833)(261,833)(259,833)(257,833)(256,833)(255,833)(253,833)
(252,833)(251,833)(250,833)(249,834)(248,834)(247,834)
\thinlines \path(247,834)(246,834)(246,834)(245,834)(244,834)(244,834)
(243,834)(242,834)(242,834)(241,834)(241,834)(240,834)(240,834)(240,834)
(239,834)(239,834)(238,834)(220,834)
\end{picture}

\caption{Critical curves for $2\leq D < 4$ .}
\label{fig:grad}
\end{figure}
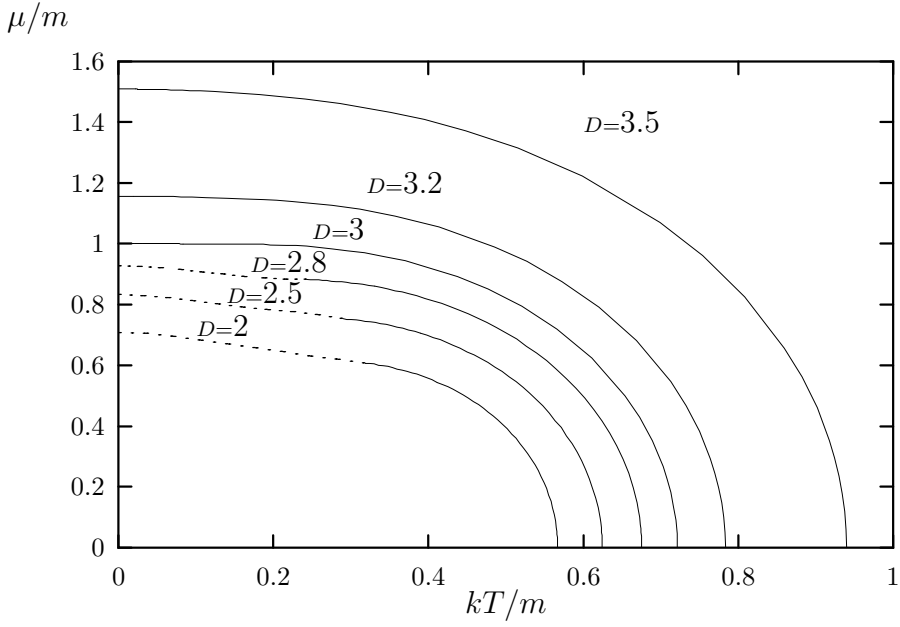

As can be seen in Fig. \ref{fig:find1st}
the critical chemical potential is given
by ${\displaystyle{\mu \over m}}=0.81$ at
\ ${\displaystyle{\sigma \over m}}=0.96$. Repeating
this kind of analysis many times we find the critical
curve for the first-order phase transition
at $D=2.5$. In Fig. \ref{fig:phase25d} is
shown the first-order critical curve by the dashed line while
the second-order critical curve is given in the full line.
We performed the full analysis of the above type
for $2 \leq D < 3$ and
obtained the critical curves as shown in Fig. \ref{fig:grad}.
In Fig. \ref{fig:grad} we clearly observe that
the first-order critical line given in the dashed line smoothly
disappears at $D=3$. We in fact checked that the phase transition is
always of second order for $3 \leq D < 4$.

\subsubsection{Specific points on the critical curve}

     It is interesting to note that we are able to find analytically
some specific points on the critical curve. Those points are shown in
Fig. \ref{fig:phase25d} by A, B, C and D.

     Let us first consider point A. Since we know that at
this point the transition is of second order, we are free to
use the formula (\ref{eq:2ndcrln}) for determining point A.
We set $\mu=0$ in Eq. (\ref{eq:2ndcrln}) to obtain
\begin{eqnarray}
	{1 \over \lambda} - {1 \over \lambda_c}
	- \frac{2(2\pi)^{D/2-2}}{\sqrt{\pi}}
	\Gamma\left({3-D \over 2}\right)(\beta\sigma_0)^{2-D}
	(2^{3-D}-1)\zeta(3-D)=0 \, ,
\label{eq:gapA}
\end{eqnarray}
where $\zeta(z)$ is the zeta function.
Using Eq. (\ref{mass:d}) we rewrite Eq. (\ref{eq:gapA})
in the following form:
\begin{equation}
	\beta m = 2\pi
		\left[
			\frac{2\Gamma\left({\displaystyle{3-D \over 2}}\right)}
			{\sqrt{\pi}\Gamma
				\left({\displaystyle{2-D \over 2}}\right)}
			(2^{3-D}-1)\zeta(3-D)
		\right]^{1/(D-2)} \, .
\label{cond:A}
\end{equation}
Eq. (\ref{cond:A}) gives the critical temperature.

\begin{figure}
\setlength{\unitlength}{0.240900pt}
\begin{picture}(1500,900)(0,0)
\tenrm
\thicklines \path(220,113)(240,113)
\thicklines \path(1436,113)(1416,113)
\put(198,113){\makebox(0,0)[r]{0}}
\thicklines \path(220,222)(240,222)
\thicklines \path(1436,222)(1416,222)
\put(198,222){\makebox(0,0)[r]{1}}
\thicklines \path(220,331)(240,331)
\thicklines \path(1436,331)(1416,331)
\put(198,331){\makebox(0,0)[r]{2}}
\thicklines \path(220,440)(240,440)
\thicklines \path(1436,440)(1416,440)
\put(198,440){\makebox(0,0)[r]{3}}
\thicklines \path(220,550)(240,550)
\thicklines \path(1436,550)(1416,550)
\put(198,550){\makebox(0,0)[r]{4}}
\thicklines \path(220,659)(240,659)
\thicklines \path(1436,659)(1416,659)
\put(198,659){\makebox(0,0)[r]{5}}
\thicklines \path(220,768)(240,768)
\thicklines \path(1436,768)(1416,768)
\put(198,768){\makebox(0,0)[r]{6}}
\thicklines \path(220,877)(240,877)
\thicklines \path(1436,877)(1416,877)
\put(198,877){\makebox(0,0)[r]{7}}
\thicklines \path(220,113)(220,133)
\thicklines \path(220,877)(220,857)
\put(220,68){\makebox(0,0){2}}
\thicklines \path(496,113)(496,133)
\thicklines \path(496,877)(496,857)
\put(496,68){\makebox(0,0){2.5}}
\thicklines \path(773,113)(773,133)
\thicklines \path(773,877)(773,857)
\put(773,68){\makebox(0,0){3}}
\thicklines \path(1049,113)(1049,133)
\thicklines \path(1049,877)(1049,857)
\put(1049,68){\makebox(0,0){3.5}}
\thicklines \path(1325,113)(1325,133)
\thicklines \path(1325,877)(1325,857)
\put(1325,68){\makebox(0,0){4}}
\thicklines \path(220,113)(1436,113)(1436,877)(220,877)(220,113)
\put(45,945){\makebox(0,0)[l]{\shortstack{$kT_c/m$}}}
\put(828,23){\makebox(0,0){$D$}}
\thinlines \path(220,175)(220,175)(231,175)(243,175)(254,176)(265,176)
(276,176)(288,176)(299,176)(310,177)(322,177)(333,177)(344,177)(355,178)
(367,178)(378,178)(389,178)(401,179)(412,179)(423,179)(434,179)(446,180)
(457,180)(468,180)(479,181)(491,181)(502,181)(513,182)(525,182)(536,182)
(547,183)(558,183)(570,183)(581,184)(592,184)(604,185)(615,185)(626,185)
(637,186)(649,186)(660,187)(671,187)(682,188)(694,188)(705,188)(716,189)
(728,190)(739,190)(750,191)(761,191)(773,192)
\thinlines \path(773,192)(773,192)(784,192)(795,193)(807,194)(818,194)
(829,195)(840,196)(852,196)(863,197)(874,198)(886,199)(897,200)(908,200)
(919,201)(931,202)(942,203)(953,204)(965,205)(976,207)(987,208)(998,209)
(1010,210)(1021,212)(1032,213)(1043,215)(1055,216)(1066,218)(1077,220)
(1089,222)(1100,224)(1111,227)(1122,229)(1134,232)(1145,235)(1156,238)
(1168,242)(1179,246)(1190,251)(1201,256)(1213,262)(1224,269)(1235,277)
(1246,287)(1258,299)(1269,315)(1275,325)(1280,337)(1286,351)(1292,368)
(1297,391)
\thinlines \path(1297,391)(1300,405)(1303,422)(1306,442)(1308,467)(1311,499)
(1313,519)(1314,542)(1316,571)(1317,606)(1318,652)(1319,680)(1320,713)
(1320,754)(1321,803)(1322,867)(1322,877)
\thinlines \dashline[-10]{25}(1325,877)(1325,113)
\end{picture}

\caption{critical temperature for $\mu=0$ as a function of
dimension $D$ .}
\label{fig:cria}
\end{figure}
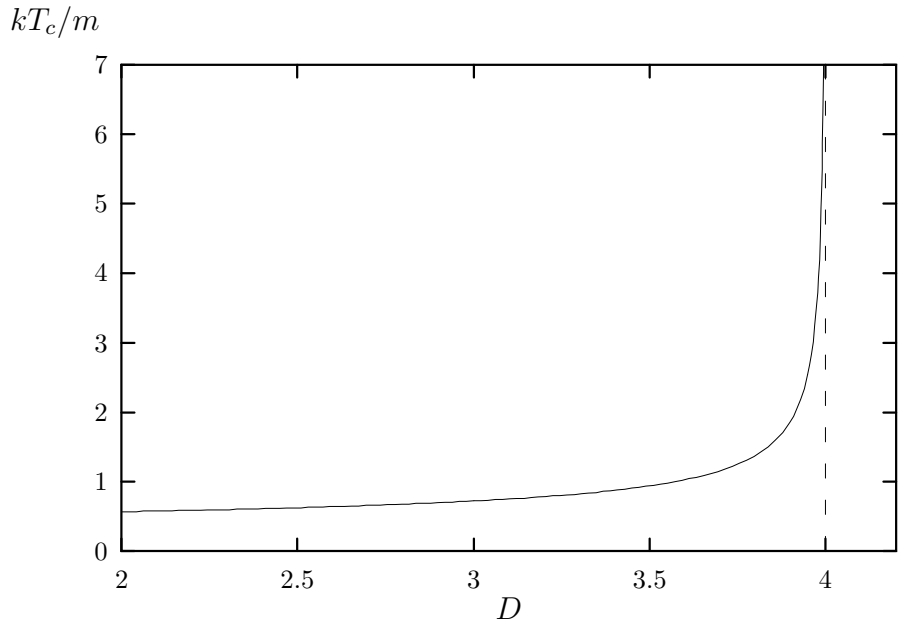

     For $D=2$ Eq. (\ref{cond:A}) reduces to the well-known formula for the
critical temperature,\cite{GNTEMP}
\begin{equation}
	\beta m = \pi {\large e}^{-\gamma} \, .
\label{cond:2dA}
\end{equation}
To see this we employ the following expansion formulae
in Eq. (\ref{cond:A}),
\begin{eqnarray}
	\Gamma(\varepsilon) &=& {1 \over \varepsilon}
		- \gamma + O(\varepsilon^2) \, ,
			\nonumber \\
	\Gamma\left(\half+\varepsilon\right)
		&=& \sqrt{\pi}\left\{
		1 - \varepsilon( 2\ln 2 + \gamma )
			+ O(\varepsilon^2)
		\right\} \, ,
			\nonumber \\
	\zeta(1+2\varepsilon) &=&
		{1 \over 2\varepsilon}
		+ \gamma + O(\varepsilon)\, ,
\end{eqnarray}
with $\gamma$ the Euler constant and  to obtain
\begin{equation}
	\beta m = \lim_{\varepsilon\rightarrow 0}
		2\pi\left\{
			1 + 2\varepsilon(\ln 2 + \gamma)
		\right\}^{-1/2\varepsilon} \, ,
\end{equation}
which clearly reproduces Eq. (\ref{cond:2dA}).
     For $D=3$ Eq. (\ref{cond:A}) again reproduces
the well-known formula\cite{D3NJL}
\begin{equation}
	\beta m = 2\ln 2 \, .
\label{cond:3dA}
\end{equation}
To derive Eq. (\ref{cond:3dA}) from Eq. (\ref{cond:A})
we only need the following expansion
formula near $D=3$,
\begin{eqnarray}
	\zeta(3-D) = - \half - \half (3-D)\ln 2\pi
	+ O((3-D)^2) \, .
\end{eqnarray}
In Fig. \ref{fig:cria} the critical temperature is plotted as a
function of dimension $D$ by the use of Eq. (\ref{cond:A}).

     We next consider point B. On the $\mu$-axis
for which $T=0\,(\beta\rightarrow\infty)$ the gap
equation (\ref{gapeqn:tm}) reads
for $\mu<m_{\beta\mu}$
\begin{equation}
	\Gamma\left({2-D \over 2}\right)
	(m^{D-2}-m_{\beta\mu}^{D-2})=0 \, ,
\label{eq:gapB}
\end{equation}
and for $\mu\geq m_{\beta\mu}$
\begin{eqnarray}
	\Gamma\left({\displaystyle{2-D \over 2}}\right)
		(m^{D-2}-m_\betamu^{D-2})
	+ \frac{2\sqrt{\pi}}
		{\Gamma\left({\displaystyle{D-1 \over 2}}\right)}
	\int_0^{\sqrt{\mu^2-m_\betamu^2}}dk
		\frac{k^{D-2}}{\sqrt{k^2+m_\betamu^2}}
	= 0 \, .
\label{eq:gapB1}
\end{eqnarray}
Eq. (\ref{eq:gapB}) gives a simple solution
\begin{equation}
	m_\betamu = m \, ,
\label{eq:gapB2}
\end{equation}
which is typically of the first-order type since it has
a gap at $\mu=\mu_c$ as is seen in Fig. \ref{fig:2D}(d).
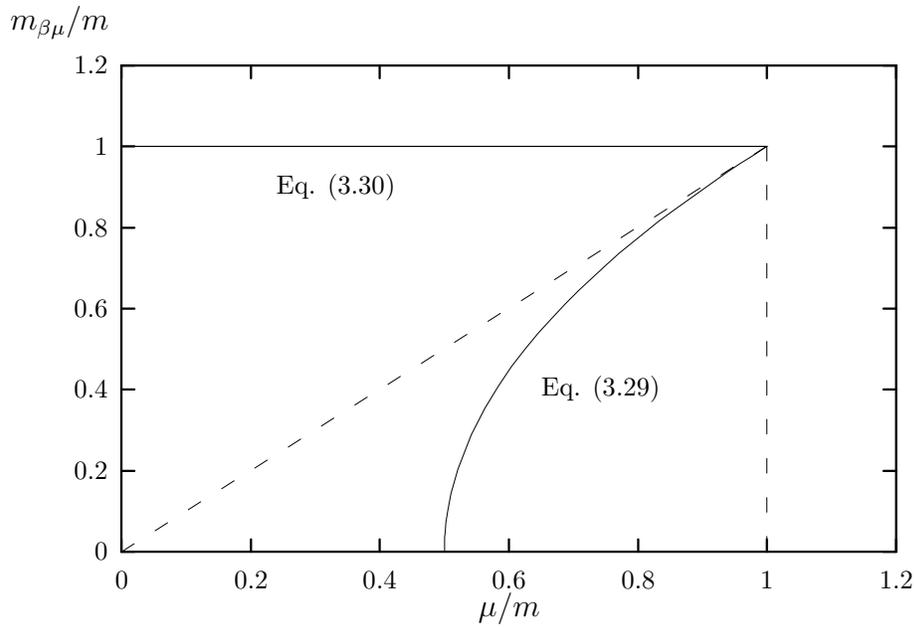
\begin{figure}
\setlength{\unitlength}{0.240900pt}
\begin{picture}(1500,900)(0,0)
\tenrm
\thicklines \path(220,113)(240,113)
\thicklines \path(1436,113)(1416,113)
\put(198,113){\makebox(0,0)[r]{0}}
\thicklines \path(220,240)(240,240)
\thicklines \path(1436,240)(1416,240)
\put(198,240){\makebox(0,0)[r]{0.2}}
\thicklines \path(220,368)(240,368)
\thicklines \path(1436,368)(1416,368)
\put(198,368){\makebox(0,0)[r]{0.4}}
\thicklines \path(220,495)(240,495)
\thicklines \path(1436,495)(1416,495)
\put(198,495){\makebox(0,0)[r]{0.6}}
\thicklines \path(220,622)(240,622)
\thicklines \path(1436,622)(1416,622)
\put(198,622){\makebox(0,0)[r]{0.8}}
\thicklines \path(220,750)(240,750)
\thicklines \path(1436,750)(1416,750)
\put(198,750){\makebox(0,0)[r]{1}}
\thicklines \path(220,877)(240,877)
\thicklines \path(1436,877)(1416,877)
\put(198,877){\makebox(0,0)[r]{1.2}}
\thicklines \path(220,113)(220,133)
\thicklines \path(220,877)(220,857)
\put(220,68){\makebox(0,0){0}}
\thicklines \path(423,113)(423,133)
\thicklines \path(423,877)(423,857)
\put(423,68){\makebox(0,0){0.2}}
\thicklines \path(625,113)(625,133)
\thicklines \path(625,877)(625,857)
\put(625,68){\makebox(0,0){0.4}}
\thicklines \path(828,113)(828,133)
\thicklines \path(828,877)(828,857)
\put(828,68){\makebox(0,0){0.6}}
\thicklines \path(1031,113)(1031,133)
\thicklines \path(1031,877)(1031,857)
\put(1031,68){\makebox(0,0){0.8}}
\thicklines \path(1233,113)(1233,133)
\thicklines \path(1233,877)(1233,857)
\put(1233,68){\makebox(0,0){1}}
\thicklines \path(1436,113)(1436,133)
\thicklines \path(1436,877)(1436,857)
\put(1436,68){\makebox(0,0){1.2}}
\thicklines \path(220,113)(1436,113)(1436,877)(220,877)(220,113)
\put(45,945){\makebox(0,0)[l]{\shortstack{$m_{\beta\mu}/m$}}}
\put(828,23){\makebox(0,0){$\mu/m$}}
\put(463,686){\makebox(0,0)[l]{Eq. (3.30)}}
\put(879,368){\makebox(0,0)[l]{Eq. (3.29)}}
\thinlines \path(727,113)(727,113)(727,136)(728,145)(729,159)(732,178)
(737,205)(748,243)(769,297)(790,338)(811,373)(832,404)(853,431)(874,457)
(896,481)(917,503)(938,524)(959,544)(980,563)(1001,582)(1022,599)(1043,616)
(1064,633)(1086,649)(1107,664)(1128,679)(1149,694)(1170,709)(1191,723)
(1212,736)(1233,750)(220,750)(262,750)(304,750)(347,750)(389,750)(431,750)
(473,750)(516,750)(558,750)(600,750)(642,750)(684,750)(727,750)(769,750)
(811,750)(853,750)(896,750)(938,750)(980,750)(1022,750)
\thinlines \path(1022,750)(1064,750)(1107,750)(1149,750)(1191,750)(1233,750)
\thinlines \dashline[-10]{25}(220,113)(1233,750)(1233,113)
\end{picture}

\caption{Dynamical fermion mass as a function of $\mu$
with $T=0$ .}
\label{fig:solofgap}
\end{figure}
On the other hand Eq. (\ref{eq:gapB1}) gives
a more complicated solution which is
shown in Fig. \ref{fig:solofgap}. This solution is typically of
the second-order type.

     For $2\leq D < 3$ the phase transition along the $\mu$-axis is of first
order as we have seen before. Hence
the solution (\ref{eq:gapB2}) gives a true vacuum and the
solution of Eq. (\ref{eq:gapB1}) corresponds to the first extremum
of the effective potential.
We adopt the solution (\ref{eq:gapB2}) to study the condition,
\begin{equation}
	\lim_{\beta\rightarrow\infty}
		V(\sigma=m)=0 \, .
\label{eq:Bcond}
\end{equation}
Using Eq. (\ref{v:full}) with Eq. (\ref{v:exp}) we rewrite
the condition (\ref{eq:Bcond}) as follows,
\begin{equation}
	{1 \over D}\,\Gamma\left({4-D \over 2}\right)m^D
	+ \frac{2\sqrt{\pi}}{\Gamma\left({\displaystyle{D-1 \over 2}}\right)}
	\int_0^\mu dk k^{D-2}(k-\mu) = 0 \, ,
\end{equation}
which provides us with the critical chemical potential
for $2\leq D < 3$ ,
\begin{equation}
	\mu = m\left\{ {3 \over 4}
		{\rm B}\left({4-D \over 2},\,{D+1 \over 2}\right)
	\right\}^{1/D} \, .
\label{cond:B}
\end{equation}
It is easy to check that Eq. (\ref{cond:B}) satisfies
the condition . For $D=2$ Eq. (\ref{cond:B}) simplifies
to{\cite{GNTM}\raisebox{.8ex}{\footnotesize ,}\cite{GNmu}
\begin{equation}
	\mu = {m \over \sqrt{2}} \, .
\label{cond:2dB}
\end{equation}

     For $3\leq D < 4$ the phase transition along the $\mu$-axis
is of second
order and hence we have to adopt the case of Eq. (\ref{eq:gapB1}).
Letting $m_\betamu \rightarrow 0$ in
Eq. (\ref{eq:gapB1}) and solving for $\mu$ we find
\begin{equation}
	\mu = m \left\{
		\half{\rm B\,}\left({4-D \over 2},\,{D-1 \over 2}\right)
	\right\}^{1/(D-2)} \, .
\label{cond:D}
\end{equation}
Obviously Eq. (\ref{cond:D}) satisfies
the condition $\mu\geq m_\betamu $.
For $D=3$ Eq. (\ref{cond:D})
simplifies to
\begin{equation}
	\mu = m \, ,
\end{equation}
which agrees with the known result.\cite{D3NJL}
In Fig. \ref{fig:crib} the the critical chemical potential is
plotted as a function of dimension $D$ by using Eqs.
(\ref{cond:B}) and (\ref{cond:D}).
\begin{figure}
\setlength{\unitlength}{0.240900pt}
\begin{picture}(1500,900)(0,0)
\tenrm
\thicklines \path(220,113)(240,113)
\thicklines \path(1436,113)(1416,113)
\put(198,113){\makebox(0,0)[r]{0}}
\thicklines \path(220,231)(240,231)
\thicklines \path(1436,231)(1416,231)
\put(198,231){\makebox(0,0)[r]{2}}
\thicklines \path(220,348)(240,348)
\thicklines \path(1436,348)(1416,348)
\put(198,348){\makebox(0,0)[r]{4}}
\thicklines \path(220,466)(240,466)
\thicklines \path(1436,466)(1416,466)
\put(198,466){\makebox(0,0)[r]{6}}
\thicklines \path(220,583)(240,583)
\thicklines \path(1436,583)(1416,583)
\put(198,583){\makebox(0,0)[r]{8}}
\thicklines \path(220,701)(240,701)
\thicklines \path(1436,701)(1416,701)
\put(198,701){\makebox(0,0)[r]{10}}
\thicklines \path(220,818)(240,818)
\thicklines \path(1436,818)(1416,818)
\put(198,818){\makebox(0,0)[r]{12}}
\thicklines \path(220,113)(220,133)
\thicklines \path(220,877)(220,857)
\put(220,68){\makebox(0,0){2}}
\thicklines \path(496,113)(496,133)
\thicklines \path(496,877)(496,857)
\put(496,68){\makebox(0,0){2.5}}
\thicklines \path(773,113)(773,133)
\thicklines \path(773,877)(773,857)
\put(773,68){\makebox(0,0){3}}
\thicklines \path(1049,113)(1049,133)
\thicklines \path(1049,877)(1049,857)
\put(1049,68){\makebox(0,0){3.5}}
\thicklines \path(1325,113)(1325,133)
\thicklines \path(1325,877)(1325,857)
\put(1325,68){\makebox(0,0){4}}
\thicklines \path(220,113)(1436,113)(1436,877)(220,877)(220,113)
\put(45,945){\makebox(0,0)[l]{\shortstack{$\mu_c/m$}}}
\put(828,23){\makebox(0,0){$D$}}
\thinlines \path(221,155)(221,155)(221,155)(223,155)(226,155)(231,155)
(243,155)(265,156)(288,156)(310,157)(333,157)(355,158)(378,159)(400,159)
(423,160)(446,161)(468,161)(491,162)(513,163)(536,163)(558,164)(581,165)
(604,165)(626,166)(649,167)(671,168)(694,169)(716,169)(739,170)(750,171)
(761,171)(784,173)(807,174)(829,176)(852,178)(874,180)(897,182)(919,185)
(942,187)(964,190)(987,192)(1010,196)(1032,199)(1055,203)(1077,207)(1100,212)
(1122,217)(1145,223)(1168,231)(1190,240)(1213,252)
\thinlines \path(1213,252)(1224,259)(1235,267)(1246,277)(1258,290)(1269,306)
(1275,316)(1280,327)(1286,341)(1292,359)(1297,381)(1300,395)(1303,412)
(1306,432)(1309,457)(1311,488)(1313,508)(1314,531)(1316,560)(1317,595)
(1318,640)(1319,668)(1320,701)(1321,741)(1321,790)(1322,854)(1322,877)
\thinlines \dashline[-10]{25}(1325,877)(1325,113)
\end{picture}

\caption{Critical chemical potential for $T=0$ as a function
of dimension $D$ .}
\label{fig:crib}
\end{figure}
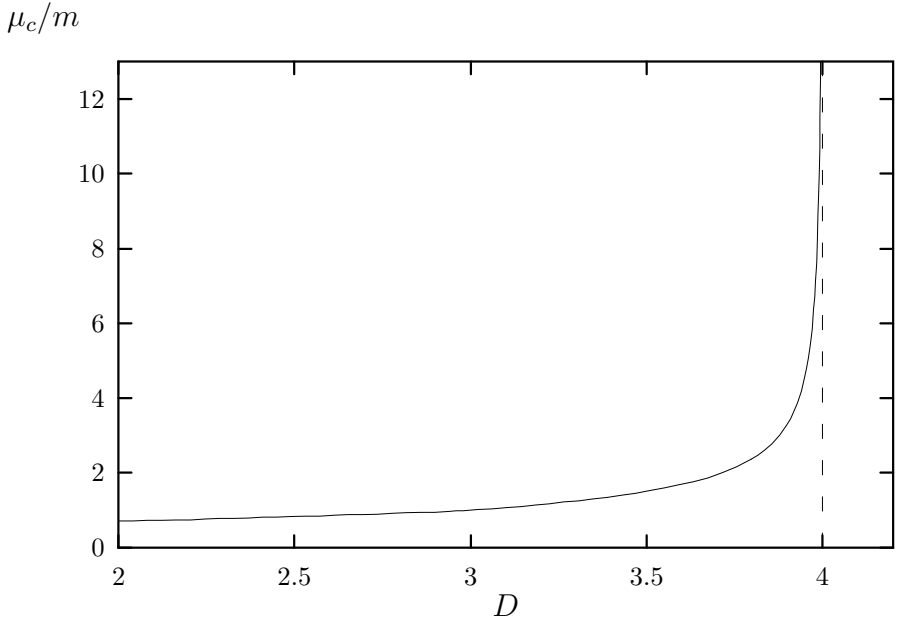

     It is worth noting here that, if we would have used the second-
order result (\ref{cond:D}) for $2\leq D < 3$ , we would have
obtained point $D$ shown
in Fig. \ref{fig:phase25d}. The value of $\mu$ at point $D$ is given
by Eq. (\ref{cond:D}) with $2\leq D < 3$ .
For $D=2$ we find
\begin{equation}
	\mu = {m \over 2} \, ,
\end{equation}
instead of Eq. (\ref{cond:2dB}). This, however,
corresponds to the first extremum of
the effective potential and has nothing to do with
the phase transition.\cite{GNmuE}

     Finally we consider point C on the critical curve. This point
appears only when $2\leq D < 3$ . At point C the first-order
critical curve meets with the second-order critical curve.
To find out the value of $T$ and $\mu$ at point C we proceed
as follows: Let us consider the gap equation
\begin{equation}
	{\partial V \over \partial\sigma}
	\equiv  \sigma f(\sigma) = 0 \, .
\end{equation}
We fix the dynamical mass at the first nonvanishing extremum . Thus we have
\begin{equation}
	f(\sigma)=0 \, .
\label{eq:gapC}
\end{equation}
By fixing $\sigma$ at $\sigma_m$ temperature $T$ is related
to chemical potential $\mu$ through Eq. (\ref{eq:gapC}).
As shown in Fig. 12 the relation between $T$ and $\mu$ is
plotted in the form of the curve G just as we have done for the critical curve.
Since the dynamical fermion mass is fixed to $\sigma_m$ on this curve
G, the behavior of the effective potential along the curve G
looks as in Fig. 13.
\begin{figure}
\hspace*{-6em}
	\begin{minipage}[t]{.47\linewidth}
\setlength{\unitlength}{0.240900pt}
\begin{picture}(1049,900)(0,0)
\tenrm
\thicklines \path(220,113)(240,113)
\thicklines \path(985,113)(965,113)
\put(198,113){\makebox(0,0)[r]{0}}
\thicklines \path(220,240)(240,240)
\thicklines \path(985,240)(965,240)
\put(198,240){\makebox(0,0)[r]{0.2}}
\thicklines \path(220,368)(240,368)
\thicklines \path(985,368)(965,368)
\put(198,368){\makebox(0,0)[r]{0.4}}
\thicklines \path(220,495)(240,495)
\thicklines \path(985,495)(965,495)
\put(198,495){\makebox(0,0)[r]{0.6}}
\thicklines \path(220,622)(240,622)
\thicklines \path(985,622)(965,622)
\put(198,622){\makebox(0,0)[r]{0.8}}
\thicklines \path(220,750)(240,750)
\thicklines \path(985,750)(965,750)
\put(198,750){\makebox(0,0)[r]{1}}
\thicklines \path(220,877)(240,877)
\thicklines \path(985,877)(965,877)
\put(198,877){\makebox(0,0)[r]{1.2}}
\thicklines \path(220,113)(220,133)
\thicklines \path(220,877)(220,857)
\put(220,68){\makebox(0,0){0}}
\thicklines \path(316,113)(316,133)
\thicklines \path(316,877)(316,857)
\put(316,68){\makebox(0,0){0.1}}
\thicklines \path(411,113)(411,133)
\thicklines \path(411,877)(411,857)
\put(411,68){\makebox(0,0){0.2}}
\thicklines \path(507,113)(507,133)
\thicklines \path(507,877)(507,857)
\put(507,68){\makebox(0,0){0.3}}
\thicklines \path(603,113)(603,133)
\thicklines \path(603,877)(603,857)
\put(603,68){\makebox(0,0){0.4}}
\thicklines \path(698,113)(698,133)
\thicklines \path(698,877)(698,857)
\put(698,68){\makebox(0,0){0.5}}
\thicklines \path(794,113)(794,133)
\thicklines \path(794,877)(794,857)
\put(794,68){\makebox(0,0){0.6}}
\thicklines \path(889,113)(889,133)
\thicklines \path(889,877)(889,857)
\put(889,68){\makebox(0,0){0.7}}
\thicklines \path(985,113)(985,133)
\thicklines \path(985,877)(985,857)
\put(985,68){\makebox(0,0){0.8}}
\thicklines \path(220,113)(985,113)(985,877)(220,877)(220,113)
\put(89,945){\makebox(0,0)[l]{\shortstack{$\mu/m$}}}
\put(602,23){\makebox(0,0){$kT/m$}}
\put(650,380){\makebox(0,0)[l]{G}}
\put(478,654){\makebox(0,0)[l]{E}}
\thinlines \dashline[-10]{25}(817,113)(817,117)(812,192)(800,265)(780,331)
(755,389)(726,439)(695,479)(664,512)(634,537)(604,556)(577,570)(552,580)
(529,587)(508,591)(489,594)(472,595)(456,596)(442,596)(430,595)(408,593)
(390,591)(375,588)(335,581)(295,575)(276,573)(257,571)(239,570)(220,570)
\thinlines \dashline[-10]{25}(220,644)(251,641)(281,637)(312,631)(343,623)
(374,616)(404,612)(435,607)(466,601)(497,593)
\thinlines \path(220,583)(220,583)(230,584)(233,584)(235,584)(241,584)
(247,584)(253,584)(265,585)(277,587)(300,590)(324,593)(347,597)(371,599)
(382,600)(388,601)(394,601)(397,601)(400,601)(403,601)(406,601)(407,601)
(409,601)(410,601)(410,601)(411,601)(412,601)(412,601)(413,601)(414,601)
(415,601)(415,601)(416,601)(417,601)(418,601)(419,601)(421,601)(424,601)
(426,601)(429,601)(435,601)(441,600)(453,599)(465,598)(476,596)(488,594)
(512,588)(535,580)(559,570)(582,558)(606,543)
\thinlines \path(606,543)(629,525)(653,504)(676,478)(700,448)(723,412)
(747,365)(758,337)(770,303)(782,260)(788,232)(794,193)(799,113)
\end{picture}

		\hspace*{5em}\parbox{15em}{\caption{The $T$-$\mu$ plot
			as given by
			Eq. (3.39) when $\sigma$ is
			fixed at $\sigma_m$ .}}
		\label{fig:solgap}
	\end{minipage}
\hfill
	\begin{minipage}[t]{.47\linewidth}
\setlength{\unitlength}{0.240900pt}
\begin{picture}(1049,900)(0,0)
\tenrm
\thinlines \dashline[-10]{25}(220,368)(985,368)
\thicklines \path(220,164)(240,164)
\thicklines \path(985,164)(965,164)
\put(198,164){\makebox(0,0)[r]{$-4\times 10^{-5}$}}
\thicklines \path(220,368)(240,368)
\thicklines \path(985,368)(965,368)
\put(198,368){\makebox(0,0)[r]{0}}
\thicklines \path(220,571)(240,571)
\thicklines \path(985,571)(965,571)
\put(198,571){\makebox(0,0)[r]{$4\times 10^{-5}$}}
\thicklines \path(220,775)(240,775)
\thicklines \path(985,775)(965,775)
\put(198,775){\makebox(0,0)[r]{$8\times 10^{-5}$}}
\thicklines \path(220,113)(220,133)
\thicklines \path(220,877)(220,857)
\put(220,68){\makebox(0,0){0}}
\thicklines \path(348,113)(348,133)
\thicklines \path(348,877)(348,857)
\put(348,68){\makebox(0,0){0.1}}
\thicklines \path(475,113)(475,133)
\thicklines \path(475,877)(475,857)
\put(475,68){\makebox(0,0){0.2}}
\thicklines \path(603,113)(603,133)
\thicklines \path(603,877)(603,857)
\put(603,68){\makebox(0,0){0.3}}
\thicklines \path(730,113)(730,133)
\thicklines \path(730,877)(730,857)
\put(730,68){\makebox(0,0){0.4}}
\thicklines \path(858,113)(858,133)
\thicklines \path(858,877)(858,857)
\put(858,68){\makebox(0,0){0.5}}
\thicklines \path(985,113)(985,133)
\thicklines \path(985,877)(985,857)
\put(985,68){\makebox(0,0){0.6}}
\thicklines \path(220,113)(985,113)(985,877)(220,877)(220,113)
\put(89,945){\makebox(0,0)[l]{\shortstack{$V/m^{2.5}$}}}
\put(602,23){\makebox(0,0){$\sigma/m$}}
\thinlines \path(220,368)(220,368)(221,368)(222,368)(223,368)(224,368)
(225,368)(227,368)(229,368)(233,368)(237,369)(245,370)(253,371)(269,375)
(285,381)(317,396)(349,416)(380,440)(412,467)(444,494)(476,520)(508,543)
(539,562)(555,569)(563,572)(571,574)(579,576)(583,577)(587,578)(591,578)
(593,578)(595,579)(596,579)(597,579)(598,579)(599,579)(600,579)(601,579)
(602,579)(603,579)(604,579)(605,579)(606,579)(607,579)(609,579)(611,578)
(615,578)(619,577)(623,577)(627,576)(635,574)
\thinlines \path(635,574)(643,571)(651,567)(667,558)(683,546)(699,532)
(730,495)(762,449)(794,395)(826,337)(858,278)(890,225)(905,203)(913,193)
(921,185)(929,178)(933,175)(937,173)(941,171)(945,169)(947,168)(949,168)
(950,168)(951,168)(952,167)(953,167)(954,167)(955,167)(956,167)(957,167)
(958,167)(959,167)(960,167)(961,167)(963,168)(965,168)(969,170)(973,172)
(977,175)(985,182)
\thinlines \path(220,368)(220,368)(221,368)(222,368)(223,368)(224,368)
(225,368)(227,368)(229,368)(233,368)(237,368)(245,369)(253,370)(269,373)
(285,377)(317,389)(349,403)(380,421)(412,440)(444,459)(476,478)(508,494)
(539,507)(555,512)(563,514)(571,515)(579,517)(583,517)(587,518)(591,518)
(593,518)(595,518)(596,518)(597,518)(598,518)(599,518)(600,518)(601,518)
(602,518)(603,518)(604,518)(605,518)(606,518)(607,518)(609,518)(611,518)
(615,518)(619,517)(627,516)(635,515)(651,511)
\thinlines \path(651,511)(667,506)(699,490)(730,470)(762,447)(794,424)
(810,414)(826,405)(834,401)(842,398)(846,397)(850,396)(854,395)(856,395)
(858,395)(859,394)(860,394)(861,394)(862,394)(863,394)(864,394)(865,394)
(866,394)(867,394)(868,394)(869,394)(870,394)(872,395)(874,395)(878,396)
(882,397)(886,398)(890,400)(897,405)(905,411)(913,419)(921,429)(937,455)
(953,491)(969,539)(985,598)
\thinlines \path(220,368)(220,368)(221,368)(222,368)(223,368)(224,368)
(225,368)(227,368)(229,368)(233,368)(237,368)(245,369)(253,369)(269,371)
(285,374)(317,381)(349,391)(380,402)(412,414)(444,425)(476,437)(508,446)
(539,453)(555,456)(563,457)(571,458)(579,458)(583,459)(587,459)(589,459)
(591,459)(593,459)(595,459)(596,459)(597,459)(598,459)(599,459)(600,459)
(601,459)(602,459)(603,459)(604,459)(605,459)(606,459)(607,459)(609,459)
(611,459)(615,459)(619,459)(627,458)(635,458)
\thinlines \path(635,458)(651,456)(667,454)(699,449)(715,447)(722,446)
(730,445)(734,445)(738,445)(742,444)(744,444)(745,444)(746,444)(747,444)
(748,444)(749,444)(750,444)(751,444)(752,444)(753,444)(754,444)(755,444)
(756,444)(758,444)(760,444)(762,445)(766,445)(770,445)(774,446)(778,447)
(786,449)(794,451)(802,454)(810,458)(826,469)(842,485)(858,506)(874,533)
(890,567)(905,610)(921,662)(953,801)(966,877)
\thinlines \path(220,368)(220,368)(221,368)(222,368)(223,368)(224,368)
(225,368)(227,368)(229,368)(233,368)(237,368)(245,368)(253,368)(269,369)
(285,371)(317,374)(349,378)(380,383)(412,388)(444,393)(476,397)(492,398)
(508,400)(524,401)(532,401)(539,401)(547,402)(555,402)(559,402)(563,402)
(567,402)(571,402)(575,402)(577,402)(579,402)(581,402)(583,402)(585,402)
(586,402)(587,402)(588,402)(589,402)(590,402)(591,402)(592,402)(593,402)
(594,402)(595,402)(596,402)(597,402)(599,402)
\thinlines \path(599,402)(600,402)(601,402)(602,402)(603,402)(604,402)
(605,402)(606,402)(607,402)(608,402)(609,402)(610,402)(611,402)(613,402)
(615,402)(617,402)(619,402)(621,402)(623,402)(627,402)(631,402)(635,402)
(639,403)(643,403)(651,403)(659,404)(667,404)(675,405)(683,406)(691,408)
(699,409)(715,414)(730,421)(746,430)(762,442)(778,457)(794,477)(810,501)
(826,531)(842,568)(858,612)(890,726)(920,877)
\thinlines \path(220,368)(220,368)(221,368)(222,368)(223,368)(224,368)
(225,368)(227,368)(229,368)(233,368)(237,368)(241,368)(245,368)(253,368)
(261,368)(269,368)(277,367)(285,367)(301,367)(317,367)(333,367)(349,366)
(364,366)(380,365)(412,363)(444,361)(476,358)(508,355)(539,351)(555,349)
(563,349)(571,348)(579,348)(583,347)(587,347)(591,347)(593,347)(595,347)
(596,347)(597,347)(598,347)(599,347)(600,347)(601,347)(602,347)(603,347)
(604,347)(605,347)(606,347)(607,347)(609,347)
\thinlines \path(609,347)(611,347)(613,347)(615,347)(619,347)(623,348)
(627,348)(635,349)(643,350)(651,352)(659,354)(667,356)(683,362)(699,371)
(715,382)(730,397)(746,416)(762,439)(778,467)(794,501)(826,590)(858,713)
(889,877)
\thinlines \path(220,368)(220,368)(221,368)(222,368)(223,368)(224,368)
(225,368)(227,368)(229,368)(233,368)(237,367)(245,367)(253,367)(269,366)
(285,364)(317,360)(349,355)(380,348)(412,340)(444,330)(476,321)(508,311)
(539,303)(555,299)(563,297)(571,296)(579,295)(583,295)(587,294)(591,294)
(593,294)(595,294)(596,294)(597,294)(598,294)(599,294)(600,294)(601,294)
(602,294)(603,294)(604,294)(605,294)(606,294)(607,294)(609,294)(611,294)
(615,294)(619,294)(623,295)(627,296)(635,297)
\thinlines \path(635,297)(643,299)(651,302)(667,309)(683,320)(699,334)
(715,351)(730,374)(746,402)(762,435)(794,523)(826,645)(858,809)(868,877)
\thinlines \dashline[-10]{25}(603,113)(603,877)
\end{picture}

		\hspace*{5em}\parbox{15em}{\caption{Behavior of
			the effective potential
			as $T$ and $\mu$ vary
			on the curve G ($\sigma_m=0.3$).}}
		\label{fig:epot}
	\end{minipage}
\end{figure}
Obviously the derivative of the effective potential, i. e.
$f(\sigma)$ , changes sign when the curve G crosses the critical
curve. On the critical curve the effective potential is flat
in the region $\sigma\leq\sigma_m$ as seen in Fig. 13.
Thus at the crossing E of
the curve G and the critical curve the second derivative of
the effective potential has to vanish,
\begin{equation}
	{\partial^2 V \over \partial\sigma^2}
	= f(\sigma) + \sigma{\partial f \over \partial \sigma}
	= 0 \, .
\label{eq:gapC2}
\end{equation}
Combining Eq. (\ref{eq:gapC}) with Eq. (\ref{eq:gapC2})
we have a condition for the point E,
\begin{equation}
	{\partial f \over \partial \sigma}
	= 0 \, .
\label{eq:gapC3}
\end{equation}
The point E divides the second-order region from the
first-order region along the curve G.
By letting $\sigma\rightarrow 0$ in
Eq. (\ref{eq:gapC2}) we reach to the point C in Fig. \ref{fig:phase25d}.
Hence the
condition to get the point C is given by
\begin{equation}
	\left.
		{\partial f \over \partial \sigma}
	\right|_{\sigma=0}
	= 0 \, .
\label{eq:gapC4}
\end{equation}
Inserting Eq. (\ref{gap:nontri}) into Eq. (\ref{eq:gapC4})
we finally obtain the condition for point C,
\begin{equation}
	{\rm Re\,}\zeta\left(
		5-D,\,\half + i {\betamu \over 2\pi}
	\right)=0 \, .
\label{cond:C}
\end{equation}
By solving Eq. (\ref{cond:C}) for $\betamu$ we have
a curve in the $T$-$\mu$ plane. Looking for the crossing
of this curve with the critical line we find
the value of $\beta$ and $\mu$ at point C.
In Fig. \ref{fig:3point} the value of $\beta\mu$ at point C is plotted
as a function of dimension $D$ by using Eq. (\ref{cond:C}).
\begin{figure}
\setlength{\unitlength}{0.240900pt}
\begin{picture}(1500,900)(0,0)
\tenrm
\thicklines \path(220,113)(240,113)
\thicklines \path(1436,113)(1416,113)
\put(198,113){\makebox(0,0)[r]{0}}
\thicklines \path(220,240)(240,240)
\thicklines \path(1436,240)(1416,240)
\put(198,240){\makebox(0,0)[r]{0.1}}
\thicklines \path(220,368)(240,368)
\thicklines \path(1436,368)(1416,368)
\put(198,368){\makebox(0,0)[r]{0.2}}
\thicklines \path(220,495)(240,495)
\thicklines \path(1436,495)(1416,495)
\put(198,495){\makebox(0,0)[r]{0.3}}
\thicklines \path(220,622)(240,622)
\thicklines \path(1436,622)(1416,622)
\put(198,622){\makebox(0,0)[r]{0.4}}
\thicklines \path(220,750)(240,750)
\thicklines \path(1436,750)(1416,750)
\put(198,750){\makebox(0,0)[r]{0.5}}
\thicklines \path(220,877)(240,877)
\thicklines \path(1436,877)(1416,877)
\put(198,877){\makebox(0,0)[r]{0.6}}
\thicklines \path(220,113)(220,133)
\thicklines \path(220,877)(220,857)
\put(220,68){\makebox(0,0){2}}
\thicklines \path(423,113)(423,133)
\thicklines \path(423,877)(423,857)
\put(423,68){\makebox(0,0){2.2}}
\thicklines \path(625,113)(625,133)
\thicklines \path(625,877)(625,857)
\put(625,68){\makebox(0,0){2.4}}
\thicklines \path(828,113)(828,133)
\thicklines \path(828,877)(828,857)
\put(828,68){\makebox(0,0){2.6}}
\thicklines \path(1031,113)(1031,133)
\thicklines \path(1031,877)(1031,857)
\put(1031,68){\makebox(0,0){2.8}}
\thicklines \path(1233,113)(1233,133)
\thicklines \path(1233,877)(1233,857)
\put(1233,68){\makebox(0,0){3}}
\thicklines \path(1436,113)(1436,133)
\thicklines \path(1436,877)(1436,857)
\put(1436,68){\makebox(0,0){3.2}}
\thicklines \path(220,113)(1436,113)(1436,877)(220,877)(220,113)
\put(45,945){\makebox(0,0)[l]{\shortstack{$kT/\mu$}}}
\put(916,23){\makebox(0,0){$D$}}
\thinlines \path(220,779)(220,779)(262,766)(304,752)(347,738)(389,724)
(431,709)(473,695)(516,680)(558,665)(600,650)(642,634)(684,618)(727,602)
(769,585)(811,567)(853,549)(895,530)(938,511)(980,490)(1022,468)(1064,444)
(1107,417)(1149,385)(1170,366)(1191,344)(1202,330)(1212,314)(1217,304)
(1223,291)(1225,283)(1228,273)(1229,267)(1231,259)(1232,247)(1233,216)
(1233,113)
\thinlines \dashline[-10]{25}(1233,113)(1233,877)
\end{picture}

\caption{The value of $kT/\mu$ at point C as a function of
dimension $D$ .}
\label{fig:3point}
\end{figure}
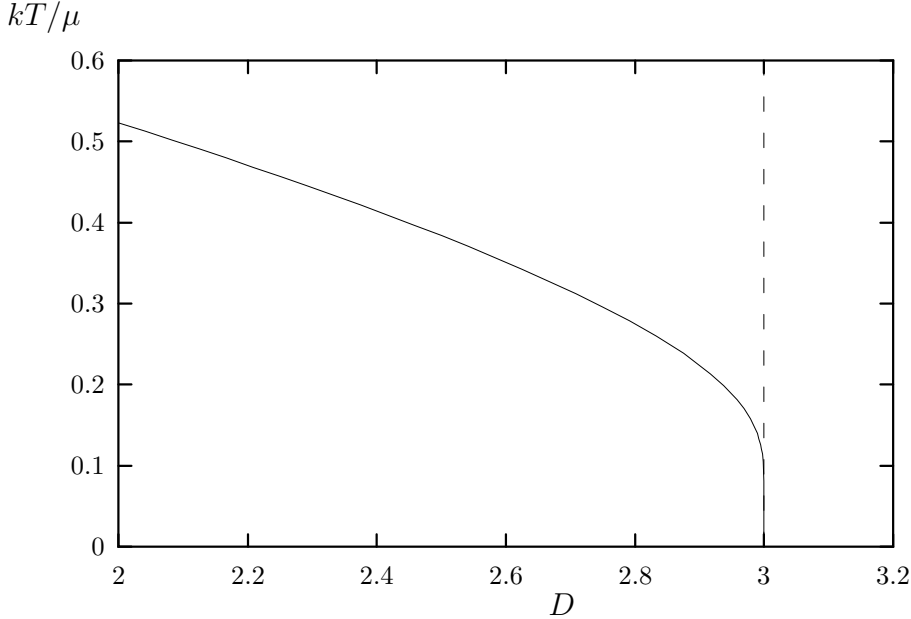

\subsubsection{Dynamical fermion mass at finite $T$ and $\mu$}

     We have seen that the broken chiral symmetry at low temperature and
chemical potential is restored at certain critical temperature and
chemical potential. The symmetry restoration is of second order for
$3\leq D < 4$ . For $2\leq D < 3$ two types of the phase transition,
first and second order, coexist.

     The dynamically generated fermion mass at low temperature and
chemical potential disappears above certain critical temperature and
chemical potential. In the following we shall examine the behavior of
the dynamical fermion mass as a function of the temperature and chemical
potential by numerically solving the gap equation. As typical dimensions
we choose $D = 2.5, 3.0, 3.5$ .

     In Fig. 15 we show the behavior of
the dynamical fermion mass as a
function of the chemical potential for some fixed temperatures.
We clearly observe that for $D=2.5$ the mass gap appears
at the critical chemical
potential as $T$ increases while no mass gap shows up
above $D=3$ . This
behavior at $D=2.5$ is a reflection of the smooth transition from the
first-order to second-order phase transition.
It should be noted here that, while at $D=3$ the phase
transition is of second order, there is an exceptional case
at $T=0$ where the mass gap exists at the critical chemical
potential $\mu=m$.
In Fig. 16 we present the behavior of
the dynamical fermion mass as a
function of the temperature for some fixed chemical
potentials.
Here for $D=2.5$ the mass gap observed at low temperature
disappears for higher temperatures while no mass gap is
observed above $D=3$ .
\vspace{8ex}
\begin{figure}[H]
\hspace*{-2em}
	\begin{minipage}[t]{.47\linewidth}
\setlength{\unitlength}{0.240900pt}
\begin{picture}(1049,900)(0,0)
\tenrm
\thicklines \path(220,113)(240,113)
\thicklines \path(985,113)(965,113)
\put(198,113){\makebox(0,0)[r]{0}}
\thicklines \path(220,240)(240,240)
\thicklines \path(985,240)(965,240)
\put(198,240){\makebox(0,0)[r]{0.2}}
\thicklines \path(220,368)(240,368)
\thicklines \path(985,368)(965,368)
\put(198,368){\makebox(0,0)[r]{0.4}}
\thicklines \path(220,495)(240,495)
\thicklines \path(985,495)(965,495)
\put(198,495){\makebox(0,0)[r]{0.6}}
\thicklines \path(220,622)(240,622)
\thicklines \path(985,622)(965,622)
\put(198,622){\makebox(0,0)[r]{0.8}}
\thicklines \path(220,750)(240,750)
\thicklines \path(985,750)(965,750)
\put(198,750){\makebox(0,0)[r]{1}}
\thicklines \path(220,877)(240,877)
\thicklines \path(985,877)(965,877)
\put(198,877){\makebox(0,0)[r]{1.2}}
\thicklines \path(220,113)(220,133)
\thicklines \path(220,877)(220,857)
\put(220,68){\makebox(0,0){0}}
\thicklines \path(373,113)(373,133)
\thicklines \path(373,877)(373,857)
\put(373,68){\makebox(0,0){0.2}}
\thicklines \path(526,113)(526,133)
\thicklines \path(526,877)(526,857)
\put(526,68){\makebox(0,0){0.4}}
\thicklines \path(679,113)(679,133)
\thicklines \path(679,877)(679,857)
\put(679,68){\makebox(0,0){0.6}}
\thicklines \path(832,113)(832,133)
\thicklines \path(832,877)(832,857)
\put(832,68){\makebox(0,0){0.8}}
\thicklines \path(985,113)(985,133)
\thicklines \path(985,877)(985,857)
\put(985,68){\makebox(0,0){1}}
\thicklines \path(220,113)(985,113)(985,877)(220,877)(220,113)
\put(133,945){\makebox(0,0)[l]{\shortstack{$m_{\beta\mu}/m$}}}
\put(602,23){\makebox(0,0){$\mu/m$}}
\put(679,782){\makebox(0,0)[l]{$kT/m=0$}}
\put(373,686){\makebox(0,0)[l]{$kT/m=0.2$}}
\put(297,559){\makebox(0,0)[l]{$kT/m=0.4$}}
\thinlines \path(220,750)(220,750)(247,750)(273,750)(300,750)(326,750)
(353,750)(380,750)(406,750)(433,750)(459,750)(486,750)(512,750)(539,750)
(566,750)(592,750)(619,750)(645,750)(672,750)(699,750)(725,750)(752,750)
(778,750)(805,750)(832,750)(858,750)
\thinlines \dashline[-10]{25}(814,580)(814,113)
\thinlines \dashline[-10]{25}(858,750)(858,113)
\thinlines \path(220,746)(220,746)(228,746)(231,746)(233,746)(239,746)
(245,746)(251,746)(263,746)(274,746)(286,746)(298,746)(309,746)(321,746)
(324,746)(325,746)(327,746)(327,746)(328,746)(328,746)(329,746)(329,746)
(330,746)(333,746)(344,745)(356,745)(368,745)(379,745)(391,744)(403,744)
(414,744)(426,743)(437,743)(449,742)(461,742)(472,741)(484,740)(496,740)
(507,739)(519,738)(531,737)(542,736)(554,735)(566,734)(577,732)(589,731)
(601,729)(612,727)(624,725)(636,723)(647,720)
\thinlines \path(647,720)(659,718)(671,715)(682,711)(694,707)(706,703)
(717,698)(729,692)(741,686)(752,678)(764,669)(776,658)(787,644)(799,624)
(814,580)
\thinlines \path(220,671)(220,671)(228,671)(231,671)(234,671)(240,671)
(246,670)(252,670)(265,670)(277,669)(289,668)(302,668)(314,666)(326,665)
(339,664)(351,662)(363,660)(376,658)(388,656)(400,654)(413,651)(425,648)
(437,645)(450,642)(462,638)(474,634)(487,630)(499,626)(511,621)(524,615)
(536,609)(548,603)(561,596)(573,589)(585,580)(598,571)(610,561)(622,550)
(635,538)(647,525)(659,509)(672,492)(684,472)(696,448)(709,419)(721,383)
(727,361)(733,335)(739,301)(743,282)(746,258)
\thinlines \path(746,258)(747,244)(749,228)(750,209)(751,204)(751,199)
(752,184)(753,141)(754,113)
\end{picture}

		\hspace*{10em}\mbox{(a) $D=2.5$}
		\vspace{5ex}
\setlength{\unitlength}{0.240900pt}
\begin{picture}(1049,900)(0,0)
\tenrm
\thicklines \path(220,113)(240,113)
\thicklines \path(985,113)(965,113)
\put(198,113){\makebox(0,0)[r]{0}}
\thicklines \path(220,240)(240,240)
\thicklines \path(985,240)(965,240)
\put(198,240){\makebox(0,0)[r]{0.2}}
\thicklines \path(220,368)(240,368)
\thicklines \path(985,368)(965,368)
\put(198,368){\makebox(0,0)[r]{0.4}}
\thicklines \path(220,495)(240,495)
\thicklines \path(985,495)(965,495)
\put(198,495){\makebox(0,0)[r]{0.6}}
\thicklines \path(220,622)(240,622)
\thicklines \path(985,622)(965,622)
\put(198,622){\makebox(0,0)[r]{0.8}}
\thicklines \path(220,750)(240,750)
\thicklines \path(985,750)(965,750)
\put(198,750){\makebox(0,0)[r]{1}}
\thicklines \path(220,877)(240,877)
\thicklines \path(985,877)(965,877)
\put(198,877){\makebox(0,0)[r]{1.2}}
\thicklines \path(220,113)(220,133)
\thicklines \path(220,877)(220,857)
\put(220,68){\makebox(0,0){0}}
\thicklines \path(316,113)(316,133)
\thicklines \path(316,877)(316,857)
\put(316,68){\makebox(0,0){0.2}}
\thicklines \path(411,113)(411,133)
\thicklines \path(411,877)(411,857)
\put(411,68){\makebox(0,0){0.4}}
\thicklines \path(507,113)(507,133)
\thicklines \path(507,877)(507,857)
\put(507,68){\makebox(0,0){0.6}}
\thicklines \path(603,113)(603,133)
\thicklines \path(603,877)(603,857)
\put(603,68){\makebox(0,0){0.8}}
\thicklines \path(698,113)(698,133)
\thicklines \path(698,877)(698,857)
\put(698,68){\makebox(0,0){1}}
\thicklines \path(794,113)(794,133)
\thicklines \path(794,877)(794,857)
\put(794,68){\makebox(0,0){1.2}}
\thicklines \path(889,113)(889,133)
\thicklines \path(889,877)(889,857)
\put(889,68){\makebox(0,0){1.4}}
\thicklines \path(985,113)(985,133)
\thicklines \path(985,877)(985,857)
\put(985,68){\makebox(0,0){1.6}}
\thicklines \path(220,113)(985,113)(985,877)(220,877)(220,113)
\put(133,945){\makebox(0,0)[l]{\shortstack{$m_{\beta\mu}/m$}}}
\put(602,23){\makebox(0,0){$\mu /m$}}
\put(603,782){\makebox(0,0)[l]{$kT/m=0$}}
\put(373,622){\makebox(0,0)[l]{$kT/m=0.4$}}
\thinlines \path(220,750)(220,750)(698,750)(702,748)(705,747)(709,745)
(712,743)(716,741)(719,739)(723,737)(726,735)(730,732)(733,730)(737,727)
(740,724)(744,721)(748,718)(751,715)(755,711)(758,708)(762,704)(765,700)
(769,696)(772,692)(776,688)(779,683)(783,679)(786,674)(790,669)(793,665)
(797,660)(800,655)(804,650)(808,644)(811,639)(815,633)(818,628)(822,622)
(825,616)(829,610)(832,603)(836,597)(839,590)(843,583)(846,576)(850,568)
(853,560)(857,552)(860,544)(864,536)(867,527)
\thinlines \path(867,527)(871,518)(875,509)(878,499)(882,489)(885,479)
(889,468)(892,457)(896,446)(899,433)(903,421)(906,407)(910,392)(913,376)
(917,360)(920,342)(924,322)(927,301)(931,277)(935,242)(938,189)(942,113)
\thinlines \path(220,749)(220,749)(236,749)(253,749)(269,749)(286,749)
(302,749)(319,749)(335,748)(352,748)(368,748)(385,748)(401,747)(418,747)
(434,746)(451,746)(467,745)(484,744)(500,743)(517,742)(533,741)(550,739)
(566,737)(583,735)(599,732)(616,728)(632,724)(649,720)(665,714)(682,707)
(698,700)
\thinlines \path(698,700)(698,700)(702,698)(706,695)(710,693)(714,691)
(718,689)(722,686)(726,683)(730,681)(734,678)(738,675)(742,672)(745,669)
(749,666)(753,662)(757,659)(761,655)(765,651)(769,647)(773,643)(777,639)
(781,635)(785,630)(789,626)(793,621)(797,616)(801,610)(805,605)(809,599)
(813,594)(816,588)(820,581)(824,575)(828,568)(832,561)(836,554)(840,546)
(844,538)(848,530)(852,522)(856,513)(860,504)(864,494)(868,484)(872,473)
(876,463)(880,451)(883,439)(887,426)(891,413)
\thinlines \path(891,413)(895,398)(899,383)(903,366)(907,349)(911,329)
(915,305)(919,278)(923,251)(927,213)(931,113)
\thinlines \path(220,728)(220,728)(232,728)(245,728)(257,728)(269,728)
(281,727)(294,727)(306,726)(318,725)(330,725)(343,724)(355,723)(367,722)
(379,720)(392,719)(404,718)(416,716)(428,714)(441,712)(453,710)(465,708)
(477,706)(490,703)(502,700)(514,697)(526,694)(539,691)(551,687)(563,683)
(576,678)(588,674)(600,669)(612,663)(625,657)(637,651)(649,644)(661,637)
(674,629)(686,620)(698,611)
\thinlines \path(698,611)(698,611)(701,608)(705,606)(708,603)(711,600)
(715,597)(718,594)(721,591)(724,588)(728,585)(731,582)(734,579)(738,575)
(741,572)(744,568)(747,565)(751,561)(754,557)(757,553)(760,550)(764,545)
(767,541)(770,537)(774,533)(777,528)(780,524)(783,519)(787,514)(790,509)
(793,504)(797,499)(800,493)(803,487)(806,482)(810,476)(813,470)(816,463)
(820,457)(823,450)(826,443)(829,435)(833,428)(836,420)(839,412)(843,403)
(846,394)(849,385)(852,375)(856,365)(859,354)
\thinlines \path(859,354)(862,342)(866,330)(869,316)(872,302)(875,286)
(879,269)(882,247)(885,225)(889,195)(892,113)
\end{picture}
		\hspace*{10em}(c) $D=3.5$
	\end{minipage}
\hfill
	\begin{minipage}[t]{.47\linewidth}
\setlength{\unitlength}{0.240900pt}
\begin{picture}(1049,900)(0,0)
\tenrm
\thicklines \path(220,113)(240,113)
\thicklines \path(985,113)(965,113)
\put(198,113){\makebox(0,0)[r]{0}}
\thicklines \path(220,240)(240,240)
\thicklines \path(985,240)(965,240)
\put(198,240){\makebox(0,0)[r]{0.2}}
\thicklines \path(220,368)(240,368)
\thicklines \path(985,368)(965,368)
\put(198,368){\makebox(0,0)[r]{0.4}}
\thicklines \path(220,495)(240,495)
\thicklines \path(985,495)(965,495)
\put(198,495){\makebox(0,0)[r]{0.6}}
\thicklines \path(220,622)(240,622)
\thicklines \path(985,622)(965,622)
\put(198,622){\makebox(0,0)[r]{0.8}}
\thicklines \path(220,750)(240,750)
\thicklines \path(985,750)(965,750)
\put(198,750){\makebox(0,0)[r]{1}}
\thicklines \path(220,877)(240,877)
\thicklines \path(985,877)(965,877)
\put(198,877){\makebox(0,0)[r]{1.2}}
\thicklines \path(220,113)(220,133)
\thicklines \path(220,877)(220,857)
\put(220,68){\makebox(0,0){0}}
\thicklines \path(348,113)(348,133)
\thicklines \path(348,877)(348,857)
\put(348,68){\makebox(0,0){0.2}}
\thicklines \path(475,113)(475,133)
\thicklines \path(475,877)(475,857)
\put(475,68){\makebox(0,0){0.4}}
\thicklines \path(603,113)(603,133)
\thicklines \path(603,877)(603,857)
\put(603,68){\makebox(0,0){0.6}}
\thicklines \path(730,113)(730,133)
\thicklines \path(730,877)(730,857)
\put(730,68){\makebox(0,0){0.8}}
\thicklines \path(858,113)(858,133)
\thicklines \path(858,877)(858,857)
\put(858,68){\makebox(0,0){1}}
\thicklines \path(985,113)(985,133)
\thicklines \path(985,877)(985,857)
\put(985,68){\makebox(0,0){1.2}}
\thicklines \path(220,113)(985,113)(985,877)(220,877)(220,113)
\put(133,945){\makebox(0,0)[l]{\shortstack{$m_{\beta\mu}/m$}}}
\put(602,23){\makebox(0,0){$\mu/m$}}
\put(622,782){\makebox(0,0)[l]{$kT/m=0$}}
\put(297,591){\makebox(0,0)[l]{$kT/m=0.4$}}
\thinlines \path(221,748)(221,748)(221,748)(222,748)(222,748)(222,748)
(223,748)(224,748)(226,748)(228,748)(231,748)(235,748)(242,748)(249,748)
(256,748)(264,748)(278,748)(292,748)(306,748)(321,747)(335,747)(349,747)
(364,747)(378,746)(392,746)(407,746)(421,745)(435,745)(449,744)(464,744)
(478,743)(492,742)(507,741)(521,740)(535,739)(549,738)(564,736)(578,735)
(592,733)(607,730)(621,728)(635,725)(650,722)(664,718)(678,714)(692,709)
(707,703)(721,696)(735,688)(750,678)(764,666)
\thinlines \path(764,666)(778,652)(793,633)(800,621)(807,608)(814,592)
(821,572)(828,547)(832,532)(835,515)(839,494)(843,468)(846,435)(848,414)
(850,389)(851,372)(851,362)(852,357)(852,332)(853,305)(854,292)(854,272)
(855,220)(856,172)(856,113)(856,113)(857,113)(859,113)(860,113)(864,113)
(878,113)(893,113)(907,113)(921,113)
\thinlines \path(221,702)(221,702)(221,702)(222,702)(222,702)(222,702)
(223,702)(224,701)(226,701)(228,701)(231,701)(235,701)(242,701)(249,701)
(256,701)(264,701)(278,700)(292,699)(306,699)(321,697)(335,696)(349,695)
(364,693)(378,691)(392,689)(407,687)(421,684)(435,681)(449,678)(464,675)
(478,671)(492,667)(507,663)(521,658)(535,653)(549,647)(564,641)(578,634)
(592,626)(607,618)(621,609)(635,599)(650,588)(664,575)(678,561)(692,546)
(707,528)(721,507)(735,483)(750,454)(764,419)
\thinlines \path(764,419)(778,374)(785,346)(793,310)(794,300)(795,295)
(796,295)(796,291)(797,285)(798,278)(800,265)(803,234)(805,215)(807,188)
(809,153)(809,135)(810,113)
\thinlines \path(220,750)(220,750)(858,750)
\thinlines \dashline[-10]{25}(858,750)(858,113)
\end{picture}

		\hspace*{10em}(b) $D=3$
		\vglue 10ex
		\hspace*{5em}
		\parbox{15em}{\caption{Dynamical fermion mas $m_\betamu$ as
			a function of the chemical
			potential $\mu$ with temperature $kT/m$ fixed
			at 0,0.2,0.4.}}
	\end{minipage}
\vspace{1.5ex}
\label{fig:chemmass}
\end{figure}
\begin{figure}[H]
\hspace*{-2em}
	\begin{minipage}[t]{.47\linewidth}
\setlength{\unitlength}{0.240900pt}
\begin{picture}(1049,900)(0,0)
\tenrm
\thicklines \path(220,113)(240,113)
\thicklines \path(985,113)(965,113)
\put(198,113){\makebox(0,0)[r]{0}}
\thicklines \path(220,240)(240,240)
\thicklines \path(985,240)(965,240)
\put(198,240){\makebox(0,0)[r]{0.2}}
\thicklines \path(220,368)(240,368)
\thicklines \path(985,368)(965,368)
\put(198,368){\makebox(0,0)[r]{0.4}}
\thicklines \path(220,495)(240,495)
\thicklines \path(985,495)(965,495)
\put(198,495){\makebox(0,0)[r]{0.6}}
\thicklines \path(220,622)(240,622)
\thicklines \path(985,622)(965,622)
\put(198,622){\makebox(0,0)[r]{0.8}}
\thicklines \path(220,750)(240,750)
\thicklines \path(985,750)(965,750)
\put(198,750){\makebox(0,0)[r]{1}}
\thicklines \path(220,877)(240,877)
\thicklines \path(985,877)(965,877)
\put(198,877){\makebox(0,0)[r]{1.2}}
\thicklines \path(220,113)(220,133)
\thicklines \path(220,877)(220,857)
\put(220,68){\makebox(0,0){0}}
\thicklines \path(329,113)(329,133)
\thicklines \path(329,877)(329,857)
\put(329,68){\makebox(0,0){0.1}}
\thicklines \path(439,113)(439,133)
\thicklines \path(439,877)(439,857)
\put(439,68){\makebox(0,0){0.2}}
\thicklines \path(548,113)(548,133)
\thicklines \path(548,877)(548,857)
\put(548,68){\makebox(0,0){0.3}}
\thicklines \path(657,113)(657,133)
\thicklines \path(657,877)(657,857)
\put(657,68){\makebox(0,0){0.4}}
\thicklines \path(766,113)(766,133)
\thicklines \path(766,877)(766,857)
\put(766,68){\makebox(0,0){0.5}}
\thicklines \path(876,113)(876,133)
\thicklines \path(876,877)(876,857)
\put(876,68){\makebox(0,0){0.6}}
\thicklines \path(985,113)(985,133)
\thicklines \path(985,877)(985,857)
\put(985,68){\makebox(0,0){0.7}}
\thicklines \path(220,113)(985,113)(985,877)(220,877)(220,113)
\put(133,945){\makebox(0,0)[l]{\shortstack{$m_{\beta\mu}/m$}}}
\put(602,23){\makebox(0,0){$kT/m$}}
\put(712,686){\makebox(0,0)[l]{$\mu/m =0$}}
\put(526,431){\makebox(0,0)[l]{$\mu/m =0.4$}}
\put(395,571){\makebox(0,0)[l]{$\mu/m =0.8$}}
\thinlines \path(220,750)(220,750)(234,749)(248,749)(262,749)(276,749)
(290,749)(304,749)(318,749)(388,749)(402,749)(416,748)(430,747)(444,746)
(458,745)(472,743)(486,741)(500,738)(514,735)(528,731)(542,727)(556,723)
(570,718)(584,712)(598,705)(612,698)(626,691)(640,682)(654,673)(668,663)
(682,652)(696,640)(710,627)(724,613)(738,598)(752,581)(766,563)
\thinlines \path(766,563)(766,563)(774,553)(781,543)(788,532)(795,520)
(802,508)(809,496)(817,483)(824,468)(831,453)(838,437)(845,420)(852,402)
(859,383)(867,361)(874,335)(881,303)(888,273)(895,232)(902,113)
\thinlines \path(220,750)(220,750)(326,749)(341,749)(356,748)(371,747)
(386,746)(401,744)(416,742)(431,739)(446,735)(461,731)(476,727)(491,722)
(506,716)(521,709)(537,702)(552,694)(567,686)(582,677)(597,666)(612,655)
(627,643)(642,629)(657,614)
\thinlines \path(657,614)(657,614)(664,608)(670,601)(676,594)(683,586)
(689,579)(696,571)(702,562)(708,554)(715,545)(721,536)(728,526)(734,517)
(740,506)(747,495)(753,484)(760,472)(766,459)(772,445)(779,431)(785,416)
(792,400)(798,382)(804,364)(811,344)(817,320)(824,291)(830,258)(836,219)
(843,113)
\thinlines \path(220,750)(220,750)(242,750)(264,749)(286,746)(307,738)
(329,726)(340,717)(351,706)(357,700)(358,698)(360,695)(361,694)(362,693)
(363,691)(364,691)(364,691)
\thinlines \dashline[-10]{25}(364,691)(364,113)
\end{picture}

		\hspace*{10em}\mbox{(a) $D=2.5$}
		\vspace{5ex}
\setlength{\unitlength}{0.240900pt}
\begin{picture}(1049,900)(0,0)
\tenrm
\thicklines \path(220,113)(240,113)
\thicklines \path(985,113)(965,113)
\put(198,113){\makebox(0,0)[r]{0}}
\thicklines \path(220,240)(240,240)
\thicklines \path(985,240)(965,240)
\put(198,240){\makebox(0,0)[r]{0.2}}
\thicklines \path(220,368)(240,368)
\thicklines \path(985,368)(965,368)
\put(198,368){\makebox(0,0)[r]{0.4}}
\thicklines \path(220,495)(240,495)
\thicklines \path(985,495)(965,495)
\put(198,495){\makebox(0,0)[r]{0.6}}
\thicklines \path(220,622)(240,622)
\thicklines \path(985,622)(965,622)
\put(198,622){\makebox(0,0)[r]{0.8}}
\thicklines \path(220,750)(240,750)
\thicklines \path(985,750)(965,750)
\put(198,750){\makebox(0,0)[r]{1}}
\thicklines \path(220,877)(240,877)
\thicklines \path(985,877)(965,877)
\put(198,877){\makebox(0,0)[r]{1.2}}
\thicklines \path(220,113)(220,133)
\thicklines \path(220,877)(220,857)
\put(220,68){\makebox(0,0){0}}
\thicklines \path(359,113)(359,133)
\thicklines \path(359,877)(359,857)
\put(359,68){\makebox(0,0){0.2}}
\thicklines \path(498,113)(498,133)
\thicklines \path(498,877)(498,857)
\put(498,68){\makebox(0,0){0.4}}
\thicklines \path(637,113)(637,133)
\thicklines \path(637,877)(637,857)
\put(637,68){\makebox(0,0){0.6}}
\thicklines \path(776,113)(776,133)
\thicklines \path(776,877)(776,857)
\put(776,68){\makebox(0,0){0.8}}
\thicklines \path(915,113)(915,133)
\thicklines \path(915,877)(915,857)
\put(915,68){\makebox(0,0){1}}
\thicklines \path(220,113)(985,113)(985,877)(220,877)(220,113)
\put(133,945){\makebox(0,0)[l]{\shortstack{$m_{\beta\mu}/m$}}}
\put(602,23){\makebox(0,0){$kT/m$}}
\put(700,686){\makebox(0,0)[l]{$\mu /m=0$}}
\put(422,444){\makebox(0,0)[l]{$\mu /m=0.8$}}
\thinlines \path(220,750)(220,750)(231,750)(241,749)(252,749)(263,750)
(273,750)(338,749)(348,749)(359,749)(370,749)(380,748)(391,747)(402,746)
(413,745)(423,744)(434,743)(445,741)(455,739)(466,737)(477,734)(487,731)
(498,728)(509,725)(520,721)(530,717)(541,713)(552,708)(562,703)(573,697)
(584,691)(594,685)(605,678)(616,671)(627,664)(637,656)
\thinlines \path(637,656)(637,656)(641,652)(645,649)(649,646)(653,643)
(657,639)(661,636)(665,632)(669,628)(673,625)(677,621)(681,617)(685,613)
(689,609)(693,605)(697,601)(701,596)(705,592)(709,588)(713,583)(717,578)
(721,574)(725,569)(729,564)(733,559)(737,553)(741,548)(745,542)(750,537)
(754,531)(758,525)(762,518)(766,512)(770,505)(774,498)(778,491)(782,484)
(786,477)(790,469)(794,461)(798,453)(802,444)(806,436)(810,427)(814,418)
(818,409)(822,399)(826,389)(830,378)(834,367)
\thinlines \path(834,367)(838,354)(842,340)(846,325)(850,309)(854,291)
(858,273)(862,254)(866,229)(870,185)(874,113)
\thinlines \path(220,750)(220,750)(234,750)(249,750)(292,750)(306,749)
(321,749)(335,749)(349,748)(364,747)(378,745)(393,743)(407,741)(421,739)
(436,735)(450,732)(465,728)(479,723)(493,719)(508,713)(522,707)(537,700)
(551,693)(565,685)(580,676)(594,667)(608,656)(623,645)(637,633)
\thinlines \path(637,633)(637,633)(641,630)(645,627)(648,624)(652,620)
(656,617)(659,613)(663,610)(667,606)(670,603)(674,599)(678,595)(681,591)
(685,587)(689,583)(692,579)(696,574)(700,570)(703,566)(707,561)(711,556)
(714,552)(718,547)(722,542)(725,537)(729,532)(733,526)(736,521)(740,515)
(744,510)(747,504)(751,498)(755,491)(758,485)(762,478)(766,472)(769,465)
(773,458)(777,451)(780,444)(784,437)(788,429)(791,421)(795,412)(799,403)
(802,394)(806,383)(810,372)(813,360)(817,348)
\thinlines \path(817,348)(821,335)(824,323)(828,311)(832,299)(835,287)
(839,273)(843,252)(846,220)(850,174)(854,113)
\thinlines \path(220,750)(220,750)(268,750)(280,750)(292,748)(304,746)
(316,743)(328,740)(340,737)(352,733)(364,729)(376,725)(388,721)(400,716)
(412,712)(424,707)(436,701)(448,695)(460,689)(472,683)(484,676)(496,669)
(508,662)(520,654)(532,645)(544,637)(556,627)(568,617)
\thinlines \path(568,617)(568,617)(571,614)(575,611)(579,608)(583,604)
(586,601)(590,598)(594,594)(598,590)(601,587)(605,583)(609,579)(613,575)
(616,571)(620,567)(624,563)(628,559)(631,555)(635,551)(639,546)(642,542)
(646,537)(650,532)(654,527)(657,522)(661,517)(665,512)(669,507)(672,502)
(676,496)(680,491)(684,485)(687,479)(691,473)(695,467)(699,460)(702,454)
(706,447)(710,440)(714,432)(717,425)(721,417)(725,409)(728,401)(732,392)
(736,384)(740,375)(743,366)(747,356)(751,345)
\thinlines \path(751,345)(755,334)(758,321)(762,307)(766,292)(770,275)
(773,258)(777,242)(781,222)(785,183)(788,113)
\end{picture}

		\hspace*{10em}(c) $D=3.5$
	\end{minipage}
\hfill
	\begin{minipage}[t]{.47\linewidth}
\setlength{\unitlength}{0.240900pt}
\begin{picture}(1049,900)(0,0)
\tenrm
\thicklines \path(220,113)(240,113)
\thicklines \path(985,113)(965,113)
\put(198,113){\makebox(0,0)[r]{0}}
\thicklines \path(220,240)(240,240)
\thicklines \path(985,240)(965,240)
\put(198,240){\makebox(0,0)[r]{0.2}}
\thicklines \path(220,368)(240,368)
\thicklines \path(985,368)(965,368)
\put(198,368){\makebox(0,0)[r]{0.4}}
\thicklines \path(220,495)(240,495)
\thicklines \path(985,495)(965,495)
\put(198,495){\makebox(0,0)[r]{0.6}}
\thicklines \path(220,622)(240,622)
\thicklines \path(985,622)(965,622)
\put(198,622){\makebox(0,0)[r]{0.8}}
\thicklines \path(220,750)(240,750)
\thicklines \path(985,750)(965,750)
\put(198,750){\makebox(0,0)[r]{1}}
\thicklines \path(220,877)(240,877)
\thicklines \path(985,877)(965,877)
\put(198,877){\makebox(0,0)[r]{1.2}}
\thicklines \path(220,113)(220,133)
\thicklines \path(220,877)(220,857)
\put(220,68){\makebox(0,0){0}}
\thicklines \path(316,113)(316,133)
\thicklines \path(316,877)(316,857)
\put(316,68){\makebox(0,0){0.1}}
\thicklines \path(411,113)(411,133)
\thicklines \path(411,877)(411,857)
\put(411,68){\makebox(0,0){0.2}}
\thicklines \path(507,113)(507,133)
\thicklines \path(507,877)(507,857)
\put(507,68){\makebox(0,0){0.3}}
\thicklines \path(603,113)(603,133)
\thicklines \path(603,877)(603,857)
\put(603,68){\makebox(0,0){0.4}}
\thicklines \path(698,113)(698,133)
\thicklines \path(698,877)(698,857)
\put(698,68){\makebox(0,0){0.5}}
\thicklines \path(794,113)(794,133)
\thicklines \path(794,877)(794,857)
\put(794,68){\makebox(0,0){0.6}}
\thicklines \path(889,113)(889,133)
\thicklines \path(889,877)(889,857)
\put(889,68){\makebox(0,0){0.7}}
\thicklines \path(985,113)(985,133)
\thicklines \path(985,877)(985,857)
\put(985,68){\makebox(0,0){0.8}}
\thicklines \path(220,113)(985,113)(985,877)(220,877)(220,113)
\put(133,945){\makebox(0,0)[l]{\shortstack{$m_{\beta\mu}/m$}}}
\put(602,23){\makebox(0,0){$kT/m$}}
\put(736,686){\makebox(0,0)[l]{$\mu/m=0$}}
\put(335,444){\makebox(0,0)[l]{$\mu/m=0.8$}}
\thinlines \path(220,750)(220,750)(233,749)(246,749)(260,749)(273,749)
(286,749)(299,749)(312,750)(365,750)(378,749)(391,749)(405,748)(418,748)
(431,747)(444,745)(457,744)(471,742)(484,740)(497,737)(510,734)(523,731)
(537,727)(550,723)(563,718)(576,713)(589,707)(603,701)
\thinlines \path(603,701)(603,701)(610,697)(618,693)(626,689)(634,685)
(642,680)(650,675)(658,670)(666,664)(673,658)(681,652)(689,646)(697,639)
(705,631)(713,623)(721,615)(729,607)(736,598)(744,588)(752,579)(760,569)
(768,559)(776,548)(784,538)(792,527)(799,515)(807,504)(815,491)(823,477)
(831,462)(839,446)(847,429)(855,410)(863,389)(870,366)(878,339)(886,309)
(894,277)(902,222)(910,113)
\thinlines \path(220,750)(220,750)(233,750)(246,750)(312,750)(326,749)
(339,749)(352,748)(365,748)(378,747)(391,745)(405,744)(418,742)(431,740)
(444,737)(457,734)(471,731)(484,727)(497,723)(510,719)(523,713)(537,708)
(550,701)(563,695)(576,688)(589,680)(603,672)
\thinlines \path(603,672)(603,672)(609,668)(616,663)(623,659)(630,654)
(637,649)(644,643)(650,637)(657,631)(664,624)(671,617)(678,610)(685,602)
(692,595)(698,587)(705,580)(712,573)(719,565)(726,558)(733,549)(740,541)
(746,531)(753,521)(760,510)(767,499)(774,488)(781,476)(788,463)(794,449)
(801,435)(808,420)(815,403)(822,386)(829,367)(835,345)(842,320)(849,293)
(856,265)(863,213)(870,113)
\thinlines \path(220,750)(220,750)(266,750)(278,749)(289,747)(301,745)
(312,741)(324,738)(335,733)(347,728)(358,723)(370,717)(382,710)(393,703)
(405,696)(416,688)(428,679)(439,671)(451,661)(462,652)(474,642)(485,631)
(497,620)(509,609)(520,597)(532,585)(543,572)(555,557)
\thinlines \path(555,557)(555,557)(559,552)(563,547)(567,541)(571,535)
(575,529)(579,523)(583,517)(588,511)(592,504)(596,498)(600,491)(604,485)
(608,478)(612,472)(616,465)(620,458)(625,451)(629,444)(633,437)(637,430)
(641,422)(645,414)(649,406)(653,398)(657,389)(661,380)(666,370)(670,360)
(674,349)(678,338)(682,325)(686,312)(690,297)(694,281)(698,264)(703,247)
(707,222)(711,180)(715,113)
\end{picture}

		\hspace*{10em}(b) $D=3$
		\vglue 10ex
		\hspace*{5em}
		\parbox{15em}{\caption{Dynamical fermion
			mass $m_\betamu$ as a function
			of the temperature $T$ with the chemical
			potential $\mu/m$ fixed at 0,0.4,0.8.}}
	\end{minipage}
\vspace{1.5ex}
\label{fig:tmass}
\end{figure}

\subsection{CONCLUSION}

     We have investigated the phase structure of four-fermion theories
at finite temperature $T$ and chemical potential $\mu$ in arbitrary
dimensions by using the effective potential and gap equation in the
leading order of the $1/N$ expansion. The theory under consideration is
renormalizable below 4 dimensions and give an insight into the phase
structure of the theory in 4 dimensions.

     Starting from the theory with broken chiral symmetry
at vanishing $T$ and $\mu$ we calculated the renormalized
effective potential for finite $T$ and $\mu$ in the leading
order of the $1/N$ expansion in arbitrary dimensions
$2\leq D < 4$ . We found that the broken chiral symmetry was restored at a
certain critical temperature and chemical potential.
The phase transition from the broken phase to the symmetric phase
is either of first
order or of second order for $2\leq D < 3$ and is of second order
for $3\leq D < 4$ .
We found the boundary curve dividing the symmetric phase and broken
phase in the $T$-$\mu$ plane. The dynamical mass generated in the broken
phase is studied as a function of the temperature and chemical potential.

     At $D=2$ and $D=3$ formulae derived in the present investigation mostly
reduce to the known results in the preceding works. Some results
obtained in the present work are not known previously. On the critical
curve we found analytic expressions for some specific points.

     Although the present work is restricted mostly to the analysis of
the mathematical properties of the four fermion theory, we are
interested in applying our results to physical problems in the early stage of
the Universe. We will continue our work further and hope to publish
reports on these problems.

\subsection*{ACKNOWLEDGMENTS}

     The authors would like to thank Emilio Elizalde, Teiji Kunihiro and
Akira Niegawa for useful conversations and Kozo Mukai for a preliminary
contribution at the early stage of this work. We are indebted to members
of our Laboratory for encouragements and discussions.

\newpage

\appendix
\renewcommand{\thesubsection}{APPENDIX \Alph{subsection}}
\subsection{}
\renewcommand{\thesubsection}{\Alph{subsection}}
\noindent
EFFECTIVE POTENTIAL FOR $T=\mu=0$ \\[3mm]
In this appendix we present details of the calculation
of the effective potential given in Eq. (\ref{v:nonren}).
After the Fourier transformation Eq. (\ref{v:gn}) becomes
\begin{equation}
     V_{0}(\sigma)  =  \frac{1}{2\lambda_0}\sigma^2
                   +i\int \frac{d^{D}k}{(2\pi)^{D}}
                   \mbox{ln det}\left(\frac{\ks-\sigma}{\ks}\right)\, .
\label{v:lndet2}
\end{equation}
It should be noted that the effective potential is
normalized so that $V_{0}(0)=0$.
The second term of the right-hand side of Eq. (\ref{v:lndet2}) is
rewritten as.
\begin{eqnarray}
     &&i\int \frac{d^{D}k}{(2\pi)^{D}}
     \mbox{ln det}\left(\frac{\ks-\sigma}{\ks}\right)
                                                            \nonumber \\
     &&=-\frac{1}{2}\int \frac{d^{D}k}{(2\pi)^{D}i}
     \left[\mbox{ln det}\left(\frac{\ks-\sigma}{\ks}\right)
     +\mbox{ln det}\left(\frac{-\ks-\sigma}{-\ks}\right)\right]
                                                            \nonumber \\
     &&=-\frac{1}{2}\mbox{tr} \int \frac{d^{D}k}{(2\pi)^{D}i}
     \mbox{ln}\left(\frac{-k^{2}+\sigma^{2}}{-k^{2}}\right)\, .
\end{eqnarray}
Performing the Wick rotation $k^{0}\rightarrow iK^{0}$ we get
\begin{eqnarray}
     &&=-\frac{1}{2}\mbox{tr}\int \frac{d^{D}K}{(2\pi)^{D}}
     \mbox{ln}\left(1+\frac{\sigma^{2}}{K^{2}}\right)
                                                            \nonumber \\
     &&=-\frac{1}{2}\mbox{tr}\int \frac{d^{D}K}{(2\pi)^{D}}
     \int^{1}_{0}dx\frac{1}{x+K^{2}/\sigma^{2}}
                                                            \nonumber \\
     &&=-\frac{1}{(2\pi)^{D/2}D}
     \Gamma \left(1-\frac{D}{2} \right)\sigma^{D}\, .
\label{eqn:int2gamma}
\end{eqnarray}
Inserting Eq. (\ref{eqn:int2gamma}) into Eq. (\ref{v:lndet})
the effective potential (\ref{v:nonren}) is obtained.

Next we consider the two, three and four dimensional limit of the
effective potential.
Taking the two dimensional limit $D\rightarrow 2$, we get
\begin{equation}
     \frac{V_{0}(\sigma)}{\sigma_{0}^{2}}=
     \frac{1}{2\lambda}\frac{\sigma^2}{{\sigma_{0}}^{2}}
     +\frac{1}{4\pi}\frac{\sigma^2}{{\sigma_{0}}^{2}}
     \left(-3+\mbox{ln}\frac{\sigma^2}{{\sigma_{0}}^{2}}\right)\, .
\end{equation}
where we use the renormalized coupling constant $\lambda$ defined
in Eq. (\ref{eqn:ren2})
with (\ref{eqn:ren}) .
Taking the three dimensional limit $D\rightarrow 3$, we get
\begin{equation}
     \frac{V_{0}(\sigma)}{\sigma_{0}^{3}}=
     \frac{1}{2\lambda}\frac{\sigma^2}{{\sigma_{0}}^{2}}
     -\frac{1}{\sqrt{2}\pi}\left(\frac{\sigma^{2}}{{\sigma_{0}}^{2}}
     -\frac{1}{3}\frac{\sigma^3}{{\sigma_{0}}^{3}}\right)\, .
\end{equation}
If we take the four dimensional limit $D\rightarrow 4$, we find
\begin{eqnarray}
   &&\hspace{-3em}\frac{V_{0}(\sigma)}{\sigma_{0}^{D}} =
     \frac{1}{2\lambda}\frac{\sigma^2}{{\sigma_{0}}^{2}}       \nonumber \\
     &&\hspace{-2em}-\frac{1}{(4\pi)^{2}}\left[
     6\left(\frac{1}{\epsilon_{4}}-\gamma+\mbox{ln}2\pi+\frac{1}{3}\right)
     \frac{\sigma^2}{{\sigma_{0}}^{2}}
     -\left(\frac{1}{\epsilon_{4}}-\gamma+\mbox{ln}2\pi+\frac{3}{2}
     -\mbox{ln}\frac{\sigma^2}{{\sigma_{0}}^{2}}\right)
     \frac{\sigma^4}{{\sigma_{0}}^{4}}
     \right]
					\nonumber \\
   &&\hspace{-2em}+O(\epsilon_4)\, ,
\end{eqnarray}
where
\begin{equation}
     \epsilon_{4}=\frac{4-D}{2}\, .
\end{equation}
To consider the theory in the case of $D=4$ as a low energy effective theory
stemming from the more fundamental theory
we calculate the effective potential using the cut-off regularization
in the case of $D=4$ and compare it with the one
obtained by the dimensional regularization.
We use the renormalization condition (\ref{cond:ren})
and calculate the effective potential
(\ref{v:lndet}) by the cut-off regularization
\begin{eqnarray}
     \frac{V_{0}(\sigma)}{\sigma_{0}^{4}}&=&
     \frac{1}{2\lambda}\frac{\sigma^2}{{\sigma_{0}}^{2}}
     -\frac{1}{(4\pi)^{2}}\left[
     6\left(\mbox{ln}\frac{\Lambda^2}{{\sigma_{0}}^{2}}-\frac{2}{3}\right)
     \frac{\sigma^2}{{\sigma_{0}}^{2}}
     -\left(\mbox{ln}\frac{\Lambda^2}{{\sigma_{0}}^{2}}+\frac{1}{2}
     -\mbox{ln}\frac{\sigma^2}{{\sigma_{0}}^{2}}\right)
     \frac{\sigma^4}{{\sigma_{0}}^{4}}
     \right]
					\nonumber \\
	&& + O\left({\sigma^2 \over \Lambda^2},
		{{\sigma_0}^2 \over \Lambda^2}\right)\, ,
\end{eqnarray}
where $\Lambda$ is the cut-off parameter, which is assumed
to be larger than $\sigma$ and $\sigma_{0}$.
This result show us that there is a correspondence between $\epsilon_{4}$
and $\Lambda$ in the leading order of the $1/N$ expansion,
\begin{equation}
     \frac{1}{\epsilon_{4}}-\gamma+\mbox{ln}2\pi+1\, \leftrightarrow\,
     \ln\,\frac{\Lambda^2}{{\sigma_{0}}^{2}}\, .
\end{equation}

\renewcommand{\thesubsection}{APPENDIX \Alph{subsection}}
\subsection{}
\renewcommand{\thesubsection}{\Alph{subsection}}
\noindent
EFFECTIVE POTENTIAL AT FINITE TEMPERATURE AND CHEMICAL POTENTIAL \\[3mm]
We shall show two types of representations for the effective potential
and the gap equation at finite temperature and density in
this appendix.
One represented by a momentum integration is convenient for numerical
calculations and the other represented by a summation is convenient for
analytical calculations.
In the Feynman functional-integral formalism, thermal Green function
is obtained by the following way.
\begin{eqnarray}
     &&\langle \alpha \mid {\mbox{\large e}}^{-\beta (H-\mu N)}
     \mid \alpha \rangle
     =\langle \alpha , -i\beta \mid \alpha , 0 \rangle         \nonumber \\
     &&=C\int [d\psi][d\bar{\psi}]\ \mbox{exp}\
      \left(i\int^{-i\beta}_{0}dt\int d^{D-1} x ({\cal L}+\mu N)\right)\, ,
\end{eqnarray}
where the number operator $N$ is expressed
\begin{equation}
     N=\int d^{D-1} x\, \bar{\psi} \gamma^{0} \psi \, .
\end{equation}
Following Eq. (\ref{def:gfunc}), we obtain the generating functional at finite
temperature and density.
\begin{equation}
     Z^{\beta\mu}[0]=\int [d\psi][d\bar{\psi}]\ \mbox{exp}\
                    \left(i\int^{-i\beta}_{0}dt\int d^{D-1} x
                    ({\cal L}+\mu N)\right)\, .
\end{equation}
Thus the effective potential (\ref{v:lndet}) is modified as
\begin{equation}
     V(\sigma)=\frac{1}{2\lambda_0}\sigma^2
              -\frac{1}{\beta}\sum^{\infty}_{n=-\infty}
               \int \frac{d^{D-1}k}{(2\pi)^{D-1}}\mbox{ln det}
               \frac{
               (-i w_n\gamma^{0}-i k_{i}\gamma^{i}-\sigma+\mu\gamma^{0})}
               {
               (-i w_n\gamma^{0}-i k_{i}\gamma^{i}+\mu\gamma^{0})}\, ,
\end{equation}
where the roman index $i$ is taken over the space components $(i=1 \sim 3)$.
We calculate the integrand in the second term.
\begin{eqnarray}
     &&\mbox{ln det}
     \frac{[-i(w_n+i\mu)\gamma^{0}-ik_{i}\gamma^{i}-\sigma]}
     {[-i(w_n+i\mu)\gamma^{0}-ik_{i}\gamma^{i}]}
     =\mbox{tr}\int^{\sigma}_{0} ds \frac{1}
     {i(w_n+i\mu)\gamma^{0}+ik_{i}\gamma^{i}+s}                \nonumber \\
     &&=\mbox{tr}\int^{\sigma}_{0} ds \frac{s}
     {(w_n+i\mu)^{2}+k_{i}k^{i}+s^2}                           \nonumber \\
     &&=\frac{1}{2}\mbox{tr ln}
     \frac{(w_n+i\mu)^{2}+k_{i}k^{i}+\sigma^2}
          {(w_n+i\mu)^{2}+k_{i}k^{i}}\, .
\end{eqnarray}
To obtain the second line we use the well-known property that the
gamma matrix is traceless.
Hence we get the effective potential at finite temperature and density.
\begin{equation}
     V(\sigma)=\frac{1}{2\lambda_{0}}\sigma^{2}
     -\frac{1}{2\beta}\sum^{\infty}_{n=-\infty}
     \int \frac{d^{D-1}k}{(2\pi)^{D-1}}
     \mbox{tr ln}
     \frac{(w_n+i\mu)^{2}+k_{i}k^{i}+\sigma^2}
          {(w_n+i\mu)^{2}+k_{i}k^{i}}\, .
\label{v:sumint}
\end{equation}
Next we explain how to deal with the summation and integration in the
effective potential (\ref{v:sumint}).
We can perform a summation in the following way.
We divide the summation into two parts.
\begin{eqnarray}
     &&\sum^{\infty}_{n=-\infty}\mbox{ln}\left[\left(\frac{2n+1}{\beta}+i\mu
     \right)^{2}+b^{2}\right]                      \nonumber \\
     &&=\sum^{\infty}_{n=0}\mbox{ln}\left[\left(\frac{2n+1}{\beta}+i\mu
     \right)^{2}+b^{2}\right]
     +\sum^{\infty}_{n=0}\mbox{ln}\left[\left(\frac{-2n-1}{\beta}+i\mu
     \right)^{2}+b^{2}\right]
                                                        \nonumber \\
     &&=\sum^{\infty}_{n=0}\mbox{ln}\left[
     \left(
     \left(\frac{2n+1}{\beta}\right)^{2}+(b-\mu)^{2}
     \right)
     \left(
     \left(\frac{2n+1}{\beta}\right)^{2}+(b+\mu)^{2}
     \right)
     \right]\, .
\end{eqnarray}
Using the formula
\begin{equation}
     \sum^{\infty}_{n=0}\mbox{ln}\left(a^{2}(2n+1)^{2}+b^{2}\right)
     =\mbox{ln cosh}\frac{\pi b}{2a}\, ,
\end{equation}
we can perform the summation and get
\begin{eqnarray}
     \lefteqn{V(\sigma)=\frac{1}{2\lambda_{0}}\sigma^{2}
     -2^{D/2-1}\int \frac{d^{D-1}k}{(2\pi)^{D-1}}
     (\sqrt{k^{i}k_{i}+\sigma^{2}}-\mid \vec{k} \mid)}
                                                              \nonumber \\
     &&-2^{D/2-1}\frac{1}{\beta}\int \frac{d^{D-1}k}{(2\pi)^{D-1}}
     \left[\mbox{ln}
     \frac{1+\mbox{\large e}^{-\beta(\sqrt{k^{i}k_{i}+\sigma^{2}}+\mu)}}
     {1+\mbox{\large e}^{-\beta(\mid\vec{k}\mid+\mu)}}
     +\mbox{ln}
     \frac{1+\mbox{\large e}^{-\beta(\sqrt{k^{i}k_{i}+\sigma^{2}}-\mu)}}
     {1+\mbox{\large e}^{-\beta(\mid\vec{k}\mid-\mu)}}\right]\, .
\label{v:int}
\end{eqnarray}
We can also integrate over space components in Eq. (\ref{v:sumint})
\begin{eqnarray}
     &&\int \frac{d^{D-1}k}{(2\pi)^{D-1}}
     \mbox{ln}
     \frac{(w_n+i\mu)^{2}+k_{i}k^{i}+\sigma^2}
          {(w_n+i\mu)^{2}+k_{i}k^{i}}
			\nonumber \\
    && =\int \frac{d^{D-1}k}{(2\pi)^{D-1}}
     \int^{1}_{0}dx \frac{\sigma^{2}}{\sigma^{2}x+(w_n+i\mu)^{2}+k_{i}k^{i}}
                                                           \nonumber \\
     &&=-\frac{1}{(4\pi)^{(D-1)/2}}\Gamma\left(\frac{1-D}{2}\right)
							\nonumber \\
     &&\hspace{3em}\times\left[\{(w_n+i\mu)^{2}+\sigma^2\}^{(D-1)/2}
     -\{(w_n+i\mu)^{2}\}^{(D-1)/2}\right]\, .
\end{eqnarray}
Hence we get
\begin{eqnarray}
     V(\sigma)=\frac{1}{2\lambda_{0}}\sigma^{2}
     &+&\frac{1}{\beta}\frac{1}{\sqrt{2}(2\pi)^{(D-1)/2}}
     \Gamma\left(\frac{1-D}{2}\right)                       \nonumber \\
     &&\hspace{-2em}\times\sum^{\infty}_{n=-\infty}
     \left[\{(w_n+i\mu)^{2}+\sigma^2\}^{(D-1)/2}
     -\{(w_n+i\mu)^{2}\}^{(D-1)/2}\right]\, .
\label{v:sum2}
\end{eqnarray}
{}From the two representations
of the effective potential (\ref{v:int}) and (\ref{v:sum2})
we can write the gap equation in the two forms.
\begin{eqnarray}
     \frac{1}{\lambda_0}
     &-&2^{D/2-2}\int \frac{d^{D-1}k}{(2\pi)^{D-1}}
     \frac{1}{\sqrt{k_{i}k^{i}+\sigma^{2}}}                 \nonumber \\
     &&\times\left[
     \frac{1-\mbox{\large e}^{-\beta(\sqrt{k^{i}k_{i}+\sigma^{2}}+\mu)}}
     {1+\mbox{\large e}^{-\beta(\sqrt{k^{i}k_{i}+\sigma^{2}}+\mu)}}
     +\frac{1-\mbox{\large e}^{-\beta(\sqrt{k^{i}k_{i}+\sigma^{2}}-\mu)}}
     {1+\mbox{\large e}^{-\beta(\sqrt{k^{i}k_{i}+\sigma^{2}}-\mu)}}
     \right]=0\, ,
                                                                      \\[1ex]
     \frac{1}{\lambda_0}
     &-&\frac{1}{\beta}\frac{\sqrt{2}}{(2\pi)^{(D-1)/2}}
     \Gamma\left(\frac{3-D}{2}\right)
     \sum^{\infty}_{n=-\infty}
     \{(w_n+i\mu)^{2}+\sigma^2\}^{(D-3)/2}
     =0\, .
\end{eqnarray}

\end{document}